\documentclass[letter,twocolumn]{jpsj3}
\usepackage{txfonts}
\usepackage{bm}
\usepackage{color}
\usepackage{graphicx}
\usepackage{amsmath}
\usepackage{amsmath,amssymb}
\usepackage{mathrsfs}
\def\agt{>\kern -9truept\lower 4.0truept\hbox{$\displaystyle\sim$}}
\def\alt{<\kern -9truept\lower 4.0truept\hbox{$\displaystyle\sim$}}

\title{Polarized Raman Response of Two-Dimensional Quasiperiodic Antiferromagnets:
       Configuration-Interaction versus Green's Function Approaches}

\author{Takashi Inoue and Shoji Yamamoto${}^{*}$}

\inst{Department of Physics, Hokkaido University, Sapporo 060-0810, Japan} 

\abst{
We study Raman response of Heisenberg antiferromagnets on
the $\mathbf{C}_\mathrm{5v}$ Penrose and $\mathbf{C}_\mathrm{8v}$ Ammann-Beenker lattices
within and beyond the Loudon-Fleury second-order perturbation scheme
intending to explore optical features peculiar to quasiperiodic magnets.
Within the Loudon-Fleury mechanism, we find one and only Raman-active mode of $\mathrm{E}_2$
symmetry without any dependence on linear incident and scattered polarizations.
Beyond the Loudon-Fleury mechanism,
two more symmetry species $\mathrm{A}_1$ and $\mathrm{A}_2$ are activated via dynamic
ring exchange and chiral spin fluctuations, respectively, which can be extracted by the use of
circular as well as linear polarizations.
We employ Green's functions on one hand and configuration-interaction wavefunctions
on the other hand to calculate the multimagnon contributions to inelastic light scatterings.
Demonstrating the great advantage of the configuration-interaction scheme,
we reveal that a major portion of the Shastry-Shraiman fourth-order Raman intensity is
mediated by multimagnon fluctuations.
\vspace*{-6mm}}

\begin{document}
\maketitle

   A variety of quasiperiodic magnetic crystals \cite{C106,F308,S476,S054417} have renewed
the theoretical exploration of novel magnetism of geometric origin.
The Penrose and Ammann-Beenker lattices in two dimensions
attract much interest in this context.
Their Ising \cite{O16,K044004} and
Heisenberg \cite{W177205,W104427,J212407,S104427} models
were calculated by the linear spin-wave (SW) theory and Monte Carlo (MC) methods,
whereas their Hubbard models \cite{K214402,K115125}
were investigated within a mean-field approximation.
In the context of inelastic-neutron-scattering experiments,
SW findings for the dynamic structure factor of the Ammann-Beenker Heisenberg antiferromagnet
reveal intriguing excitation features possibly due to the quasiperiodicity \cite{W104427};
the coexistence of linear soft modes near the magnetic Bragg peaks at low frequencies,
self-similar structures at intermediate frequencies, and
flat bands at high frequencies.
We may take further interest in inelastic light scatterings in quasiperiodic
magnets, which strongly reflect their lattice symmetry and potentially bring brandnew information
by virtue of the light polarization degrees of freedom.

   Recent technical progress of manipulating optical lattices \cite{G3363,V110404,S200604} also
deserves special mention.
Two-dimensional (2D) quasiperiodic potentials
with five- \cite{SP053607} or eight-fold \cite{J66003,J149} rotational symmetry were
theoretically designed in terms of standing-wave lasers and indeed observed
via Bragg diffraction \cite{SP053607,V110404}.
By trapping ultracold bosonic or fermionic atoms in optically tunable potentials,
we can change the on-site interaction and tunneling energy to obtain
various effective spin models \cite{D090402}.
There may be an extended-to-localized phase transition \cite{J053609,S200604} of a Bose-Einstein
condensate, for instance.

   Thus motivated, we study Raman responses of 2D quasiperiodic spin-$\frac{1}{2}$ Heisenberg
antiferromagnets on the ${\mathbf{C}}_{\mathrm{5v}}$ Penrose \cite{dB39,dB53} and
${\mathbf{C}}_{\mathrm{8v}}$ Ammann-Beenker \cite{S10519,B8091} lattices, described by
the Hamiltonian
$
   \mathcal{H}
  =J\sum_{\langle i,j \rangle}{\bm S}_{i}\cdot{\bm S}_{j}\ (J>0),
$
where $\sum_{\langle i,j \rangle}$ runs over all pairs of connected vertices.
With their bipartite structures in mind,
we introduce Holstein-Primakoff bosons \cite{H1098} and expand the bosonic Hamiltonian
in powers of the inverse spin magnitude $1/S$,
denoting the terms of order $S^m$ by $\mathcal{H}^{(m)}$.
When we decompose the quartic boson operators $\mathcal{H}^{(0)}$ into quadratic terms
$\mathcal{H}_{\mathrm{BL}}^{(0)}$ and residual normal-ordered quartic interactions
$:\mathcal{H}^{(0)}:$
using Wick's theorem \cite{N034714, Y094412},
the up-to-$O(S^0)$ bosonic Hamiltonian reads
\vspace*{-1mm}
\begin{align}
   \mathcal{H}
  =\mathcal{H}^{(2)}
  +\mathcal{H}^{(1)}
  +\mathcal{H}^{(0)}_{\mathrm{BL}}
  +:\mathcal{H}^{(0)}:
  \equiv
   \mathcal{H}_{\mathrm{BL}}
  +:\mathcal{H}^{(0)}:.
   \label{E:H}
   \\[-8mm]\nonumber
\end{align}
First we diagonalize the bilinear Hamiltonian $\mathcal{H}_{\mathrm{BL}}$
via the generalized Bogoliubov transformation \cite{W104427,W450,SSW},
\vspace*{-1mm}
\begin{align}
   \mathcal{H}_{\mathrm{BL}}
  =\sum_{m=0}^{2}E^{(m)}
  +\sum_{\sigma=\pm}
   \sum_{l_\sigma=1}^{L_\sigma}\varepsilon_{l_\sigma}^\sigma
   \alpha_{l_\sigma}^{\sigma\dagger}\alpha_{l_\sigma}^\sigma;\ 
   \sum_{\sigma=\pm}L_\sigma
  \equiv
   L,
   \label{E:HBLdiag}
   \\[-8mm]\nonumber
\end{align}
where $E^{(2)}$ is the classical ground-state energy and $E^{(m)}$ ($m=1,0$) are the $O(S^m)$
quantum corrections to it, while $\alpha_{l_\sigma}^{\sigma\dagger}$ creates
an antiferromagnetic ($\sigma=+$) or ferromagnetic ($\sigma=-$) SW \cite{Y11033,Y064426}
of energy $\varepsilon_{l_\sigma}^\sigma$
for the vacuum state $|0\rangle_{\mathrm{BL}}$,
enhancing or reducing the ground-state magnetization,
respectively \cite{B3921,Y14008}.
Then we take account of the two-body interactions $:\mathcal{H}^{(0)}:$ in two ways, i.e.,
Green's function (GF) perturbative and configuration-interaction (CI) variational approaches.

   Shastry and Shraiman \cite{S1068,S365} formulated magnetic Raman scatterings in
an antiferromagnetic insulator by perturbatively treating the single-band Hubbard model in
the limit of sufficiently strong correlation $U$ compared to hopping $t$ \cite{SRO}.
With the electron band being half filled, the second-order vertex
${}^{[2]}\mathcal{R}$ results in the well-known Loudon-Fleury operator \cite{F514},
consisting of pair exchange interactions ${\bm S}_{i}\cdot{\bm S}_{j}$, whereas
the fourth-order vertices ${}^{[4]}\mathcal{R}$ contain
three-spin scalar-chirality terms $\propto\bm{S}_{i}\cdot(\bm{S}_{j}\times\bm{S}_{k})$ and
four-spin ring-exchange terms $\propto(\bm{S}_{i}\cdot\bm{S}_{j})(\bm{S}_{k}\cdot\bm{S}_{l})$
\cite{K024414,M184424,I2000118}.
With varying incident photon frequency $\omega_{\mathrm{in}}$,
${}^{[2]}\mathcal{R}$ predominates in the far-resonant regime,
$t \ll |U-\hbar\omega_{\mathrm{in}}|$, while
${}^{[4]}\mathcal{R}$'s are also of major importance in
the near-resonant regime, $t \simeq |U-\hbar\omega_{\mathrm{in}}|$.
The Raman operators are classified according to the number of their constituent spin operators,
$
   {}^{[2n]}\mathcal{R}
  =\sum_{\tau=2}^{2n}
   {}_{\;\;\;\;\tau}^{[2n]}\mathcal{R}
$
$(n=1,2)$,
and each component can be expanded in $1/S$ in terms of the bosonic language,
\vspace*{-1mm}
\begin{align}
   {}_{\;\;\;\;\tau}^{[2n]}\mathcal{R}
  =\sum_{m=0}^\infty
   {}_{\;\;\;\;\tau}^{[2n]}\mathcal{R}^{(\tau-m)}
  =\sum_{m=0}^\infty\sum_{l=0}^m
   {}_{\;\;\;\;\tau}^{[2n]}\mathcal{R}_{2l\mathrm{M}}^{(\tau-m)},
   \label{E:Rfull}
   \\[-8mm]\nonumber
\end{align}
where ${}_{\;\;\;\;\tau}^{[2n]}\mathcal{R}^{(\tau-m)}$, consisting of
$2l\,(0\leq l\leq m)$-magnon ($2l\mathrm{M}$) vertices, is of order $S^{\tau-m}$.
Via the Bogoliubov transformation, \eqref{E:Rfull} truncated at $m=2$, i.e.,
the up-to-$O(S^0)$ vertices become
\vspace*{-1mm}
\begin{align}
   \mathcal{R}
  \simeq
   \sum_{n=1}^{1\,\mathrm{or}\,2}
   \sum_{l=0}^{2}
   {}^{[2n]}\mathcal{R}_{2l\mathrm{M}}
  =\sum_{n=1}^{1\,\mathrm{or}\,2}
   \sum_{\tau=2}^{2n}
   \sum_{l=0}^{2}
   \sum_{m=l}^2
   {}_{\;\;\;\;\tau}^{[2n]}\mathcal{R}_{2l\mathrm{M}}^{(\tau-m)},
   \label{E:RuptoO(S^0)}
   \\[-8mm]\nonumber
\end{align}
where
${}_{\;\;\;\;\tau}^{[2n]}\mathcal{R}_{2l\mathrm{M}}^{(\tau-m)}$ are normal-ordered
with respect to the quasiparticle magnon operators.
${}^{[2n]}\mathcal{R}_{0\mathrm{M}}$ merely contribute to elastic (Rayleigh) scatterings and
are thus omitted hereafter.

\begin{figure*}[ht]
\centering
\includegraphics[width=160mm]{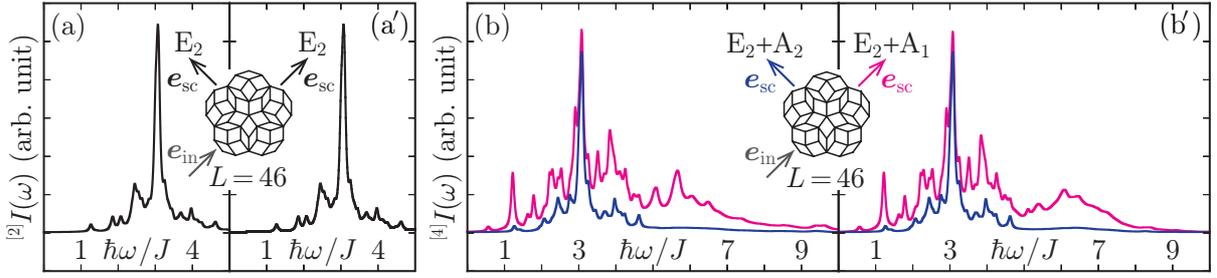}
\vspace*{-3mm}
\caption{
         GF calculations of the Raman intensities
         ${}^{[p]}I(\omega) \equiv \sum_{l=1}^2 {}^{[p]}I_{2l\mathrm{M}}(\omega)$
         for the $L=46$ 2D Penrose lattice of ${\mathbf{C}}_{5\mathrm{v}}$
         point symmetry in the Loudon-Fleury second-order ($p=2$) [(a) and (a$'$)] and
         Shastry-Shraiman fourth-order ($p=4$) [(b) and (b$'$)] perturbation schemes.
         ${}^{[p]}I_{2\mathrm{M}}(\omega)$ is obtained from \eqref{E:1to4MGF(t)b} \cite{SGF},
         while ${}^{[p]}I_{4\mathrm{M}}(\omega)$ is calculated in two ways, by the use of
         \eqref{E:4MGF22} \cite{SGF} [(a) and (b)] and
         \eqref{E:4MGF31} \cite{SGF} [(a$'$) and (b$'$)].
         Two combinations of linear incident and scattered polarizations,
         $\phi_-=0,\pi/2$, are simulated.
         The Shastry-Shraiman perturbation parameter $t/(U-\hbar\omega_{\mathrm{in}})$ is set
         equal to $1/9$ and $9/10$ within [(a) and (a$'$)] and beyond [(b) and (b$'$)]
         the Loudon-Fleury scheme, respectively.
         Every spectral line is Lorentzian-broadened by a width of $0.1J$.}
\label{F:PenroseGF}
\vspace*{-8.5mm}
\end{figure*}
\noindent
Putting
$
   \mathcal{R}(t)
  \equiv
   e^{ i\mathcal{H}t/\hbar}
   \mathcal{R}
   e^{-i\mathcal{H}t/\hbar}
$
for any operator $\mathcal{R}$,
we define the $2l\mathrm{M}$-mediated Raman intensities at absolute zero as
\cite{GSRamanIntensity}
\vspace*{-1mm}
\begin{align}
   \int_{-\infty}^\infty
   \frac{dt\,e^{i\omega t}}{2\pi\hbar L}
   \sum_{n,n'=1}^{p/2}
   \langle 0|
    {}^{[2n ]}\mathcal{R}_{2l\mathrm{M}}^{\dagger}(t)\,
    {}^{[2n']}\mathcal{R}_{2l\mathrm{M}}
   |0\rangle
  \equiv
   {}^{[p]}I_{2l\mathrm{M}}(\omega).
   \label{E:I(w)2lM}
   \\[-8mm]\nonumber
\end{align}
Once we find the exact ground state of the Heisenberg Hamiltonian,
$\mathcal{H}|0\rangle=E_0|0\rangle$, the Loudon-Fleury ($p=2$) and Shastry-Shraiman ($p=4$)
intensities can be exactly evaluated as \cite{L15295,E1468,N553,S8478}
\vspace*{-5mm}
\begin{align}
   &
   {}^{[p]}I(\omega)
  =\int_{-\infty}^\infty
   \frac{dt\,e^{i\omega t}}{2\pi\hbar L}
   \sum_{n,n'=1}^{p/2}
   \langle 0|
    {}^{[2n ]}\mathcal{R}^\dagger(t)\,
    {}^{[2n']}\mathcal{R}
   |0\rangle
  =\frac{-1}{\pi L}
   \allowdisplaybreaks
   \nonumber \\[-3mm]
   &\quad\times
   \sum_{n,n'=1}^{p/2}
   \mathop{\mathrm{lim}}_{\eta\rightarrow +0}
   \mathrm{Im}
   \langle 0|
    {}^{[2n ]}\mathcal{R}^\dagger
    \frac{1}{\hbar\omega+E_{0}+i\eta-\mathcal{H}}
    {}^{[2n']}\mathcal{R}
   |0\rangle
   \label{E:[p]I(w)}
   \\[-9mm]\nonumber
\end{align}
and may be approximated by the sum
$
   \sum_{l=1}^2
   {}^{[p]}I_{2l\mathrm{M}}(\omega)
$.

   Magnon-magnon interactions significantly modify the Raman spectra
\cite{S8478,C7127,C9760,P174412}, because pair-exchange and multiple-spin cyclic-exchange
Raman vertices, emergent within and beyond the Loudon-Fleury scheme, respectively,
play qualitatively different roles in inelastic photon scatterings.
First we demonstrate this by a renormalized perturbation theory \cite{SGF}.
In terms of the magnon GFs \eqref{E:1to4MGF(t)b} and \eqref{E:1to4MGF(t)d} \cite{SGF},
the $2l\mathrm{M}$-mediated Raman scattering intensities are calculated as \cite{SRO}
\vspace*{-1mm}
\begin{align}
   &\!\!\!
   {}^{[p]}I_{2l\mathrm{M}}(\omega)
 =-\mathrm{Im}
   \int_{-\infty}^{\infty}
   \frac{dt\,e^{i\omega t}}{i\pi\hbar L}
   \sum_{n,n'=1}^{p/2}
   \langle 0|\mathcal{T}\,
    {}^{[2n ]}\mathcal{R}_{2l\mathrm{M}}^{\dagger}(t)\,
    {}^{[2n']}\mathcal{R}_{2l\mathrm{M}}
   |0\rangle
   \allowdisplaybreaks
   \nonumber \\[-2mm]
   &\!\!\!\qquad\quad\ \ \ 
  \equiv
  -\mathrm{Im}
   \int_{-\infty}^{\infty}
   \frac{dt\,e^{i\omega t}}{\pi\hbar L}
   {}^{[p]}\mathcal{G}_{2l\mathrm{M}}(t);
   \allowdisplaybreaks
   \nonumber \\
   &\!\!\!
   {}^{[p]}\mathcal{G}_{2\mathrm{M}}(t)
  =\sum_{k_{+},k_{-}'}\sum_{l_{+},l_{-}'}
   {}^{[p]}W_{k_{+}k_{-}'}^{(2)\,*}
   G^{k_{+}k_{-}'}_{l_{+}l_{-}'}(t)\,
   {}^{[p]}W_{l_{+}l_{-}'}^{(2)},\ 
   {}^{[p]}\mathcal{G}_{4\mathrm{M}}(t)
   \allowdisplaybreaks
   \nonumber \\[-2mm]
   &\!\!\!\ \ 
  =\sum_{k_{+},k_{+}',k_{-}'',k_{-}'''}
   \sum_{l_{+},l_{+}',l_{-}'',l_{-}'''}
   {}^{[p]}X_{k_{+}k_{+}'k_{-}''k_{-}'''}^{(7)\,*}
   G^{k_{+}k_{+}'k_{-}''k_{-}'''}_{l_{+}l_{+}'l_{-}''l_{-}'''}(t)\,
   {}^{[p]}X_{l_{+}l_{+}'l_{-}''l_{-}'''}^{(7)},
   \label{E:[p]I2lM(w)GF}
   \\[-8mm]\nonumber
\end{align}
where we numerically obtain the coefficients
${}^{[p]}W_{l_{+}l_{-}'}^{(2)}$ and
${}^{[p]}X_{l_{+}l_{+}'l_{-}''l_{-}'''}^{(7)}$.
Since any perturbative renormalization is hardly tractable for more-than-3M GFs,
we decompose 4M GFs into 2M GFs as \eqref{E:4MGFsimeq2+2} \cite{SGF} on one hand
and into 3M and 1M GFs as \eqref{E:4MGFsimeq3+1} \cite{SGF} on the other hand.
The renormalized 1M GFs reduce to the Hartree-Fock solutions \eqref{E:Dyson} \cite{SGF},
whereas the 2M and 3M ones are calculated through ladder-approximation Bethe-Salpeter equations
\eqref{E:2lBS} \cite{SGF} and \eqref{E:3lBS} \cite{SGF,C054326}, respectively.
\begin{figure*}[t]
\centering
\includegraphics[width=160mm]{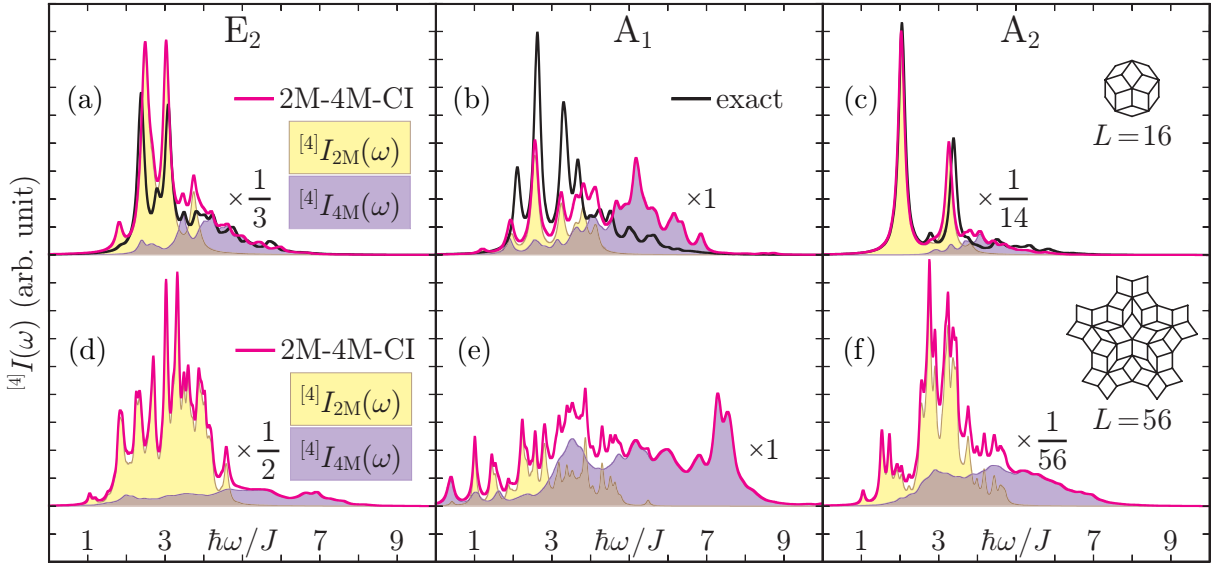}
\vspace*{-3mm}
\caption{
         2M-4M-CI calculations of the Shastry-Shraiman fourth-order Raman intensities
         ${}^{[4]}I(\omega) \equiv \sum_{l=1}^2 {}^{[4]}I_{2l\mathrm{M}}(\omega)$ for
         the $L=16$ [(a) to (c)] and $L=56$ [(d) to (f)]
         2D Penrose lattice of ${\mathbf{C}}_{5\mathrm{v}}$ point symmetry,
         the above three of which are compared with the exact solutions,
         where the perturbation parameter $t/(U-\hbar\omega_{\mathrm{in}})$ is set to $9/10$
         and every spectral line is Lorentzian-broadened by a width of $0.1J$.
         The pure symmetry components are extracted from three polarization combinations,
         \eqref{E:[4]I(w)LP} with $\phi_-=0,\pi/2$ and
         \eqref{E:[p]I(w)CP} with $\sigma_{\mathrm{in}}\sigma_{\mathrm{sc}}=-1$.
         All the SW calculations of $I_{2l\mathrm{M}}(\omega)$ each
         are distinguishably colored.}
\label{F:PenroseCI}
\vspace*{-9mm}
\end{figure*}

   The Raman operator is written as a rank-$2$ tensor dotted with the polarization vectors of
the incident ($\bm{e}_{\mathrm{in}}$) and scattered ($\bm{e}_{\mathrm{sc}}$) photons
\cite{P174412,P094439},
\vspace*{-2mm}
\begin{align}
   {}^{[p]}\mathcal{R}
  =\sum_{\mu,\nu=x,y}
   e_{\mathrm{in}}^{\mu}
   {}^{[p]}\mathcal{R}^{\mu\nu}
   e_{\mathrm{sc}}^{\nu\,*},
   \label{E:[p]R_munu}
   \\[-8mm]\nonumber
\end{align}
where we set $\bm{e}_{\mathrm{in}}$ and $\bm{e}_{\mathrm{sc}}$ parallel to the lattice plane
($e_{\mathrm{in}}^z=e_{\mathrm{sc}}^z=0$) to reduce ${}^{[p]}\mathcal{R}$ to a $2\times 2$
matrix.
In terms of the point symmetry group $\mathbf{P}$ of the lattice, this is rewritten as
\cite{K214411,Y063701}
\vspace*{-2mm}
\begin{align}
   {}^{[p]}\mathcal{R}
  =\mathop{{\sum_i}'}\sum_{\mu=1}^{d_{\varXi_{i}}^{\mathbf{P}}}
   E_{\varXi_i:\mu}^{\mathbf{P}}\,
   {}^{[p]}\mathcal{R}_{\varXi_i:\mu}^{\mathbf{P}},
   \label{E:[p]R_Xi}
   \\[-9mm]\nonumber
\end{align}
where $\sum_{i}'$ runs over the Raman-active irreducible representations $\varXi_i$ of
$\mathbf{P}$, each with dimensionality $d_{\varXi_{i}}^{\mathbf{P}}$, and
$E_{\varXi_i:\mu}^{\mathbf{P}}$ and $\mathcal{R}_{\varXi_i:\mu}^{\mathbf{P}}$
are the $\mu$th polarization-vector basis function and Raman vertex for
$\varXi_i$, respectively.
Since the ground state $|0\rangle$ is invariant under every symmetry operation of $\mathbf{P}$,
any expectation value between Raman vertices of different symmetry species for it goes to zero
\cite{P094439,K214411,Y063701,D175}.
We can therefore classify the Raman intensities as to symmetry species,
\vspace*{-1.5mm}
\begin{align}
   \int_{-\infty}^\infty
   \frac{dt\,e^{i\omega t}}{2\pi\hbar L}
   \sum_{n,n'=1}^{p/2}
   \langle 0|
    {}^{[2n ]}\mathcal{R}_{\varXi_i:\mu}^{\mathbf{P}\,\dagger}(t)\,
    {}^{[2n']}\mathcal{R}_{\varXi_i:\mu}^{\mathbf{P}}
   |0\rangle
  \equiv
   {}^{[p]}I_{\varXi_i:\mu}^{\mathbf{P}}(\omega).
   \label{E:[p]I_Xi:mu^P(w)}
   \\[-8mm]\nonumber
\end{align}
Considering that
${}^{[p]}I_{\varXi_i:1  }^{\mathbf{P}}(\omega)
={}^{[p]}I_{\varXi_i:\mu}^{\mathbf{P}}(\omega)$
$(\mu=2,\cdots,d_{\varXi_{i}}^{\mathbf{P}})$ for any multidimensional representation
$\varXi_i$ \cite{P094439,K214411}, we find
\vspace*{-1mm}
\begin{align}
   {}^{[p]}I(\omega)
  =\mathop{{\sum_i}'}
   {}^{[p]}I_{\varXi_{i}:1}^{\mathbf{P}}(\omega)
   \sum_{\mu=1}^{d_{\varXi_{i}}^{\mathbf{P}}}
   \bigl|E_{\varXi_{i}:\mu}^{\mathbf{P}}\bigr|^{2}.
   \label{E:sum_i&mu[p]I_Xi:mu^P(w)}
   \\[-8mm]\nonumber
\end{align}

   When $\mathbf{P}=\mathbf{C}_{\mathrm{5v}}$,
\eqref{E:[p]R_Xi} contains two 1D and one 2D symmetry species \cite{SIRD},
whose basis functions and vertices are given by
\vspace*{-1mm}
\begin{align}
   &
   E_{\mathrm{A}_{1}:1}^{\mathbf{C}_{\mathrm{5v}}}
  =e_{\mathrm{in}}^{x}e_{\mathrm{sc}}^{x\,*}
  +e_{\mathrm{in}}^{y}e_{\mathrm{sc}}^{y\,*},\ 
   E_{\mathrm{A}_{2}:1}^{\mathbf{C}_{\mathrm{5v}}}
  =e_{\mathrm{in}}^{x}e_{\mathrm{sc}}^{y\,*}
  -e_{\mathrm{in}}^{y}e_{\mathrm{sc}}^{x\,*},
   \allowdisplaybreaks
   \nonumber \\
   &
   E_{\mathrm{E}_{2}:1}^{\mathbf{C}_{\mathrm{5v}}}
  =e_{\mathrm{in}}^{x}e_{\mathrm{sc}}^{y\,*}
  +e_{\mathrm{in}}^{y}e_{\mathrm{sc}}^{x\,*},\ 
   E_{\mathrm{E}_{2}:2}^{\mathbf{C}_{\mathrm{5v}}}
  =e_{\mathrm{in}}^{x}e_{\mathrm{sc}}^{x\,*}
  -e_{\mathrm{in}}^{y}e_{\mathrm{sc}}^{y\,*},
   \label{E:PSBF}
   \allowdisplaybreaks
   \\
   &
   {}^{[p]}\mathcal{R}_{\mathrm{A}_{1}:1}^{\mathbf{C}_{\mathrm{5v}}}
  =\frac{{}^{[p]}\mathcal{R}^{xx}+{}^{[p]}\mathcal{R}^{yy}}{2},\ 
   {}^{[p]}\mathcal{R}_{\mathrm{A}_{2}:1}^{\mathbf{C}_{\mathrm{5v}}}
  =\frac{{}^{[p]}\mathcal{R}^{xy}-{}^{[p]}\mathcal{R}^{yx}}{2},
   \allowdisplaybreaks
   \nonumber \\
   &
   {}^{[p]}\mathcal{R}_{\mathrm{E}_{2}:1}^{\mathbf{C}_{\mathrm{5v}}}
  =\frac{{}^{[p]}\mathcal{R}^{xy}+{}^{[p]}\mathcal{R}^{yx}}{2},\ 
   {}^{[p]}\mathcal{R}_{\mathrm{E}_{2}:2}^{\mathbf{C}_{\mathrm{5v}}}
  =\frac{{}^{[p]}\mathcal{R}^{xx}-{}^{[p]}\mathcal{R}^{yy}}{2}.
   \label{E:PSRV}
   \\[-8mm]\nonumber
\end{align}
For the linear polarizations
$\bm{e}_{\mathrm{in}/\mathrm{sc}}
=(\cos\phi_{\mathrm{in}/\mathrm{sc}},\sin\phi_{\mathrm{in}/\mathrm{sc}},0)$,
\eqref{E:PSBF} reads
$
   E_{\mathrm{A}_{1}:1}^{\mathbf{C}_{\mathrm{5v}}}
  =\cos\phi_-
$,
$
   E_{\mathrm{A}_{2}:1}^{\mathbf{C}_{\mathrm{5v}}}
  =\sin\phi_-
$,
$
   E_{\mathrm{E}_{2}:1}^{\mathbf{C}_{\mathrm{5v}}}
  =\sin\phi_+
$, and
$
   E_{\mathrm{E}_{2}:2}^{\mathbf{C}_{\mathrm{5v}}}
  =\cos\phi_+
$
with $\phi_\pm\equiv\phi_{\mathrm{sc}}\pm\phi_{\mathrm{in}}$.
Since ${}^{[2]}\mathcal{R}_{\mathrm{A}_{1}:1}^{\mathbf{C}_{\mathrm{5v}}}$ commutes with
the Heisenberg Hamiltonian,
the $\mathrm{A}_1$ species is Raman inactive within the Loudon-Fleury scheme.
The $\mathrm{A}_2$ species is also Loudon-Fleury-Raman inactive.
Since the Raman operator is time-reversal-invariant \cite{S365},
the $(\bm{e}_{\mathrm{in}}\leftrightarrow\bm{e}_{\mathrm{sc}}^{*})$-exchange-antisymmetric
basis function $E_{\mathrm{A}_{2}:1}^{\mathbf{C}_{\mathrm{5v}}}$ demands that
${}^{[p]}\mathcal{R}_{\mathrm{A}_{2}:1}^{\mathbf{C}_{\mathrm{5v}}}$ be also
time-reversal-antisymmetric.
The second-order pair-exchange Raman vertices are all time-reversal-invariant.
${}^{[2]}\mathcal{R}$ for the 2D Penrose lattice thus consists of one and only Raman-active
species ${\mathrm{E}_{2}}$ to yield depolarized spectra,
\vspace*{-1mm}
\begin{align}
   {}^{[2]}I(\omega)
  =\left(
     \bigl|E_{\mathrm{E}_{2}:1}^{\mathbf{C}_{\mathrm{5v}}}\bigr|^{2}
    +\bigl|E_{\mathrm{E}_{2}:2}^{\mathbf{C}_{\mathrm{5v}}}\bigr|^{2}
   \right)\,
   {}^{[2]}I_{\mathrm{E}_{2}:1}^{\mathbf{C}_{\mathrm{5v}}}(\omega)
     ={}^{[2]}I_{\mathrm{E}_{2}:1}^{\mathbf{C}_{\mathrm{5v}}}(\omega),
   \label{E:[2]I(w)LP}
   \\[-8mm]\nonumber
\end{align}
as is shown in Figs. \ref{F:PenroseGF}(a) and \ref{F:PenroseGF}(a$'$).
This is no longer the case with higher-order Raman vertices.
${}^{[4]}\mathcal{R}_{\mathrm{A}_{1}:1}^{\mathbf{C}_{\mathrm{5v}}}$ contains ring-exchange
spin fluctuations $(\bm{S}_{i}\cdot\bm{S}_{j})(\bm{S}_{k}\cdot\bm{S}_{l})$ incommutable with
the Heisenberg Hamiltonian, whereas
${}^{[4]}\mathcal{R}_{\mathrm{A}_{2}:1}^{\mathbf{C}_{\mathrm{5v}}}$ comprises
chiral spin fluctuations ${\bm S}_{i}\cdot({\bm S}_{j}\times{\bm S}_{k})$ breaking
the time-reversal symmetry,
both of which drive the fourth-order Raman response to depend on the light polarization,
\begin{align}
   {}^{[4]}I(\omega)
  ={}^{[4]}I_{\mathrm{E}_{2}:1}^{\mathbf{C}_{\mathrm{5v}}}(\omega)
  +{}^{[4]}I_{\mathrm{A}_{1}:1}^{\mathbf{C}_{\mathrm{5v}}}(\omega)
   \cos^{2}\phi_-
  +{}^{[4]}I_{\mathrm{A}_{2}:1}^{\mathbf{C}_{\mathrm{5v}}}(\omega)
   \sin^{2}\phi_-,
   \label{E:[4]I(w)LP}
   \\[-9.5mm]\nonumber
\end{align}
as is demonstrated in Figs. \ref{F:PenroseGF}(b) and \ref{F:PenroseGF}(b$'$).
\begin{figure*}[htb]
\centering
\includegraphics[width=160mm]{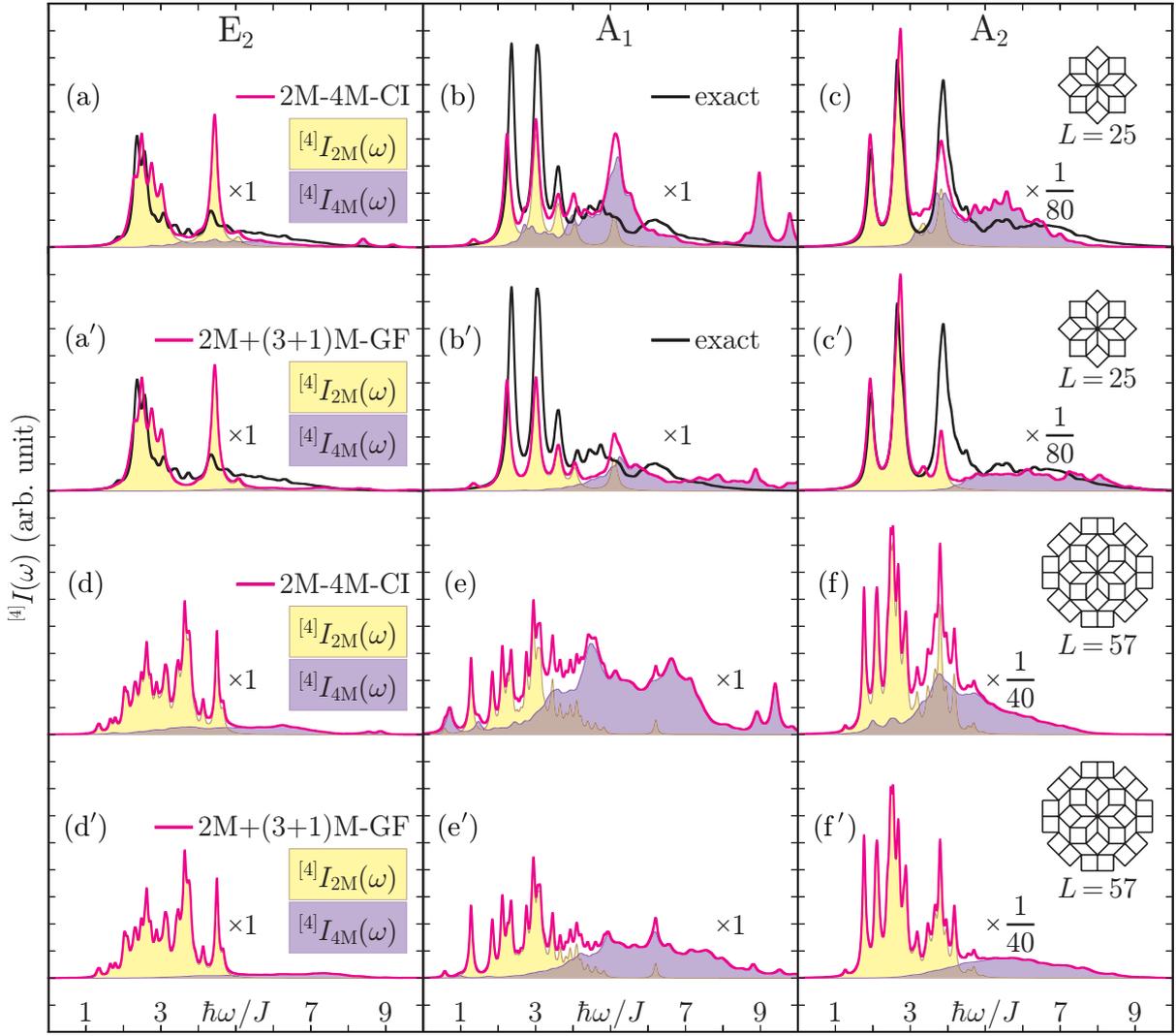}
\vspace*{-3mm}
\caption{
         2M-4M-CI [(a) to (f)] and
         $2\mathrm{M}+4\mathrm{M}$ [approximated by \eqref{E:4MGF31}]-GF [(a$'$) to (f$'$)]
         calculations of ${}^{[4]}I(\omega) \equiv \sum_{l=1}^2 {}^{[4]}I_{2l\mathrm{M}}(\omega)$
         for the $L=25$ and $L=57$ 2D Ammann-Beenker lattice of
         ${\mathbf{C}}_{8\mathrm{v}}$ point symmetry, the above six of which are compared with
         the exact solutions.
         All other details are the same as Fig. \ref{F:PenroseCI}.}
\label{F:AmmannBeenkerCIvsGF}
\vspace*{-9mm}
\end{figure*}

   Figures \ref{F:PenroseGF}(a) and \ref{F:PenroseGF}(a$'$) present almost the same
observations, while Figs. \ref{F:PenroseGF}(b) and \ref{F:PenroseGF}(b$'$) show artificial
differences at $\hbar\omega \gtrsim 5J$.
Since we cannot directly evaluate more-than-3M GFs in practice,
${}^{[4]}I(\omega)$ and more generally Raman responses beyond the Loudon-Fleury scheme,
containing significant multimagnon contributions, are much harder to reliably calculate
in terms of GFs than ${}^{[2]}I(\omega)$ obtainable from the established Bethe-Salpeter
equation.
Even though we can calculate 3M GFs, for instance, there may be some different Bethe-Salpeter-like
manners of renormalization \cite{SGF,B052503} to complicate matters further.
In order to reliably evaluate the role of multimagnon scatterings in novel Raman responses beyond
the Loudon-Fleury scheme, we propose an alternative approach utilizing CI variational
wavefunctions \cite{SCI}.
Once we proceed to the fourth-order perturbation scheme,
the polarized Raman spectra \eqref{E:[4]I(w)LP} for linearly polarized components of the incident
and scattered field amplitudes are inadequate to identify all the three Raman-active symmetry
species, and therefore, we consider circularly polarized components of them as well
\cite{S3864,L063530}.
For the incident and scattered fields of circular polarizations,
$\bm{e}_{\mathrm{in}/\mathrm{sc}}
=(1,\sigma_{\mathrm{in}/\mathrm{sc}}i,0)/\sqrt{2}$,
\eqref{E:PSBF} reads
$
   E_{\mathrm{A}_{1}:1}^{\mathbf{C}_{\mathrm{5v}}}
  =(1+\sigma_{\mathrm{in}}\sigma_{\mathrm{sc}})/2
$,
$
   E_{\mathrm{A}_{2}:1}^{\mathbf{C}_{\mathrm{5v}}}
  =-i(\sigma_{\mathrm{in}}+\sigma_{\mathrm{sc}})/2
$,
$
   E_{\mathrm{E}_{2}:1}^{\mathbf{C}_{\mathrm{5v}}}
  =i(\sigma_{\mathrm{in}}-\sigma_{\mathrm{sc}})/2
$, and
$
   E_{\mathrm{E}_{2}:2}^{\mathbf{C}_{\mathrm{5v}}}
  =(1-\sigma_{\mathrm{in}}\sigma_{\mathrm{sc}})/2
$
to yield the second- and fourth-order Raman responses
\vspace*{-1mm}
\begin{align}
   &
   {}^{[2]}I(\omega)
  =(1-\sigma_{\mathrm{in}}\sigma_{\mathrm{sc}})
   \,{}^{[2]}I_{\mathrm{E}_{2}:1}^{\mathbf{C}_{\mathrm{5v}}}(\omega),\ 
   {}^{[4]}I(\omega)
  =(1-\sigma_{\mathrm{in}}\sigma_{\mathrm{sc}})
   \nonumber
   \allowdisplaybreaks \\
   &\quad\times
   {}^{[4]}I_{\mathrm{E}_{2}:1}^{\mathbf{C}_{\mathrm{5v}}}(\omega)
  +\frac{1+\sigma_{\mathrm{in}}\sigma_{\mathrm{sc}}}{2}
   \left[
    {}^{[4]}I_{\mathrm{A}_{1}:1}^{\mathbf{C}_{\mathrm{5v}}}(\omega)
   +{}^{[4]}I_{\mathrm{A}_{2}:1}^{\mathbf{C}_{\mathrm{5v}}}(\omega)
   \right].
   \label{E:[p]I(w)CP}
   \\[-8mm]\nonumber
\end{align}
Substituting \eqref{E:[p]I(w)CP} with $\sigma_{\mathrm{in}}\sigma_{\mathrm{sc}}=-1$ into
\eqref{E:[4]I(w)LP} with $\phi_-=0$ or $\phi_-=\pi/2$ reveals all the symmetry species separately.

   A CI wavefunction generally consists of a linear combination of Hartree-Fock Slater
determinants \cite{O045122,Y235205}, including a certain set of quasiparticle excited states
as well as the ground state $|0\rangle_{\mathrm{BL}}$.
In our 2M-4M-CI scheme, the Hilbert space in which the bosonic Hamiltonian \eqref{E:H} operates
is spanned by the up-to-4M $N_{\mathrm{CI}}$ basis states
\eqref{E:CIBasisSets0M}--\eqref{E:CIBasisSets4M} \cite{SCI}.
In terms of the variationally corrected eigenstates and eigenvalues of the 2M-4M-CI Hamiltonian
\eqref{E:CIHV(1)V(4)V(9)} \cite{SCI}, the $2l\mathrm{M}$-mediated Raman scattering intensities
read
\vspace*{-2mm}
\begin{align}
   \!\!\!
   {}^{[p]}I_{2l\mathrm{M}}(\omega)
  =\frac{1}{L}\!\sum_{\nu=0}^{N_{\mathrm{CI}}-1}\!
   \left|
   \langle\nu|
    \sum_{n=1}^{p/2}
    {}^{[2n]}\mathcal{R}_{2l\mathrm{M}}
   |0\rangle
   \right|^{2}\!
   \delta(\hbar\omega-\varepsilon_{\nu}+\varepsilon_{0}).
   \label{E:[p]I2lM(w)CI}
   \\[-8mm]\nonumber
\end{align}

   We verify the CI evaluation \eqref{E:[p]I2lM(w)CI} in Fig. \ref{F:PenroseCI}.
The CI findings for the $\mathrm{E}_2$ and $\mathrm{A}_2$ symmetry species are in very good
agreement with the exact solutions obtained by a recursion method based on the Lanczos
algorithm \cite{VandMrecursion,Y224}.
While the $\mathrm{E}_2$ scattering is Loudon-Fleury-Raman active and arises chiefly from
pair-exchange spin fluctuations, a nonnegligible portion of this scattering intensity is
mediated by 4M fluctuations, as is revealed by our CI scheme.
While the $\mathrm{A}_2$ scattering intensity is small compared to the other symmetries,
it is so interesting as to allow for directly observing the chiral spin fluctuations
$\bm{S}_{i}\cdot(\bm{S}_{j}\times\bm{S}_{k})$ \cite{S3864}.
The spin-chirality-driven $\mathrm{A}_2$ Raman response is emergent in the honeycomb and kagome
lattices consisting of nonparallelograms but impossible in the square and triangular lattices
comprising rhombuses \cite{K024414}.
The present scattering of $\mathrm{A}_2$ symmetry owes to the quasiperiodic geometry \cite{L313}
whose rank $D$, i.e., the smallest number of wavevectors that can span the whole diffraction
pattern of the crystal by their integral linear combinations, is larger than the actual physical
dimension $d$.
The $d=2$ Penrose and Ammann-Beenker lattices have the same indexing dimension $D=4$.
Our CI scheme well reproduces the $\mathrm{A}_1$ exact solution as well but somewhat overestimates
its 4M spectral weight.
This is because the $\mathrm{A}_1$ symmetry species becomes Raman active due to ring exchange
interactions such as
$S_i^+S_j^-S_k^+S_l^-$, $S_i^+S_j^-S_k^zS_l^z$, and $S_i^zS_j^zS_k^zS_l^z$,
the latter two of which are essentially described by six or more Holstein-Primakoff bosons
and therefore sensitive to the Hamiltonian of $O(S^{-1})$.

   What will happen to clusters without ${\mathbf{C}}_{5\mathrm{v}}$ symmetry?
The $\mathrm{A}_1$ and $\mathrm{A}_2$ symmetry species remain Loudon-Fleury-Raman inactive,
considering that
${}^{[2]}\mathcal{R}_{\mathrm{A}_{1}:1}^{\mathbf{C}_{\mathrm{5v}}}$ is commutable with
the Heisenberg Hamiltonian and
${}^{[2]}\mathcal{R}_{\mathrm{A}_{2}:1}^{\mathbf{C}_{\mathrm{5v}}}$ is time-reversal-invariant
regardless of the size and shape of clusters,
whereas the Loudon-Fleury-Raman response
$
   {}^{[2]}I(\omega)
  =\sum_{\mu=1}^2
   \bigl|E_{\mathrm{E}_{2}:\mu}^{\mathbf{P}}\bigr|^{2}
   {}^{[2]}I_{\mathrm{E}_{2}:\mu}^{\mathbf{P}}(\omega)
$
is no longer perfectly depolarized without the equality
${}^{[p]}I_{\mathrm{E}_{2}:1}^{\mathbf{P}}(\omega)
={}^{[p]}I_{\mathrm{E}_{2}:2}^{\mathbf{P}}(\omega)$.
However, the polarization dependence disappears in the thermodynamic limit and it is already faint
in such medium-sized clusters of $L\gtrsim 50$.
Size dependence of the Raman spectra is further demonstrated in Supplemental Material
\cite{SupplPenroseRaman}.

   Figure \ref{F:AmmannBeenkerCIvsGF} shows comparative CI and GF calculations of
the 2D Ammann-Beenker lattice.
Even though we set $\mathbf{P}$ to $\mathbf{C}_{\mathrm{8v}}$ in \eqref{E:sum_i&mu[p]I_Xi:mu^P(w)},
the second- and fourth-order Raman responses remain the same as
\eqref{E:[2]I(w)LP}, \eqref{E:[4]I(w)LP} and \eqref{E:[p]I(w)CP}.
Depolarization of ${}^{[2]}I(\omega)$ occurs in a certain class of periodic planar magnets as well,
including Heisenberg antiferromagnets on the triangular \cite{P174412} and kagome
\cite{K024414,C172406} lattices and Kitaev spin liquids on the pure \cite{K187201} and decorated
\cite{Y063701} honeycomb lattices.
In these lattice geometries,
the two polarization-vector basis functions
$e_{\mathrm{in}}^{x}e_{\mathrm{sc}}^{y\,*}+e_{\mathrm{in}}^{y}e_{\mathrm{sc}}^{x\,*}$
and
$e_{\mathrm{in}}^{x}e_{\mathrm{sc}}^{x\,*}-e_{\mathrm{in}}^{y}e_{\mathrm{sc}}^{y\,*}$
span such a 2D irreducible representation as to be one and only Raman-active symmetry species.
This criterion is met by 2D lattices with certain rotational symmetry \cite{SIRD},
including periodic ones of triangular geometry and
all noncrystallographic---in the conventional sense---ones \cite{S1951,S79}.

   Figure \ref{F:AmmannBeenkerCIvsGF} reveals our CI scheme to be much superior to
the conventional GF approach.
Indirect evaluation of the 4M GFs cannot reproduce the significant 4M-mediated scattering
intensity in general.
The spin-chirality-driven $\mathrm{A}_2$ scattering intensity is especially misunderstood by
the GF approach.
Indeed low-energy peaks are almost of 2M character, but high-energy ones clearly owe to
both 2M and 4M scatterings, as is revealed by the CI calculations.
The inaccurate 4M GFs are totally ignorant of the mixed character of these $\mathrm{A}_2$
scattering peaks and such is the case with every symmetry species.
The essential features of all the calculational schemes developed are further demonstrated in
Supplemental Material \cite{SupplPenroseRaman}.

   We have demonstrated the advantages of CI over GF in analyzing Raman scattering intensities
according to mediating magnons.
Once we go beyond the Loudon-Fleury second-order perturbation scheme,
or sometimes even within it,
we will find out any spectral weight of multimagnon character correctly only by
evaluating the Raman correlation functions ${}^{[p]}\mathcal{G}_{2l\mathrm{M}}(t)$ directly.
Real-frequency dynamic quantities cannot be obtained directly from path-integral calculations
such as quantum MC (QMC) findings.
Relevant imaginary-time correlation functions have to be first calculated and then continued to
real frequency.
Laplace transforms are difficult to invert numerically and maximum-entropy analytic continuation,
for instance, of QMC data is not necessarily successful even for small periodic clusters
\cite{S8478}.
Under such circumstances, our elaborate CI approach can open up a new path of calculating
dynamic properties.
In terms of its ability to clarify what kind of intermediate states are essential in which
scattering channel, we note that a newly developed representation theory of the quantum affine
Lie algebra for the spin-$\frac{1}{2}$ \textit{XXZ} infinite chain \cite{D89,A733} opened
the door to a full understanding of its dynamics.
We are aware of how much percentage of its total structure factor intensity two- and four-spinon
intermediate states contribute \cite{C12013,M435}.
Dynamic spin structure factors of quasiperiodic magnets \cite{W104427} are also analyzable in
full detail with the present CI scheme.

\begin{acknowledgment}
We are grateful to J. Ohara for his critical comments on our coding.
This work is supported by JSPS KAKENHI Grant Number 22K03502.
\end{acknowledgment}


\end{document}


\maketitle

\vspace*{-18mm}

\section*{S1. Spin-Wave Hamiltonian}
   We discuss antiferromagnetic Heisenberg models on
the two-dimensional (2D) Penrose and Ammann-Beenker lattices of point symmetry
$\mathbf{C}_{\mathrm{5v}}$ and $\mathbf{C}_{\mathrm{8v}}$, respectively, both of which are
described by the Hamiltonian
\begin{align}
   \mathcal{H}
  =J\sum_{\langle i,j \rangle}
   \bm{S}_{i}\cdot\bm{S}_{j}\ (J>0),
   \label{E:Heisenberg}
\end{align}
where $\bm{S}_{i}$ is the vector spin-$\frac{1}{2}$ operator at site $i$
and $\sum_{\langle i,j \rangle}$ runs over all pairs of connected vertices.
Since the Penrose and Ammann-Beenker lattices are both bipartite,
we divide them each into two sublattices,
A with $L_{\mathrm{A}}$ sites and B with $L_{\mathrm{B}}\,(\equiv L-L_{\mathrm{A}})$
sites, respectively.
We introduce the Holstein-Primakoff (HP) bosonic spin deviation operators
\cite{H1098}
\begin{align}
   &
   S_{i}^{+}
  =\left(2S-a_{i}^{\dagger}a_{i}\right)^{\frac{1}{2}}
   a_{i},\ 
   S_{i}^{-}
  =a_{i}^{\dagger}
   \left(2S-a_{i}^{\dagger}a_{i}\right)^{\frac{1}{2}},\ 
   \allowdisplaybreaks
   \nonumber \\
   &
   S_{i}^{z}
  =S-a_{i}^{\dagger}a_{i};\ 
   \allowdisplaybreaks
   \nonumber \\
   &
   S_{j}^{+}
  =b_{j}^{\dagger}
   \left(2S-b_{j}^{\dagger}b_{j}\right)^{\frac{1}{2}},\ 
   S_{j}^{-}
  =\left(2S-b_{j}^{\dagger}b_{j}\right)^{\frac{1}{2}}
   b_{j},\ 
   \allowdisplaybreaks
   \nonumber \\
   &
   S_{j}^{z}
  =b_{j}^{\dagger}b_{j}-S,
   \label{E:HPboson}
\end{align}
where the site indices are understood as $i\in\mathrm{A}$ and $j\in\mathrm{B}$.
We expand the Hamiltonian \eqref{E:Heisenberg}
in powers of the inverse spin magnitude $1/S$,
\begin{align}
    \mathcal{H}
   = \mathcal{H}^{(2)}
    +\mathcal{H}^{(1)}
    +\mathcal{H}^{(0)}
    +O(S^{-1}),
\end{align}
where $\mathcal{H}^{(m)}$, on the order of $S^m$, reads
\begin{align}
   &
   \mathcal{H}^{(2)}
 =-JS^2
   \sum_{i\in\mathrm{A}}\sum_{j\in\mathrm{B}}
   l_{i,j},\ 
   \mathcal{H}^{(1)}
  =JS
   \sum_{i\in\mathrm{A}}\sum_{j\in\mathrm{B}}
   l_{i,j}
   \allowdisplaybreaks
   \nonumber \\
  &\times
   \left(
    a_{i}^{\dagger}a_{i}+b_{j}^{\dagger}b_{j}
   +a_{i}b_{j}+a_{i}^{\dagger}b_{j}^{\dagger}
   \right),\ 
   \mathcal{H}^{(0)}
 =-J
   \sum_{i\in\mathrm{A}}\sum_{j\in\mathrm{B}}
   l_{i,j}
   \allowdisplaybreaks
   \nonumber \\
  &\times
   \left[
    a_{i}^{\dagger}a_{i}b_{j}^{\dagger}b_{j}
   +\frac{1}{4}
    \left(
     a_{i}^{\dagger}a_{i}a_{i}b_{j}
    +a_{i}^{\dagger}b_{j}^{\dagger}b_{j}^{\dagger}b_{j}
    +\mathrm{H.c.}
    \right)
   \right],
\end{align}
with $l_{i,j}$ being $1$ for connected vertices $i$ and $j$, otherwise $0$.
We decompose the $O(S^{0})$ quartic Hamiltonian $\mathcal{H}^{(0)}$ into
quadratic terms $\mathcal{H}^{(0)}_{\mathrm{BL}}$
and normal-ordered quartic terms $:\mathcal{H}^{(0)}:$
through Wick's theorem \cite{N034714,Y094412},
\begin{align}
   a_{i}^{\dagger}a_{i}b_{j}^{\dagger}b_{j}
  =&:a_{i}^{\dagger}a_{i}b_{j}^{\dagger}b_{j}:
   \allowdisplaybreaks
   \nonumber \\
  +&{}_{\mathrm{BL}}\langle 0|a_{i}^{\dagger}a_{i}|0\rangle_{\mathrm{BL}}
    b_{j}^{\dagger}b_{j}
   +{}_{\mathrm{BL}}\langle 0|b_{j}^{\dagger}b_{j}|0\rangle_{\mathrm{BL}}
    a_{i}^{\dagger}a_{i}
   \allowdisplaybreaks
   \nonumber \\
  +&{}_{\mathrm{BL}}\langle 0|a_{i}^{\dagger}b_{j}^{\dagger}|0\rangle_{\mathrm{BL}}
    a_{i}b_{j}
   +{}_{\mathrm{BL}}\langle 0|a_{i}b_{j}|0\rangle_{\mathrm{BL}}
    a_{i}^{\dagger}b_{j}^{\dagger}
   \allowdisplaybreaks
   \nonumber \\
  -&{}_{\mathrm{BL}}\langle 0|a_{i}^{\dagger}a_{i}|0\rangle_{\mathrm{BL}}\,
    {}_{\mathrm{BL}}\langle 0|b_{j}^{\dagger}b_{j}|0\rangle_{\mathrm{BL}}
   \allowdisplaybreaks
   \nonumber \\
  -&{}_{\mathrm{BL}}\langle 0|a_{i}^{\dagger}b_{j}^{\dagger}|0\rangle_{\mathrm{BL}}\,
    {}_{\mathrm{BL}}\langle 0|a_{i}b_{j}|0\rangle_{\mathrm{BL}},
   \allowdisplaybreaks
   \nonumber \\
   a_{i}^{\dagger}a_{i}a_{i}b_{j}
  =&:a_{i}^{\dagger}a_{i}a_{i}b_{j}:
   \allowdisplaybreaks
   \nonumber \\
  +&2\left(
    {}_{\mathrm{BL}}\langle 0|a_{i}^{\dagger}a_{i}|0\rangle_{\mathrm{BL}}
    a_{i}b_{j}
   +{}_{\mathrm{BL}}\langle 0|a_{i}b_{j}|0\rangle_{\mathrm{BL}}
    a_{i}^{\dagger}a_{i}
   \right)
   \allowdisplaybreaks
   \nonumber \\
  -&2{}_{\mathrm{BL}}\langle 0|a_{i}^{\dagger}a_{i}|0\rangle_{\mathrm{BL}}\,
     {}_{\mathrm{BL}}\langle 0|a_{i}b_{j}|0\rangle_{\mathrm{BL}},
   \allowdisplaybreaks
   \nonumber \\
   a_{i}^{\dagger}b_{j}^{\dagger}b_{j}^{\dagger}b_{j}
  =&:a_{i}^{\dagger}b_{j}^{\dagger}b_{j}^{\dagger}b_{j}:
   \allowdisplaybreaks
   \nonumber \\
  +&2\left( 
    {}_{\mathrm{BL}}\langle 0|a_{i}^{\dagger}b_{j}^{\dagger}|0\rangle_{\mathrm{BL}}
    b_{j}^{\dagger}b_{j}
   +{}_{\mathrm{BL}}\langle 0|b_{j}^{\dagger}b_{j}|0\rangle_{\mathrm{BL}}
    a_{i}^{\dagger}b_{j}^{\dagger}
   \right)
   \allowdisplaybreaks
   \nonumber \\
  -&2{}_{\mathrm{BL}}\langle 0|a_{i}^{\dagger}b_{j}^{\dagger}|0\rangle_{\mathrm{BL}}\,
     {}_{\mathrm{BL}}\langle 0|b_{j}^{\dagger}b_{j}|0\rangle_{\mathrm{BL}},
\end{align}
where $|0\rangle_{\mathrm{BL}}$ denotes the quasiparticle magnon vacuum.
The up-to-$O(S^{0})$ bosonic Hamiltonian reads
\begin{align}
   \mathcal{H}
  =\mathcal{H}^{(2)}
  +\mathcal{H}^{(1)}
  +\mathcal{H}^{(0)}_{\mathrm{BL}}
  +:\mathcal{H}^{(0)}:
  \equiv
   \mathcal{H}_{\mathrm{BL}}+:\mathcal{H}^{(0)}:.
   \tag{\ref{E:H}}
\end{align}

   Let us express the bilinear Hamiltonian $\mathcal{H}_{\mathrm{BL}}$ as
\begin{align}
   \mathcal{H}_{\mathrm{BL}}
  =\bm{c}^{\dagger}\mathcal{M}\bm{c}+\sum_{m=0}^{2}\tilde{E}^{(m)};\ 
   \mathcal{M}
   \equiv
   \left[
   \begin{array}{c|c}
    \mathbf{A}           & \mathbf{C}
   \\ \hline
    \mathbf{C}^{\dagger} & \mathbf{B}
   \end{array}
   \right],
   \label{E:BLH1}
\end{align}
where we define the row vectors $\bm{a}^{\dagger}$ and $\bm{b}^{\dagger}$ of dimension
$L_{\mathrm{A}}$ and $L_{\mathrm{B}}$, respectively,
\begin{align}
   \bm{c}^{\dagger}
  =\left[
    a_{1}^{\dagger}, \cdots, a_{L_{\mathrm{A}}}^{\dagger},
    b_{1},\cdots, b_{L_{\mathrm{B}}}
   \right]
  \equiv
   \left[\bm{a}^{\dagger}, {}^{t}\bm{b}\right],
\end{align}
the matrices $\mathbf{A}$, $\mathbf{B}$, and $\mathbf{C}$ of dimension
$L_{\mathrm{A}}\times L_{\mathrm{A}}$,
$L_{\mathrm{B}}\times L_{\mathrm{B}}$, and
$L_{\mathrm{A}}\times L_{\mathrm{B}}$, respectively,
\begin{align}
   \left[\mathbf{A}\right]_{i,i'}
  =&J\delta_{i,i'}\sum_{j\in\mathrm{B}}l_{i,j}
   \left[\vphantom{-\frac{1}{2}}
     S
    -{}_{\mathrm{BL}}\langle 0|b_{j}^{\dagger}b_{j}|0\rangle_{\mathrm{BL}}
   \right.
   \allowdisplaybreaks
   \nonumber \\
   &\left.
    -\frac{1}{2}\left(
      {}_{\mathrm{BL}}\langle 0|a_{i}b_{j}|0\rangle_{\mathrm{BL}}
     +{}_{\mathrm{BL}}\langle 0|a_{i}^{\dagger}b_{j}^{\dagger}|0\rangle_{\mathrm{BL}}
     \right)
   \right],
   \allowdisplaybreaks
   \nonumber \\
   \left[\mathbf{B}\right]_{j,j'}
  =&J\delta_{j,j'}\sum_{i\in\mathrm{A}}l_{i,j}
   \left[\vphantom{-\frac{1}{2}}
     S
    -{}_{\mathrm{BL}}\langle 0|a_{i}^{\dagger}a_{i}|0\rangle_{\mathrm{BL}}
   \right.
   \allowdisplaybreaks
    \nonumber \\
   &\left.
    -\frac{1}{2}\left(
      {}_{\mathrm{BL}}\langle 0|a_{i}b_{j}|0\rangle_{\mathrm{BL}}
     +{}_{\mathrm{BL}}\langle 0|a_{i}^{\dagger}b_{j}^{\dagger}|0\rangle_{\mathrm{BL}}
    \right)
   \right],
   \allowdisplaybreaks
   \nonumber \\
   \left[\mathbf{C} \right]_{i,j}
  =&Jl_{i,j}
   \left[\vphantom{-\frac{1}{2}}
     S
    -{}_{\mathrm{BL}}\langle 0|a_{i}b_{j}|0\rangle_{\mathrm{BL}}
   \right.
   \allowdisplaybreaks
   \nonumber \\
   &\left.
    -\frac{1}{2}\left(
      {}_{\mathrm{BL}}\langle 0|a_{i}^{\dagger}a_{i}|0\rangle_{\mathrm{BL}}
     +{}_{\mathrm{BL}}\langle 0|b_{j}^{\dagger}b_{j}|0\rangle_{\mathrm{BL}}
    \right)
   \right],
\end{align}
and the constants
\begin{align}
   &
   \tilde{E}^{(2)}
  =\mathcal{H}^{(2)}
  \equiv
   E^{(2)},\ 
   \tilde{E}^{(1)}
 =-JS
   \sum_{i\in\mathrm{A}}\sum_{j\in\mathrm{B}}
   l_{i,j},
   \allowdisplaybreaks
   \nonumber \\
   &
   \tilde{E}^{(0)}
  =J
   \sum_{i\in\mathrm{A}}\sum_{j\in\mathrm{B}}
   l_{i,j}
   \biggl[
    {}_{\mathrm{BL}}\langle 0|a_{i}^{\dagger}a_{i}|0\rangle_{\mathrm{BL}}
   \allowdisplaybreaks
   \nonumber \\
   &\quad
   +\frac{1}{2}
    \left(
     {}_{\mathrm{BL}}\langle 0|a_{i}b_{j}|0\rangle_{\mathrm{BL}}
    +{}_{\mathrm{BL}}\langle 0|a_{i}^{\dagger}b_{j}^{\dagger}|0\rangle_{\mathrm{BL}}
    \right)
   \allowdisplaybreaks
   \nonumber \\
   &\quad
   +{}_{\mathrm{BL}}\langle 0|a_{i}^{\dagger}a_{i}|0\rangle_{\mathrm{BL}}\,
    {}_{\mathrm{BL}}\langle 0|b_{j}^{\dagger}b_{j}|0\rangle_{\mathrm{BL}}
   \allowdisplaybreaks
   \nonumber \\
   &\quad
   +{}_{\mathrm{BL}}\langle 0|a_{i}^{\dagger}b_{j}^{\dagger}|0\rangle_{\mathrm{BL}}\,
    {}_{\mathrm{BL}}\langle 0|a_{i}b_{j}|0\rangle_{\mathrm{BL}}
   \allowdisplaybreaks
   \nonumber \\
  &\quad
   +\frac{1}{2}
    \left( {}_{\mathrm{BL}}\langle 0|a_{i}^{\dagger}a_{i}|0\rangle_{\mathrm{BL}}
          +{}_{\mathrm{BL}}\langle 0|b_{j}^{\dagger}b_{j}|0\rangle_{\mathrm{BL}}\right)
   \allowdisplaybreaks
   \nonumber \\
   &\quad\times
    \left( {}_{\mathrm{BL}}\langle 0|a_{i}^{\dagger}b_{j}^{\dagger}|0\rangle_{\mathrm{BL}}
          +{}_{\mathrm{BL}}\langle 0|a_{i}b_{j}|0\rangle_{\mathrm{BL}} \right)
   \biggr].
\end{align}
We carry out the Bogoliubov transformation
\begin{align}
   \bm{c}=\mathbf{X}\bm{\alpha};\ 
   \mathbf{X}
  \equiv
   \left[
   \begin{array}{c|c}
    \mathbf{S} & \mathbf{U} \\ \hline
    \mathbf{V} & \mathbf{T}
   \end{array}
   \right],
   \label{E:BT}
\end{align}
where we define the matrices
$\mathbf{S}$, $\mathbf{T}$, $\mathbf{U}$, and $\mathbf{V}$ of dimension
$L_{\mathrm{A}}\times L_{-}$,
$L_{\mathrm{B}}\times L_{+}$,
$L_{\mathrm{A}}\times L_{+}$, and
$L_{\mathrm{B}}\times L_{-}$, respectively,
to obtain the ferromagnetic and antiferromagnetic magnon operators
\begin{align}
   \left[\alpha_{1}^{-\dagger},\cdots,\alpha_{L_{-}}^{-\dagger},
         \alpha_{1}^{+},\cdots,\alpha_{L_{+}}^{+}\right]
  \equiv
   \bm{\alpha}^{\dagger}.
\end{align}
By virtue of the bosonic commutation relations,
the Bogoliubov transformation matrix $\mathbf{X}$ satisfies
\cite{W450,W104427,S104427}
\begin{align}
   &
   \mathbf{X}\mathbf{\Gamma}'\mathbf{X}^{\dagger}=\mathbf{\Gamma}; \ 
   \mathbf{\Gamma}
  \equiv
   \left[
   \begin{array}{cc}
   -\mathbf{I}(L_{\mathrm{A}}) & 0 \\
    0 & \mathbf{I}(L_{\mathrm{B}})
   \end{array}
   \right],
   \allowdisplaybreaks
   \nonumber \\
   &
   \mathbf{X}^{\dagger}\mathbf{\Gamma}\mathbf{X} =\mathbf{\Gamma}';\ 
   \mathbf{\Gamma}'
  \equiv
   \left[
   \begin{array}{cc}
   -\mathbf{I}(L_{-}) & 0 \\
    0 & \mathbf{I}(L_{+})
   \end{array}
   \right],
\end{align}
where $\mathbf{I}(L)$ denotes the $L\times L$ identity matrix.
Demanding that $\mathbf{X}$ should diagonalize $\mathcal{M}$, we obtain
\begin{align}
   \mathbf{X}^{\dagger}\mathcal{M}\mathbf{X}
  =\mathrm{diag}\left[
    \varepsilon_{1}^{-},\cdots,\varepsilon_{L_{-}}^{-},
    \varepsilon_{1}^{+},\cdots,\varepsilon_{L_{+}}^{+}\right]
  \equiv
   \mathbf{E},
   \label{E:MATDIAG}
\end{align}
where the eigenvalues $\varepsilon_{l_{-}}^{-}$
and $\varepsilon_{l_{+}}^{+}$ are non-negative.
Multiplying \eqref{E:MATDIAG} by $\mathbf{X}\mathbf{\Gamma}'$ from the left yields
\begin{align}
   \mathbf{\Gamma} \mathcal{M} \mathbf{X}
  =\mathbf{X}\mathbf{\Gamma}'\mathbf{E}.
\end{align}
The column vectors of $\mathbf{X}$ and the diagonal elements of $\mathbf{\Gamma}'\mathbf{E}$
are the right eigenvectors and their eigenvalues for $\mathbf{\Gamma}\mathcal{M}$, respectively.
\begin{figure}[tb]
\centering
\includegraphics[width=\linewidth]{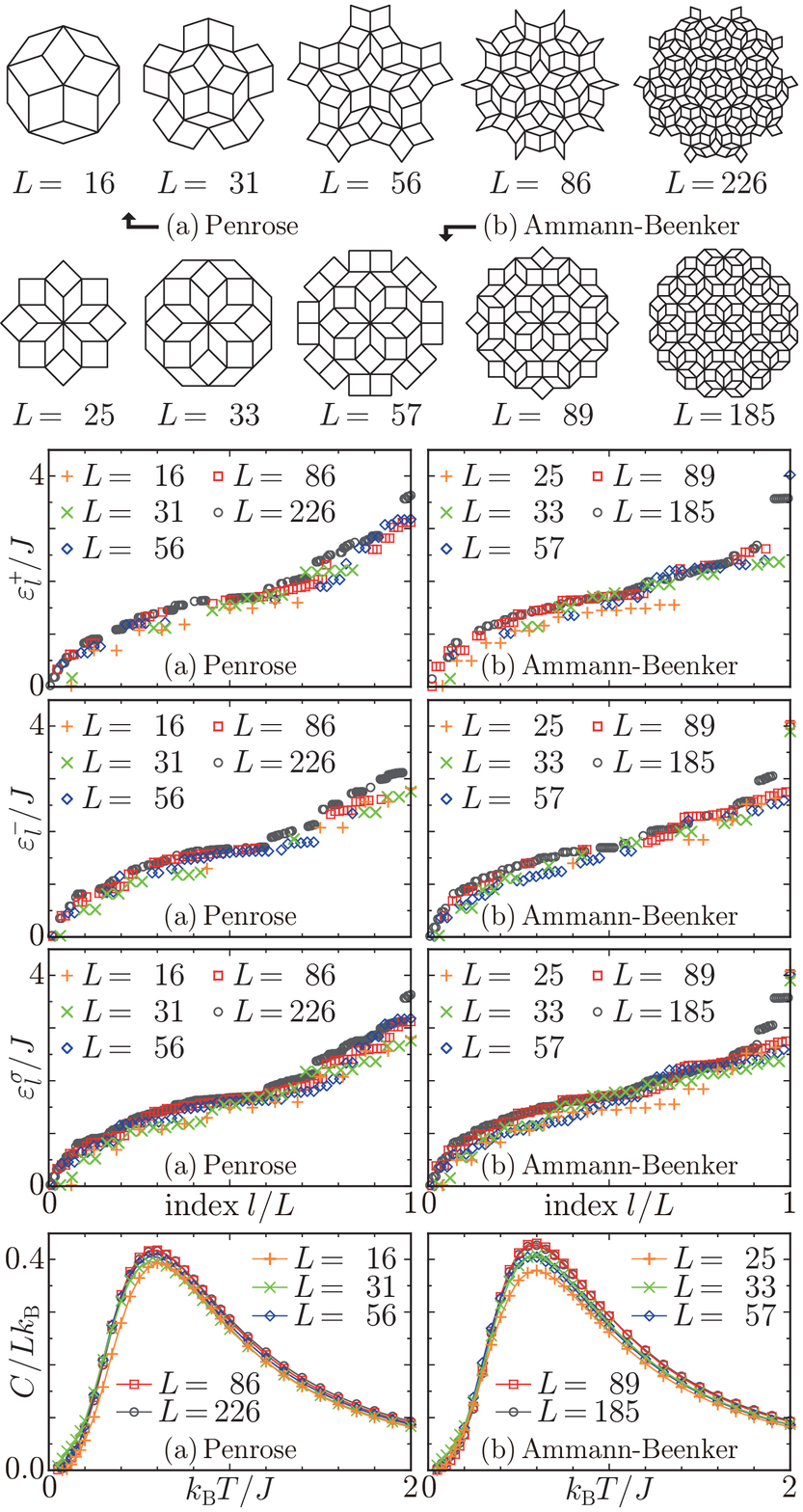}
\vspace*{-7mm}
\caption{The eigenvalues $\varepsilon_l^\sigma$ $(\sigma=\pm,\,1\leq l\leq L)$ of
         the bosonic Hamiltonian \eqref{E:HBLdiag} and quantum Monte Carlo calculations of
         the specific heat $C$ as a function of temperature $T$ of the antiferromagnetic
         Heisenberg Hamiltonian \eqref{E:Heisenberg} for various 2D Penrose (a) and Ammann-Beenker (b)
         clusters of point symmetry $\mathbf{C}_{\mathrm{5v}}$ and $\mathbf{C}_{\mathrm{8v}}$,
         respectively.}
\vspace*{-8mm}
\label{F:EV&C}
\end{figure}

   The eigenvalues of $\mathbf{\Gamma}\mathcal{M}$ comprise
$L_{-}$ negative and $L_{+}$ positive eigenvalues \cite{W104427,S104427},
\begin{align}
   \mathbf{\Gamma}'\mathbf{E}
  =\mathbf{E}'
  \equiv
   \mathrm{diag}\left[
    \varepsilon_{1}^{\prime -},\cdots,\varepsilon_{L_{-}}^{\prime -},
    \varepsilon_{1}^{\prime +},\cdots,\varepsilon_{L_{+}}^{\prime +}
   \right].
\end{align}
Having in mind that $(\mathbf{\Gamma}')^{2}=\mathbf{I}(L)$, we find
$\mathbf{E}=\mathbf{\Gamma}'\mathbf{E}'$.
Then the non-negative eigenvalues
$\varepsilon_{l_{-}}^{-}$ and $\varepsilon_{l_{+}}^{+}$ read
\begin{align}
   \varepsilon_{l_{-}}^{-}
  =-\varepsilon_{l_{-}}^{\prime -},\ 
   \varepsilon_{l_{+}}^{+}
  =\varepsilon_{l_{+}}^{\prime +}
\end{align}
to yield the diagonal one-body Hamiltonian
\begin{align}
   \mathcal{H}_{\mathrm{BL}}
  =\sum_{m=0}^{2}\tilde{E}^{(m)}
  +\sum_{l_{-}=1}^{L_{-}}
   \varepsilon_{l_{-}}^{-}
   \alpha_{l_{-}}^{-\dagger}\alpha_{l_{-}}^{-}
  +\sum_{l_{+}=1}^{L_{+}}
   \varepsilon_{l_{+}}^{+}
   \alpha_{l_{+}}^{+}\alpha_{l_{+}}^{+\dagger}.
\end{align}
Denoting the $O(S^m)$ component of $\varepsilon_{l_{\sigma}}^{\sigma}$ by
$\varepsilon_{l_{\sigma}}^{\sigma(m)}$,
we express the $O(S^0)$ quantum corrections to the classical ground-state energy $E^{(2)}$ as
\begin{align}
   E^{(m)}
  =\tilde{E}^{(m)}+\sum_{l_{+}=1}^{L_{+}}\varepsilon_{l_{+}}^{+(m)}\ 
   (m=1,0).
\end{align}
and rewrite the Hamiltonian as
\begin{align}
   \mathcal{H}_{\mathrm{BL}}
  =\sum_{m=0}^{2}E^{(m)}
  +\sum_{\sigma=\pm}
   \sum_{l_{\sigma}=1}^{L_{\sigma}}\varepsilon_{l_{\sigma}}^{\sigma}
   \alpha_{l_{\sigma}}^{\sigma\dagger}\alpha_{l_{\sigma}}^{\sigma}.
   \tag{\ref{E:HBLdiag}}
\end{align}

   We show in Fig. \ref{F:EV&C}
the magnon eigenvalues $\varepsilon_{l_\sigma}^\sigma$ of the bosonic Hamiltonian \eqref{E:HBLdiag},
together with the rigorous specific heat curves.
The major portion of the eigenvalues continuously distributes in almost the same energy region
with almost the same pattern at each system size.
The rest are isolated from them and strongly localized to sites with relatively high coordination
numbers \cite{W104427,S104427}.
Since there is not much of a difference in the temperature profile of the specific heat between
various clusters, sites of the highest and/or second highest coordination numbers,
which are possibly absent from small clusters, do not seriously affect the thermodynamic
properties.

\begin{figure}[t]
\centering
\includegraphics[width=0.92\linewidth]{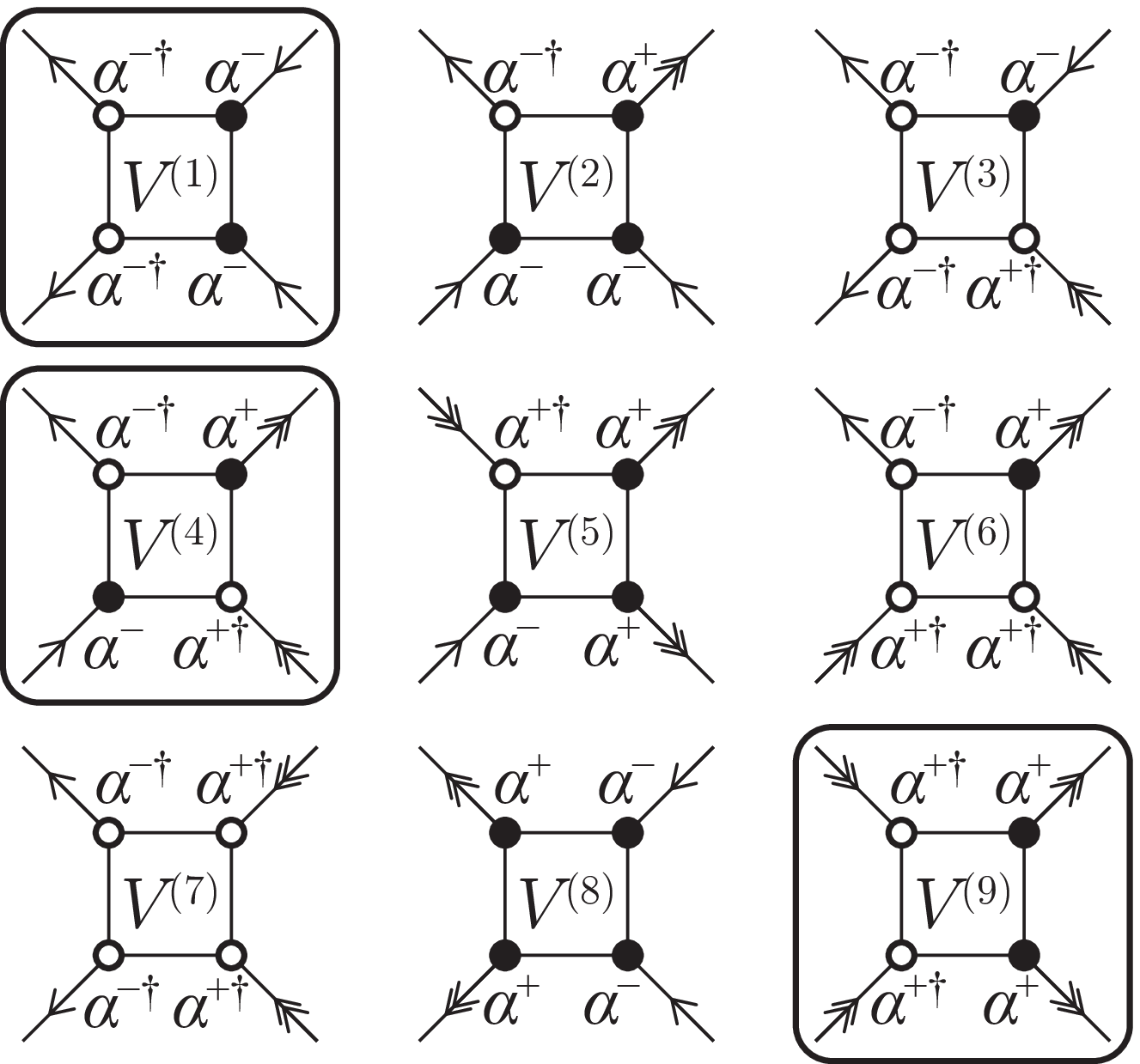}
\vspace*{-3mm}
\caption{Diagrammatic representation of the magnon-magnon interactions
         $V_{ij;l_{\sigma}l_{\sigma'}'l_{\sigma''}''l_{\sigma'''}'''}^{(m)}$ 
         emergent in \eqref{E::H(0):=V(1)toV(9)}.
         The open and closed circles each signify creating and annihilating a magnon,
         whether it is antiferromagnetic or ferromagnetic.
         The single and double arrowheads are to form antiferromagnetic and
         ferromagnetic magnon propagators, respectively,
         both of which enhance and reduce the magnetization when they are
         incoming and outgoing, respectively.
         The enclosed diagrams keep the number of magnons constant.}
\vspace*{-8mm}
\label{F:INTERACTIONS}
\end{figure}

   The quartic interaction $:\mathcal{H}^{(0)}:$ is given by
\begin{align}
   &
   :\mathcal{H}^{(0)}:
  =-J
   \sum_{i\in\mathrm{A}}\sum_{j\in\mathrm{B}}l_{i,j}
   \left\{
    \sum_{l_{-},l_{-}',l_{-}'',l_{-}'''}
    V_{ij;l_{-} l_{-}' l_{-}'' l_{-}'''}^{(1)}
    \alpha_{l_{-}}^{-\dagger}
    \alpha_{l_{-}'}^{-\dagger}
    \alpha_{l_{-}''}^{-}
    \alpha_{l_{-}'''}^{-}
   \right.
   \allowdisplaybreaks
   \nonumber \\
   &\qquad\qquad
  +\sum_{l_{-},l_{+}',l_{-}'',l_{-}'''}
    V_{ij;l_{-} l_{+}' l_{-}'' l_{-}'''}^{(2)}
    \alpha_{l_{-}}^{-\dagger}
    \alpha_{l_{+}'}^{+}
    \alpha_{l_{-}''}^{-}
    \alpha_{l_{-}'''}^{-}
   \allowdisplaybreaks
   \nonumber \\
   &\qquad\qquad
  +\sum_{l_{+},l_{-}',l_{-}'',l_{-}'''}
    V_{ij;l_{-} l_{-}' l_{+}'' l_{-}'''}^{(3)}
    \alpha_{l_{+}}^{+\dagger}
    \alpha_{l_{-}'}^{-\dagger}
    \alpha_{l_{-}''}^{-\dagger}
    \alpha_{l_{-}'''}^{-}
   \allowdisplaybreaks
   \nonumber \\
   &\qquad\qquad
  +\sum_{l_{+},l_{+}',l_{-}'',l_{-}'''}
    V_{ij;l_{+} l_{+}' l_{-}'' l_{-}'''}^{(4)}
    \alpha_{l_{+}}^{+\dagger}
    \alpha_{l_{+}'}^{+}
    \alpha_{l_{-}''}^{-\dagger}
    \alpha_{l_{-}'''}^{-}
   \allowdisplaybreaks
   \nonumber \\
   &\qquad\qquad
  +\sum_{l_{+},l_{+}',l_{+}'',l_{-}'''}
    V_{ij;l_{+} l_{+}' l_{+}'' l_{-}'''}^{(5)}
    \alpha_{l_{+}}^{+\dagger}
    \alpha_{l_{+}'}^{+}
    \alpha_{l_{+}''}^{+}
    \alpha_{l_{-}'''}^{-}
   \allowdisplaybreaks
   \nonumber \\
   &\qquad\qquad
  +\sum_{l_{+},l_{+}' l_{-}'' l_{+}'''}
    V_{ij;l_{+} l_{+}' l_{-}'' l_{+}'''}^{(6)}
    \alpha_{l_{+}}^{+\dagger}
    \alpha_{l_{+}'}^{+\dagger}
    \alpha_{l_{-}''}^{-\dagger}
    \alpha_{l_{+}'''}^{+}
   \allowdisplaybreaks
   \nonumber \\
   &\qquad\qquad
  +\sum_{l_{+},l_{+}',l_{-}'',l_{-}'''}
    V_{ij;l_{+} l_{+}' l_{-}'' l_{-}'''}^{(7)}
    \alpha_{l_{+}}^{+\dagger}
    \alpha_{l_{+}'}^{+\dagger}
    \alpha_{l_{-}''}^{-\dagger}
    \alpha_{l_{-}'''}^{-\dagger}
   \allowdisplaybreaks
   \nonumber \\
   &\qquad\qquad
  +\sum_{l_{+},l_{+}',l_{-}'',l_{-}'''}
    V_{ij;l_{+} l_{+}' l_{-}'' l_{-}'''}^{(8)}
    \alpha_{l_{+}}^{+}
    \alpha_{l_{+}'}^{+}
    \alpha_{l_{-}''}^{-}
    \alpha_{l_{-}'''}^{-}
   \allowdisplaybreaks
   \nonumber \\
   &\qquad\qquad
   \left.
  +\sum_{l_{+},l_{+}',l_{+}'',l_{+}'''}
    V_{ij;l_{+} l_{+}' l_{+}'' l_{+}'''}^{(9)}
    \alpha_{l_{+}}^{+\dagger}
    \alpha_{l_{+}'}^{+\dagger}
    \alpha_{l_{+}''}^{+}
    \alpha_{l_{+}'''}^{+}
   \right\}
   \label{E::H(0):=V(1)toV(9)}
\end{align}
with
$
   V_{ij;l_{-} l_{+}' l_{-}'' l_{-}'''}^{(2)}
  =V_{ij;l_{+}' l_{-}''' l_{-}'' l_{-}}^{(3)\,*}
$,
$
   V_{ij;l_{+} l_{+}' l_{+}'' l_{-}'''}^{(5)}
  =V_{ij;l_{+}'' l_{+}' l_{+} l_{-}'''}^{(6)\,*}
$,
and
$
   V_{ij;l_{+} l_{+}' l_{-}'' l_{-}'''}^{(7)}
  =V_{ij;l_{+} l_{+}' l_{-}'' l_{-}'''}^{(8)\,*}
$.
Figure \ref{F:INTERACTIONS} shows the magnon-magnon interactions
$V_{ij;l_{\sigma}l_{\sigma'}'l_{\sigma''}''l_{\sigma'''}'''}^{(m)}$ diagrammatically.
We give the magnon-number-conserving interactions explicitly in particular,
\begin{align}
   V_{ij;l_{-}l_{-}'l_{-}''l_{-}'''}^{(1)}
   &
   =\frac{1}{4}
   \left(
    s_{i,l_{-}}^{*}s_{i,l_{-}'''}v_{j,l_{-}''}v_{j,l_{-}'}^{*}
   +s_{i,l_{-}'}^{*}s_{i,l_{-}'''}v_{j,l_{-}''}v_{j,l_{-}}^{*}
   \right.
   \allowdisplaybreaks
   \nonumber \\
   &\qquad\ 
   \left.
   +s_{i,l_{-}}^{*}s_{i,l_{-}''}v_{j,l_{-}'''}v_{j,l_{-}'}^{*}
   +s_{i,l_{-}'}^{*}s_{i,l_{-}''}v_{j,l_{-}'''}v_{j,l_{-}}^{*}
   \right)
   \allowdisplaybreaks
   \nonumber \\
  &
   +\frac{1}{8}
   \left(
    s_{i,l_{-}}^{*}s_{i,l_{-}'''}s_{i,l_{-}''}v_{j,l_{-}'}^{*}
   +s_{i,l_{-}}^{*}v_{j,l_{-}'''}v_{j,l_{-}''}v_{j,l_{-}'}^{*}
   \right.
   \allowdisplaybreaks
   \nonumber \\
  &\qquad\ 
   +s_{i,l_{-}''}v_{j,l_{-}'''}v_{j,l_{-}}^{*}v_{j,l_{-}'}^{*}
   +s_{i,l_{-}}^{*}s_{i,l_{-}'}^{*}s_{i,l_{-}''}v_{j,l_{-}'''}
   \allowdisplaybreaks
   \nonumber \\
  &\qquad\ 
   +s_{i,l_{-}'}^{*}s_{i,l_{-}'''}s_{i,l_{-}''}v_{j,l_{-}}^{*}
   +s_{i,l_{-}'}^{*}v_{j,l_{-}'''}v_{j,l_{-}''}v_{j,l_{-}}^{*}
   \allowdisplaybreaks
   \nonumber \\
  &\qquad\ 
   \left.
   +s_{i,l_{-}'''}v_{j,l_{-}''}v_{j,l_{-}}^{*}v_{j,l_{-}'}^{*}
   +s_{i,l_{-}}^{*}s_{i,l_{-}'}^{*}s_{i,l_{-}'''}v_{j,l_{-}''}
   \right),
   \allowdisplaybreaks
   \nonumber \\
   V_{ij;l_{+} l_{+}' l_{-}'' l_{-}'''}^{(4)}
   &
   =s_{i,l_{-}''}^{*}s_{i,l_{-}'''}t_{i,l_{+}}t_{i,l_{+}'}^{*}
   +s_{i,l_{-}''}^{*}u_{i,l_{+}}v_{i,l_{-}'''}t_{i,l_{+}'}^{*}
   \allowdisplaybreaks
   \nonumber \\
  &\qquad\ 
   +u_{i,l_{+}'}^{*}s_{i,l_{-}'''}t_{i,l_{+}}v_{i,l_{-}''}^{*}
   +u_{i,l_{+}'}^{*}u_{i,l_{+}}v_{i,l_{-}'''}v_{i,l_{-}''}^{*}
   \allowdisplaybreaks
   \nonumber \\
  &
  +\frac{1}{4}
   \left(
    s_{i,l_{-}''}^{*}s_{i,l_{-}'''}u_{i,l_{+}}t_{i,l_{+}'}^{*}
   +s_{i,l_{-}''}^{*}u_{i,l_{+}}s_{i,l_{-}'''}t_{i,l_{+}'}^{*}
   \right.
   \allowdisplaybreaks
   \nonumber \\
  &\qquad\ 
   +u_{i,l_{+}'}^{*}s_{i,l_{-}'''}u_{i,l_{+}}v_{i,l_{-}''}^{*}
   +u_{i,l_{+}'}^{*}u_{i,l_{+}}s_{i,l_{-}'''}v_{i,l_{-}''}^{*}
   \allowdisplaybreaks
   \nonumber \\
  &\qquad\ 
   +s_{i,l_{-}''}^{*}v_{i,l_{-}'''}t_{i,l_{+}}t_{i,l_{+}'}^{*}
   +s_{i,l_{-}''}^{*}t_{i,l_{+}}v_{i,l_{-}'''}t_{i,l_{+}'}^{*}
   \allowdisplaybreaks
   \nonumber \\
  &\qquad\ 
   \left.
   +u_{i,l_{+}'}^{*}v_{i,l_{-}'''}t_{i,l_{+}}v_{i,l_{-}''}^{*}
   +u_{i,l_{+}'}^{*}t_{i,l_{+}}v_{i,l_{-}'''}v_{i,l_{-}''}^{*}
   \right)
   \allowdisplaybreaks
   \nonumber \\
  &\quad
  +\frac{1}{2}
   \left(
    s_{i,l_{-}'''}t_{i,l_{+}}v_{i,l_{-}''}^{*}t_{i,l_{+}'}^{*}
   +u_{i,l_{+}}v_{i,l_{-}'''}v_{i,l_{-}''}^{*}t_{i,l_{+}'}^{*}
   \right.
   \allowdisplaybreaks
   \nonumber \\
  &\qquad\ 
   \left.
   +s_{i,l_{-}''}^{*}u_{i,l_{+}'}^{*}s_{i,l_{-}'''}t_{i,l_{+}}
   +s_{i,l_{-}''}^{*}u_{i,l_{+}'}^{*}u_{i,l_{+}}v_{i,l_{-}'''}
   \right),
   \allowdisplaybreaks
   \nonumber \\
   V_{ij;l_{+}l_{+}'l_{+}''l_{+}'''}^{(9)}
  &
  =\frac{1}{4}
   \left(
    u_{i,l_{+}''}^{*}u_{i,l_{+}'}t_{j,l_{+}}t_{j,l_{+}'''}^{*}
   +u_{i,l_{+}''}^{*}u_{i,l_{+}}t_{j,l_{+}'}t_{j,l_{+}'''}^{*}
   \right.
   \allowdisplaybreaks
   \nonumber \\
  &\qquad\ 
   \left.
   +u_{i,l_{+}'''}^{*}u_{i,l_{+}'}t_{j,l_{+}}t_{j,l_{+}''}^{*}
   +u_{i,l_{+}'''}^{*}u_{i,l_{+}}t_{j,l_{+}'}t_{j,l_{+}''}^{*}
   \right)
   \allowdisplaybreaks
   \nonumber \\
  &
  +\frac{1}{8}\left(
    u_{i,l_{+}''}^{*}u_{i,l_{+}'}u_{i,l_{+}}t_{j,l_{+}'''}^{*}
   +u_{i,l_{+}''}^{*}t_{j,l_{+}'}t_{j,l_{+}}t_{j,l_{+}'''}^{*}
   \right.
   \allowdisplaybreaks
   \nonumber \\
  &\qquad\ 
   +u_{i,l_{+}}t_{j,l_{+}'}t_{j,l_{+}''}^{*}t_{j,l_{+}'''}^{*}
   +u_{i,l_{+}''}^{*}u_{i,l_{+}'''}^{*}u_{i,l_{+}}t_{j,l_{+}'}
   \allowdisplaybreaks
   \nonumber \\
  &\qquad\ 
   +u_{i,l_{+}'''}^{*}u_{i,l_{+}'}u_{i,l_{+}}t_{j,l_{+}''}^{*}
   +u_{i,l_{+}'''}^{*}t_{j,l_{+}'}t_{j,l_{+}}t_{j,l_{+}''}^{*}
   \allowdisplaybreaks
   \nonumber \\
  &\qquad\ 
   \left.
   +u_{i,l_{+}'}t_{j,l_{+}}t_{j,l_{+}''}^{*}t_{j,l_{+}'''}^{*}
   +u_{i,l_{+}''}^{*}u_{i,l_{+}'''}^{*}u_{i,l_{+}'}t_{j,l_{+}}
   \right),
\end{align}
in terms of the matrix elements
$s_{i,l_{-}}\equiv[\mathbf{S}]_{i,l_{-}}$,
$t_{j,l_{+}}\equiv[\mathbf{T}]_{j,l_{+}}$,
$u_{i,l_{+}}\equiv[\mathbf{U}]_{i,l_{+}}$, and
$v_{j,l_{-}}\equiv[\mathbf{V}]_{j,l_{-}}$
defined in \eqref{E:BT}.

\section*{S2. Magnetic Raman Scattering}
   Following the Shastry-Shraiman perturbation theory \cite{S1068, S365, K024414},
we derive spin-$\frac{1}{2}$ magnetic Raman operators from the half-filled single-band
nearest-neighbor Hubbard model
\begin{align}
   \mathscr{H}
   &
  =U\sum_{i}
   c_{i\uparrow}^\dagger c_{i\uparrow}
   c_{i\downarrow}^\dagger c_{i\downarrow}
  -t\sum_{\langle i,j \rangle}\sum_{\sigma=\uparrow,\downarrow}
   (c_{i\sigma}^{\dagger}c_{j\sigma}+\mathrm{H.c.}),
   \allowdisplaybreaks
   \nonumber \\[-1mm]
   &
  \equiv
   \mathscr{H}_{U}+\mathscr{H}_{t},
   \label{E:Hubbard}
\end{align}
where $c_{i\sigma}^{\dagger}$ creates an electron with spin $\sigma$ at site $i$.
We assume that $0<t\ll U$.
We replace all the hopping terms $c_{i\sigma}^{\dagger}c_{j\sigma}$ in $\mathscr{H}_{t}$ with
$
   c_{i\sigma}^{\dagger}c_{j\sigma} 
   \exp\left[
    \frac{ie}{\hbar c}
    \int_{\bm{r}_{j}}^{\bm{r}_{i}}\bm{A}(\bm{r})\cdot d\bm{r}
   \right]
$,
where $e$ and $c$ are the elementary electric charge and the light velocity,
respectively, and then the applied electric field $\bm{E}(t)$ reads as
$-\partial\bm{A}(t)/c\partial t$.
Suppose
$\gamma_{\bm{q},p}^{\dagger}$ creates a photon of momentum $\bm{q}$,
energy $\hbar\omega_{\bm{q}}$, and polarization $p$, then
the second-quantized vector potential is written as
\begin{align}
   \bm{A}(\bm{r})
  =\sum_{\bm{q},p}
   \sqrt{\frac{hc^{2}}{V\omega_{\bm{q}}}}
   \left(
    \bm{e}_{\bm{q},p}
    \gamma_{\bm{q},p}
    e^{i\bm{q}\cdot\bm{r}}
   +\bm{e}_{\bm{q},p}^{*}
    \gamma_{\bm{q},p}^{\dagger}
    e^{-i\bm{q}\cdot\bm{r}}
   \right)
   \label{E:A(r)}
\end{align}
with $V$ being the appropriate volume of the sample.
For visible light, we may put $e^{i\bm{q}\cdot\bm{r}}\simeq 1$
and therefore denote $\bm{A}(\bm{r})$ simply by $\bm{A}$ hereafter.
Then the electron-photon-coupled Hamiltonian reads
\vspace*{-1mm}
\begin{align}
   &
   \mathscr{H}_{\mathrm{el\mbox{-}ph}}
  =\mathscr{H}
  +\varOmega
  +\sum_{m=1}^{\infty}{}^{[m]}\!\!\!\mathscr{J};\ 
   \varOmega
  \equiv
   \sum_{\bm{q},p}
   \hbar\omega_{\bm{q}}
   \gamma_{\bm{q},p}^{\dagger}\gamma_{\bm{q},p},\ 
   \allowdisplaybreaks
   \nonumber \\
   &
   {}^{[m]}\!\!\!\mathscr{J}
  \equiv
  -t
   \sum_{\langle i,j \rangle}\sum_{\sigma=\uparrow,\downarrow}
   \frac{1}{m!}
   \left[
    c_{i\sigma}^{\dagger}c_{j\sigma}
    \left(
     \frac{-ie}{\hbar c}
     \bm{A}\cdot\bm{d}_{i,j}
    \right)^{m}
   +{\mathrm{H.c.}}
   \right]
   \allowdisplaybreaks
   \nonumber \\
   &
   \quad\ \ \,\,
  =-t
   \sum_{i,j=1}^{L}l_{i,j}\sum_{\sigma=\uparrow,\downarrow}
   \frac{1}{m!}
   c_{i\sigma}^{\dagger}c_{j\sigma}
   \left(
    \frac{-ie}{\hbar c}
    \bm{A}\cdot\bm{d}_{i,j}
   \right)^{m}
   \label{E:Hel-ph}
\end{align}
with $\bm{d}_{i,j}\equiv\bm{r}_j-\bm{r}_i$.

   Let the photoinduced current operators ${}^{[m]}\!\!\!\mathscr{J}$ be perturbations
to $\mathscr{H}+\varOmega$.
The transition between arbitrary states,
$|i\rangle$ of energy $\varepsilon_i$ and $|f\rangle$ of energy $\varepsilon_f$,
each being a product of electronic and photonic states, are rated as
\vspace*{-1mm}
\begin{align}
   W_{i,f}
  =\frac{2\pi}{\hbar}
   \left|
    \langle f|\mathscr{T}|i\rangle
   \right|^{2}
   \delta(\varepsilon_{f}-\varepsilon_{i}).
   \\[-8mm] \nonumber
\end{align}
Any Raman scattering contains two photons, starting with an incident photon and ending in
a scattered photon, where \eqref{E:A(r)} is explicitly written as
\vspace*{-1mm}
\begin{align}
   \bm{A}
  =\sqrt{\frac{hc^{2}}{V\omega_{\mathrm{in}}}}
   \bm{e}_{\mathrm{in}}
   \gamma_{\bm{q}_{\mathrm{in}},\bm{e}_{\mathrm{in}}}
  +\sqrt{\frac{hc^{2}}{V\omega_{\mathrm{sc}}}}
   \bm{e}_{\mathrm{sc}}^{*}
   \gamma_{\bm{q}_{\mathrm{sc}},\bm{e}_{\mathrm{sc}}}^{\dagger}
\end{align}
with
$\omega_{\mathrm{in}}$ $(\omega_{\mathrm{sc}})$,
$\bm{q}_{\mathrm{in}}$ $(\bm{q}_{\mathrm{sc}})$,
and $\bm{e}_{\mathrm{in}}$ $(\bm{e}_{\mathrm{sc}})$
being the frequency, momentum, and polarization of the incident (scattered) photon,
respectively.
The Raman transition matrix $\mathscr{T}$ in proportion to $\bm{A}^2$ reads
\vspace*{-1mm}
\begin{align}
   \mathscr{T}
  ={}^{[2]}\!\!\!\mathscr{J}
  +{}^{[1]}\!\!\!\mathscr{J}
    \frac{1}{\varepsilon_{i}-\varOmega-\mathscr{H}_{U}-\mathscr{H}_{t}}
   {}^{[1]}\!\!\!\mathscr{J}.
\end{align}
Every magnetic Raman scattering demands that the electronic state should belong in
the singly-occupied ground-state manifold at the beginning and end,
where ${}^{[m]}\!\!\!\mathscr{J}$, inducing a single electron transfer, singly has
no contribution to the transition rate,
$\langle f|{}^{[m]}\!\!\!\mathscr{J}|i\rangle=0$.
Relevant intermediate states obtained by operating ${}^{[1]}\!\!\!\mathscr{J}$ on
the initial state each have one doublon-holon pair together with no photon or two photons.
The photonic state is also singly occupied at the beginning and end.
Considering that $t\ll U$, we regard both ${}^{[m]}\!\!\!\mathscr{J}$ and $\mathscr{H}_{t}$ as
perturbations to $\mathscr{H}_{U}$ and therefore express the effective Raman operator as
\vspace*{-1mm}
\begin{align}
\!\!\!\!\!\!
   &
   \mathcal{R}
  =\mathcal{P}\,
    {}^{[1]}\!\!\!\mathscr{J}
    \frac{1}{\varepsilon_{i}-\varOmega-\mathscr{H}_{U}-\mathscr{H}_{t}}
    {}^{[1]}\!\!\!\mathscr{J}
   \mathcal{P}
   \allowdisplaybreaks
   \nonumber \\
   &
  =\mathcal{P}\,
    {}^{[1]}\!\!\!\mathscr{J}
    \frac{1}{\varepsilon_{i}-\varOmega-\mathscr{H}_{U}}
    \sum_{n=0}^{\infty}
    \left(
     \mathscr{H}_{t}
     \frac{1}{\varepsilon_{i}-\varOmega-\mathscr{H}_{U}}
    \right)^{n}
   {}^{[1]}\!\!\!\mathscr{J}
   \mathcal{P},\!
   \label{E:GERO}
\end{align}
where $\mathcal{P}$ is the projection operator to the singly-occupied ground-state manifold.

   No-photon and two-photon intermediate states are higher in energy than the ground state
by $U-\hbar\omega_{\mathrm{in}}$ and $U+\hbar\omega_{\mathrm{sc}}$, respectively.
Assuming that the incident photon energy is comparable to the electronic correlation energy,
$t \lesssim |U-\hbar\omega_{\mathrm{in}}| \ll U$,
we may replace $(\varepsilon_{i}-\varOmega-\mathscr{H}_{U})^{-1}$ by
$(\hbar\omega_{\mathrm{in}}-U)^{-1}$.
With the single occupancy at every site in mind, we express the electron operators
in terms of the spin operators,
\begin{align}
   \mathcal{P}c_{i\sigma_{1}}^{\dagger}c_{i\sigma_{2}}\mathcal{P}
  =\frac{1}{2}\delta_{\sigma_{1},\sigma_{2}}
  +\sum_{\mu=x,y,z}
   S_{i}^{\mu}
   \left[
    \bm{\sigma}^{\mu}
   \right]_{\sigma_{2}\sigma_{1}},
   \label{E:e-to-spin}
   \\[-8mm] \nonumber
\end{align}
where $\bm{\sigma}^\mu$'s are the Pauli matrices.
\begin{figure}[tb]
\centering
\includegraphics[width=\linewidth]{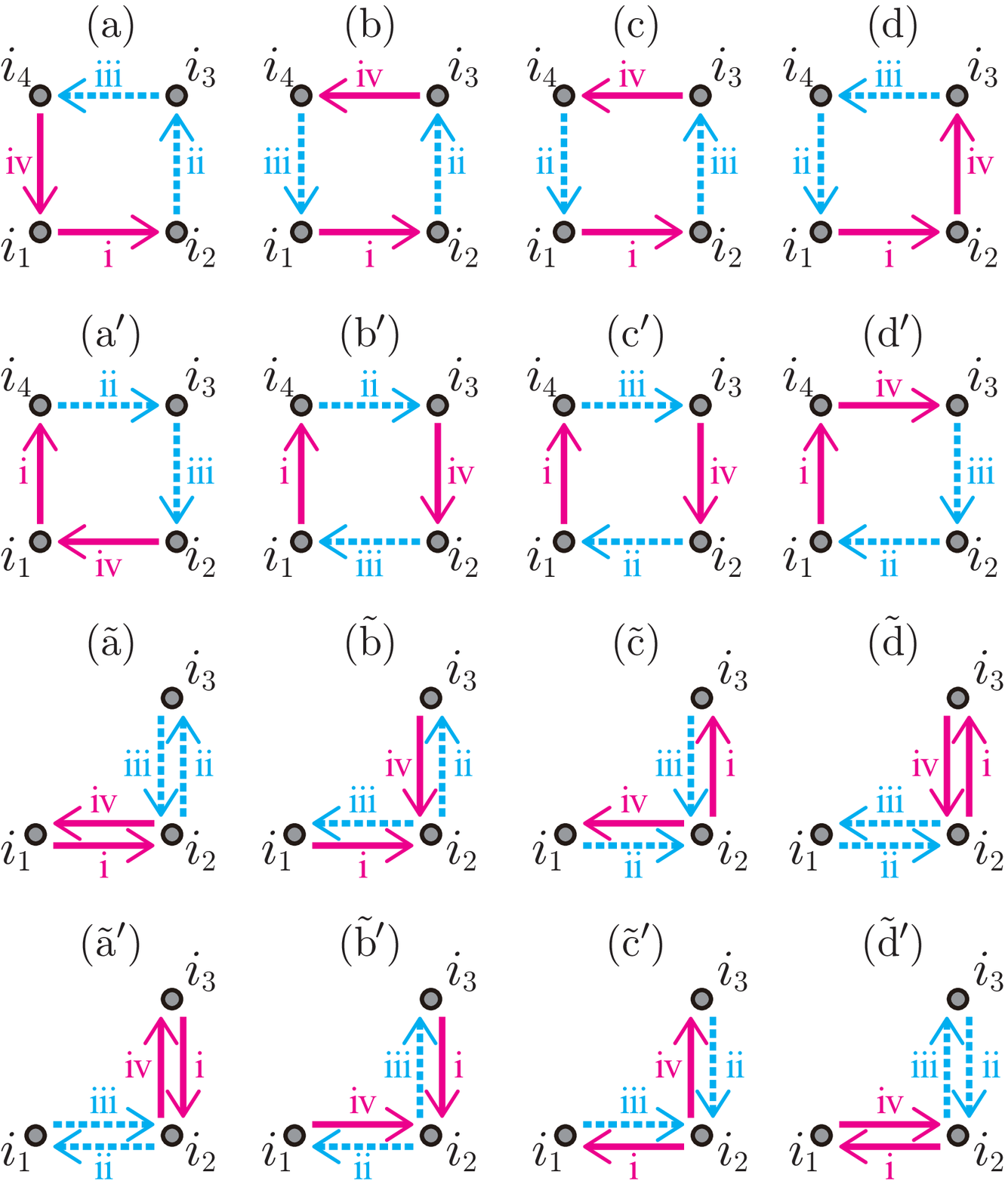}
\vspace*{-3mm}
\caption{Fourth-order electron hopping paths.
         (a)-(d) and ($\mathrm{a}^{\prime}$)-($\mathrm{d}^{\prime}$) are cyclic paths, while
         ($\tilde{\mathrm{a}}$)-($\tilde{\mathrm{d}}$) and
         ($\tilde{\mathrm{a}}^{\prime}$)-($\tilde{\mathrm{d}}^{\prime}$) are round paths.
         Solid arrows create or annihilate a doublon-holon pair arising from
         ${}^{[1]}\!\!\!\mathscr{J}$,
         whereas broken arrows correspond to electron transfer arising from $\mathscr{H}_{t}$.}
\vspace*{-8mm}
\label{F:PATH4TH}
\end{figure}

   The lowest-order $n=0$ term in \eqref{E:GERO}, which is of second order in $t$, reads
\vspace*{-1mm}
\begin{align}
   &
   {}^{[2]}\mathcal{R}
  =\frac{-1}{U-\hbar\omega_{\mathrm{in}}}
   \mathcal{P}\,
    {}^{[1]}\!\!\!\mathscr{J}{}^{[1]}\!\!\!\mathscr{J}
   \mathcal{P}
  =\frac{2\pi e^{2}}{\hbar V\sqrt{\omega_{\mathrm{in}}\omega_{\mathrm{sc}}}}
   \frac{t^{2}}{U-\hbar\omega_{\mathrm{in}}}
   \allowdisplaybreaks
   \nonumber \\
   &\ \times
   \sum_{i_{1},j_{1}=1}^{L}l_{i_{1},j_{1}}
   \sum_{i_{2},j_{2}=1}^{L}l_{i_{2},j_{2}}
   \sum_{\sigma_{1},\sigma_{2}}
   (\bm{e}_{\mathrm{sc}}^{*}\cdot\bm{d}_{i_{2},j_{2}})
   (\bm{e}_{\mathrm{in}}\cdot\bm{d}_{i_{1},j_{1}})
   \allowdisplaybreaks
   \nonumber \\
   &\ \times
   \mathcal{P}
    c_{i_{2}\sigma_{2}}^{\dagger}c_{j_{2}\sigma_{2}}
    c_{i_{1}\sigma_{1}}^{\dagger}c_{j_{1}\sigma_{1}}
   \mathcal{P}
  =\frac{2\pi e^{2}}{\hbar V\sqrt{\omega_{\mathrm{in}}\omega_{\mathrm{sc}}}}
   \frac{t^{2}}{U-\hbar\omega_{\mathrm{in}}}
   \allowdisplaybreaks
   \nonumber \\
   &\ \times
   \sum_{i_{1},j_{1}=1}^{L}l_{i_{1},j_{1}}
   \sum_{\sigma_{1},\sigma_{2}}
   (\bm{e}_{\mathrm{sc}}^{*}\cdot\bm{d}_{j_{1},i_{1}})
   (\bm{e}_{\mathrm{in}}\cdot\bm{d}_{i_{1},j_{1}})
   \allowdisplaybreaks
   \nonumber \\
   &\ \times
   \mathcal{P}
    c_{j_{1}\sigma_{2}}^{\dagger}c_{i_{1}\sigma_{2}}
    c_{i_{1}\sigma_{1}}^{\dagger}c_{j_{1}\sigma_{1}}
   \mathcal{P}.
   \label{E:R2I}
\end{align}
Applying \eqref{E:e-to-spin} to \eqref{E:R2I} and discarding Rayleigh (elastic scattering) terms,
we obtain what they call the Loudon-Fleury Raman vertex \cite{F514}
\begin{align}
   {}^{[2]}\mathcal{R}
  =\frac{2\pi e^{2}}{\hbar V\sqrt{\omega_{\mathrm{in}}\omega_{\mathrm{sc}}}}
   \frac{4t^{2}}{U-\hbar\omega_{\mathrm{in}}}
   \sum_{\langle i,j \rangle}
   (\bm{e}_{\mathrm{in}}\cdot\bm{d}_{i,j})
   (\bm{e}_{\mathrm{sc}}^{*}\cdot\bm{d}_{i,j})
   \bm{S}_{i}\cdot\bm{S}_{j}.
   \label{E:[2]R}
\end{align}
The Loudon-Fleury second-order mechanism predominates in the Raman response
when the incident photon energy $\hbar\omega_{\mathrm{in}}$ is in the far-resonant regime,
$t\ll |U-\hbar\omega_{\mathrm{in}}|$.

   The $n=1$ term in \eqref{E:GERO}, which is of third order in $t$, and the odd-integral-$n$
ones in general, vanish by virtue of the electron-hole symmetry.
Note that even though $n$ is an odd integer, the electronic state can come back again into
the singly-occupied ground-state manifold at the end in a triangular lattice, for instance.
With the electron band being half filled, creation and annihilation of an electron can be
described in terms of a hole as
\begin{align}
   d_{i\sigma}^\dagger=(-1)^{\delta_{\sigma,\uparrow}}c_{i\bar{\sigma}},\ 
   c_{i\sigma}^\dagger=(-1)^{\delta_{\sigma,\downarrow}}d_{i\bar{\sigma}}.
   \label{E:e-and-h}
\end{align}
With this in mind, the $n=1$ Raman vertex reads
{\color{white}blank blank}
\begin{widetext}
\vspace*{-3mm}
\begin{align}
   &
   {}^{[3]}\mathcal{R}
  =\frac{1}{(U-\hbar\omega_{\mathrm{in}})^{2}}
   \mathcal{P}\,
    {}^{[1]}\!\!\!\mathscr{J}
    \mathscr{H}_{t}\,
    {}^{[1]}\!\!\!\mathscr{J}
   \mathcal{P}
  =\frac{2\pi e^{2}}{\hbar V\sqrt{\omega_{\mathrm{in}}\omega_{\mathrm{sc}}}}
   \frac{t^{3}}{(U-\hbar\omega_{\mathrm{in}})^{2}}
   \sum_{i_{1},j_{1}=1}^{L}l_{i_{1},j_{1}}
   \sum_{i_{2},j_{2}=1}^{L}l_{i_{2},j_{2}}
   \sum_{i_{3},j_{3}=1}^{L}l_{i_{3},j_{3}}
   \allowdisplaybreaks
   \nonumber \\
   &\qquad\times
   \sum_{\sigma_{1},\sigma_{2},\sigma_{3}}
   (\bm{e}_{\mathrm{sc}}^{*}\cdot\bm{d}_{i_{3},j_{3}})
   (\bm{e}_{\mathrm{in}}\cdot\bm{d}_{i_{1},j_{1}})
   \mathcal{P}
    c_{i_{3}\sigma_{3}}^{\dagger}c_{j_{3}\sigma_{3}}
    c_{i_{2}\sigma_{2}}^{\dagger}c_{j_{2}\sigma_{2}}
    c_{i_{1}\sigma_{1}}^{\dagger}c_{j_{1}\sigma_{1}}
   \mathcal{P}
   \allowdisplaybreaks
   \nonumber \\[-1mm]
   &\quad\ 
  =\frac{2\pi e^{2}}{\hbar V\sqrt{\omega_{\mathrm{in}}\omega_{\mathrm{sc}}}}
   \frac{t^{3}}{(U-\hbar\omega_{\mathrm{in}})^{2}}
   \sum_{i_{1},j_{1}=1}^{L}l_{i_{1},j_{1}}
   \sum_{i_{2},j_{2}=1}^{L}l_{i_{2},j_{2}}
   \sum_{i_{3},j_{3}=1}^{L}l_{i_{3},j_{3}}
   \sum_{\sigma_{1},\sigma_{2},\sigma_{3}}
   (\bm{e}_{\mathrm{sc}}^{*}\cdot\bm{d}_{i_{3},j_{3}})
   (\bm{e}_{\mathrm{in}}\cdot\bm{d}_{i_{1},j_{1}})
   \allowdisplaybreaks
   \nonumber \\
   &\qquad\times
   \frac{1}{2}
   \mathcal{P}
   \left[
    c_{i_{3}\sigma_{3}}^{\dagger}c_{j_{3}\sigma_{3}}
    c_{i_{2}\sigma_{2}}^{\dagger}c_{j_{2}\sigma_{2}}
    c_{i_{1}\sigma_{1}}^{\dagger}c_{j_{1}\sigma_{1}}
   +(-1)^{3}
    d_{i_{3}\sigma_{3}}^{\dagger}d_{j_{3}\sigma_{3}}
    d_{i_{2}\sigma_{2}}^{\dagger}d_{j_{2}\sigma_{2}}
    d_{i_{1}\sigma_{1}}^{\dagger}d_{j_{1}\sigma_{1}}
   \right]
   \mathcal{P}.
   \label{E:R3I}
\end{align}
Substituting \eqref{E:e-and-h} into \eqref{E:e-to-spin} with the single-occupancy constraint
$\sum_\sigma d_{i\sigma}^\dagger d_{i\sigma}=1$ in mind yields
\begin{align}
   \mathcal{P}c_{i\sigma_{1}}^{\dagger}c_{i\sigma_{2}}\mathcal{P}
  =\mathcal{P}
   (-1)^{\delta_{\sigma_{1},\downarrow}+\delta_{\sigma_{2},\downarrow}}
   \left(
     \delta_{\bar{\sigma}_{1},\bar{\sigma}_{2}}
    -d_{i\bar{\sigma}_{2}}^{\dagger}d_{i\bar{\sigma}_{1}}
   \right)
   \mathcal{P}
  =\mathcal{P}
   \left[
    \delta_{\sigma_{1},\sigma_{2}}
   +(-1)^{\delta_{\sigma_{1},\sigma_{2}}}
    d_{i\bar{\sigma}_{2}}^{\dagger}d_{i\bar{\sigma}_{1}}
   \right]
   \mathcal{P}
  =\mathcal{P}d_{i\sigma_{1}}^{\dagger}d_{i\sigma_{2}}\mathcal{P},
   \label{E:h-to-spin}
\end{align}
showing that the electron and hole at each site have the same spin projection, and therefore,
\eqref{E:R3I} vanishes.
Likewise, all the odd-integral-$n$ vertices contribute nothing to the Raman intensity
of our system \eqref{E:Hel-ph}.

   The next leading $n=2$ term in \eqref{E:GERO}, which is of fourth order in $t$, reads
\vspace*{-1mm}
\begin{align}
   &
   {}^{[4]}\mathcal{R}
  =\frac{-1}{(U-\hbar\omega_{\mathrm{in}})^{3}}
   \mathcal{P}\,
    {}^{[1]}\!\!\!\mathscr{J}
    \mathscr{H}_{t}
    \mathscr{H}_{t}\,
    {}^{[1]}\!\!\!\mathscr{J}
   \mathcal{P}
  =\frac{2\pi e^{2}}{\hbar V\sqrt{\omega_{\mathrm{in}}\omega_{\mathrm{sc}}}}
   \frac{t^{4}}{(U-\hbar\omega_{\mathrm{in}})^{3}}
   \sum_{i_{1},j_{1}=1}^{L}l_{i_{1},j_{1}}
   \sum_{i_{2},j_{2}=1}^{L}l_{i_{2},j_{2}}
   \sum_{i_{3},j_{3}=1}^{L}l_{i_{3},j_{3}}
   \sum_{i_{4},j_{4}=1}^{L}l_{i_{4},j_{4}}
   \allowdisplaybreaks
   \nonumber \\
   &\qquad\times
   \sum_{\sigma_{1},\sigma_{2},\sigma_{3},\sigma_{4}}
   (\bm{e}_{\mathrm{sc}}^{*}\cdot\bm{d}_{i_{4},j_{4}})
   (\bm{e}_{\mathrm{in}}\cdot\bm{d}_{i_{1},j_{1}})
   \mathcal{P}
    c_{i_{4}\sigma_{4}}^{\dagger}c_{j_{4}\sigma_{4}}
    c_{i_{3}\sigma_{3}}^{\dagger}c_{j_{3}\sigma_{3}}
    c_{i_{2}\sigma_{2}}^{\dagger}c_{j_{2}\sigma_{2}}
    c_{i_{1}\sigma_{1}}^{\dagger}c_{j_{1}\sigma_{1}}
   \mathcal{P}.
   \label{E:R4I}
\end{align}
Figure \ref{F:PATH4TH} shows in what order how many electrons move in a variety of fourth-order
hopping paths.
After some algebra, we find the fourth-order Raman vertex as
\vspace*{-1mm}
\begin{align}
   &
   {}^{[4]}\mathcal{R}
  =\frac{2\pi e^{2}}{\hbar V\sqrt{\omega_{\mathrm{in}}\omega_{\mathrm{sc}}}}
   \frac{t^{4}}{(U-\hbar\omega_{\mathrm{in}})^{3}}
   \sum_{\langle i_{1},i_{2},i_{3},i_{4} \rangle}
   \left\{
   \vphantom{\sum_{n=1}^{2}\mathcal{D}_{n}^{\times}\bm{S}_{i_{n}}\cdot\bm{S}_{i_{n+2}}}
   -4\mathcal{Q}
    \left[\vphantom{\Bigl[}
      (\bm{S}_{i_{1}}\cdot\bm{S}_{i_{2}})(\bm{S}_{i_{3}}\cdot\bm{S}_{i_{4}})
     +(\bm{S}_{i_{1}}\cdot\bm{S}_{i_{4}})(\bm{S}_{i_{3}}\cdot\bm{S}_{i_{2}})
     -(\bm{S}_{i_{1}}\cdot\bm{S}_{i_{3}})(\bm{S}_{i_{2}}\cdot\bm{S}_{i_{4}})
    \right]
   \right.
   \allowdisplaybreaks
   \nonumber \\[-2mm]
   &\qquad\qquad\qquad\qquad\qquad\qquad\qquad\qquad \quad
   \left.
   +2i\sum_{n=1}^{4}
    \mathcal{T}_{n}
    \bm{S}_{i_{n+2}}\cdot(\bm{S}_{i_{n+1}}\times\bm{S}_{i_{n}})
   +\sum_{n=1}^{4}
    \mathcal{D}_{n}^{+}
    \bm{S}_{i_{n}}\cdot\bm{S}_{i_{n+1}}
   +\sum_{n=1}^{2}
    \mathcal{D}_{n}^{\times}
    \bm{S}_{i_{n}}\cdot\bm{S}_{i_{n+2}}
   \right\}
   \allowdisplaybreaks
   \nonumber \\[-1mm]
   &\qquad
   +\frac{2\pi e^{2}}{\hbar V\sqrt{\omega_{\mathrm{in}}\omega_{\mathrm{sc}}}}
    \frac{t^{4}}{(U-\hbar\omega_{\mathrm{in}})^{3}}
    \sum_{\langle i_{1},i_{2},i_{3} \rangle}
   \left\{
    4i\widetilde{\mathcal{T}}
    \bm{S}_{i_{3}}\cdot(\bm{S}_{i_{2}}\times\bm{S}_{i_{1}})
   +2\sum_{n=1}^{2}
    \widetilde{\mathcal{D}}_{n}^{+}
    \bm{S}_{i_{n}}\cdot\bm{S}_{i_{n+1}}
   -2\widetilde{\mathcal{D}}^{\times}
    \bm{S}_{i_{1}}\cdot\bm{S}_{i_{3}}
   \right\};
   \label{E:[4]R}
   \allowdisplaybreaks
   \\[-1mm]
   &
   \mathcal{Q}
  \equiv
   \sum_{n=1}^{4}
    (\bm{e}_{\mathrm{in}}\cdot\bm{d}_{n,n+1})
    \left[\bm{e}_{\mathrm{sc}}^{*}\cdot( \bm{d}_{n+1,n+2}
                                        +2\bm{d}_{n+2,n+3}
                                        +\bm{d}_{n+3,n})\right],
   \tag{\ref{E:[4]R}a}
   \allowdisplaybreaks
   \\[-1mm]
   &
   \mathcal{T}_{n}
  \equiv
   (\bm{e}_{\mathrm{in}}\cdot\bm{d}_{n,n+1})
   \left[\bm{e}_{\mathrm{sc}}^{*}\cdot(-\bm{d}_{n+1,n+2}
                                       -2\bm{d}_{n+2,n+3}
                                       +\bm{d}_{n+3,n})\right]
   +(\bm{e}_{\mathrm{in}}\cdot\bm{d}_{n+1,n+2})
    \left[\bm{e}_{\mathrm{sc}}^{*}\cdot(-\bm{d}_{n+2,n+3}
                                        +2\bm{d}_{n+3,n}
                                        +\bm{d}_{n,n+1})\right]
   \allowdisplaybreaks
   \nonumber \\
   &\quad
   +(\bm{e}_{\mathrm{in}}\cdot\bm{d}_{n+2,n+3})
    \left[\bm{e}_{\mathrm{sc}}^{*}\cdot( \bm{d}_{n+3,n}
                                        +2\bm{d}_{n,n+1}
                                        +\bm{d}_{n+1,n+2})\right]
   +(\bm{e}_{\mathrm{in}}\cdot\bm{d}_{n+3,n})
    \left[\bm{e}_{\mathrm{sc}}^{*}\cdot(-\bm{d}_{n,n+1}
                                        -2\bm{d}_{n+1,n+2}
                                        -\bm{d}_{n+2,n+3})\right],
   \tag{\ref{E:[4]R}b}
   \allowdisplaybreaks
   \\
   &
   \mathcal{D}_{n}^{+}
  \equiv
    (\bm{e}_{\mathrm{in}}\cdot\bm{d}_{n,n+1})
    \left[\bm{e}_{\mathrm{sc}}^{*}\cdot(-\bm{d}_{n+1,n+2}
                                        +2\bm{d}_{n+2,n+3}
                                        -\bm{d}_{n+3,n})\right]
   +(\bm{e}_{\mathrm{in}}\cdot\bm{d}_{n+1,n+2})
    \left[\bm{e}_{\mathrm{sc}}^{*}\cdot( \bm{d}_{n+2,n+3}
                                        -2\bm{d}_{n+3,n}
                                        -\bm{d}_{n,n+1})\right]
   \allowdisplaybreaks
   \nonumber \\
   &\quad
  +(\bm{e}_{\mathrm{in}}\cdot\bm{d}_{n+2,n+3})
    \left[\bm{e}_{\mathrm{sc}}^{*}\cdot( \bm{d}_{n+3,n}
                                        +2\bm{d}_{n,n+1}
                                        +\bm{d}_{n+1,n+2})\right]
   +(\bm{e}_{\mathrm{in}}\cdot\bm{d}_{n+3,n})
    \left[\bm{e}_{\mathrm{sc}}^{*}\cdot(-\bm{d}_{n,n+1}
                                        -2\bm{d}_{n+1,n+2}
                                        +\bm{d}_{n+2,n+3})\right],
   \allowdisplaybreaks
   \nonumber \\
   &
   \mathcal{D}_{n}^{\times}
  \equiv
    (\bm{e}_{\mathrm{in}}\cdot\bm{d}_{n,n+1})
    \left[\bm{e}_{\mathrm{sc}}^{*}\cdot( \bm{d}_{n+1,n+2}
                                        +2\bm{d}_{n+2,n+3}
                                        -\bm{d}_{n+3,n})\right]
   +(\bm{e}_{\mathrm{in}}\cdot\bm{d}_{n+1,n+2})
    \left[\bm{e}_{\mathrm{sc}}^{*}\cdot(-\bm{d}_{n+2,n+3}
                                        +2\bm{d}_{n+3,n}
                                        +\bm{d}_{n,n+1})\right]
   \allowdisplaybreaks
   \nonumber \\
   &\quad
  +(\bm{e}_{\mathrm{in}}\cdot\bm{d}_{n+2,n+3})
    \left[\bm{e}_{\mathrm{sc}}^{*}\cdot( \bm{d}_{n+3,n}
                                        +2\bm{d}_{n,n+1}
                                        -\bm{d}_{n+1,n+2})\right]
   +(\bm{e}_{\mathrm{in}}\cdot\bm{d}_{n+3,n})
    \left[\bm{e}_{\mathrm{sc}}^{*}\cdot(-\bm{d}_{n,n+1}
                                        +2\bm{d}_{n+1,n+2}
                                        +\bm{d}_{n+2,n+3})\right],
   \tag{\ref{E:[4]R}c}
   \allowdisplaybreaks
   \\
   &
   \widetilde{\mathcal{T}}
  \equiv
    (\bm{e}_{\mathrm{in}}\cdot\bm{d}_{1,2})
    (\bm{e}_{\mathrm{sc}}^{*}\cdot\bm{d}_{2,3})
   -(\bm{e}_{\mathrm{in}}\cdot\bm{d}_{2,3})
    (\bm{e}_{\mathrm{sc}}^{*}\cdot\bm{d}_{1,2}),
   \tag{\ref{E:[4]R}d}
   \allowdisplaybreaks
   \\
   &
   \widetilde{\mathcal{D}}_{n}^{+}
  \equiv
    (\bm{e}_{\mathrm{in}}\cdot\bm{d}_{n,n+1})
    \left[\bm{e}_{\mathrm{sc}}^{*}\cdot(\bm{d}_{1,2}+\bm{d}_{2,3})\right]
   +\left[\bm{e}_{\mathrm{in}}\cdot(\bm{d}_{1,2}+\bm{d}_{2,3})\right]
    (\bm{e}_{\mathrm{sc}}^{*}\cdot\bm{d}_{n,n+1}),
   \allowdisplaybreaks
   \nonumber \\
   &
   \widetilde{\mathcal{D}}^{\times}
  \equiv
   (\bm{e}_{\mathrm{in}}\cdot\bm{d}_{1,2})
   (\bm{e}_{\mathrm{sc}}^{*}\cdot\bm{d}_{2,3})
  +(\bm{e}_{\mathrm{in}}\cdot\bm{d}_{2,3})
   (\bm{e}_{\mathrm{sc}}^{*}\cdot\bm{d}_{1,2}),
   \tag{\ref{E:[4]R}e}
\end{align}
where
$\sum_{\langle i_{1},i_{2},i_{3},i_{4} \rangle}$ and
$\sum_{\langle i_{1},i_{2},i_{3} \rangle}$ run over
four-site-cyclic and three-site-round paths, respectively,
and
we abbreviate $\bm{d}_{i_{n},i_{n'}}\equiv\bm{r}_{i_{n'}}-\bm{r}_{i_{n}}$ as $\bm{d}_{n,n'}$
with $1\leq n,n' \pmod{4} \leq 4$.
The Shastry-Shraiman fourth-order mechanism is of major importance in the Raman response
when the incident photon energy $\hbar\omega_{\mathrm{in}}$ is in the near-resonant regime,
$t \simeq |U-\hbar\omega_{\mathrm{in}}|$.
\end{widetext}

\vspace*{-12pt}
   Intending to further express the Raman operators \eqref{E:[2]R} and \eqref{E:[4]R} in terms of
the bosonic language, we classify them according to the number of their constituent spin operators,
which we shall denote by $\tau$,
\begin{align}
   \mathcal{R}
  \equiv
   \sum_{n=1}^{1\,\mathrm{or}\,2}
   {}^{[2n]}\mathcal{R}
  =\sum_{n=1}^{1\,\mathrm{or}\,2}
   \sum_{\tau=2}^{2n}
   {}_{\;\;\;\;\tau}^{[2n]}\mathcal{R}.
\end{align}
${}_{\;\;\;\;2}^{[2n]}\mathcal{R}$,
${}_{\;\;\;\;3}^{[2n]}\mathcal{R}$, and
${}_{\;\;\;\;4}^{[2n]}\mathcal{R}$ consist of
the Heisenberg pair exchange, three-spin scalar-chirality, and four-spin ring-exchange terms,
respectively.
They also read descending powers of the spin magnitude and each further break into components
of different numbers of mediating magnons,
\begin{align}
   {}_{\;\;\;\;\tau}^{[2n]}\mathcal{R}
  =\sum_{m=0}^\infty
   {}_{\;\;\;\;\tau}^{[2n]}\mathcal{R}^{(\tau-m)}
  =\sum_{m=0}^\infty\sum_{l=0}^m
   {}_{\;\;\;\;\tau}^{[2n]}\mathcal{R}_{2l\mathrm{M}}^{(\tau-m)},
   \tag{\ref{E:Rfull}}
\end{align}
where ${}_{\;\;\;\;\tau}^{[2n]}\mathcal{R}^{(\tau-m)}$ is of order $S^{\tau-m}$ and
${}_{\;\;\;\;\tau}^{[2n]}\mathcal{R}_{2l\mathrm{M}}^{(\tau-m)}$ is a linear combination of
terms containing $2l$ magnon operators.
We truncate the inverse-spin-magnitude series \eqref{E:Rfull} at $m=2$ to have the up-to-$O(S^0)$
Raman vertices
\begin{align}
   \mathcal{R}
  \simeq
   \sum_{n=1}^{p/2}
   \sum_{l=0}^{2}
   {}^{[2n]}\mathcal{R}_{2l\mathrm{M}}
  =\sum_{n=1}^{p/2}
   \sum_{\tau=2}^{2n}
   \sum_{l=0}^{2}
   \sum_{m=l}^2
   {}_{\;\;\;\;\tau}^{[2n]}\mathcal{R}_{2l\mathrm{M}}^{(\tau-m)},
   \tag{\ref{E:RuptoO(S^0)}}
\end{align}
where setting $p$ equal to $2$ and $4$ corresponds to the Loudon-Fleury second-order and
Shastry-Shraiman fourth-order perturbation schemes, respectively.
We employ Wick's theorem so as for all the vertices
${}_{\;\;\;\;\tau}^{[2n]}\mathcal{R}_{2l\mathrm{M}}^{(\tau-m)}$
to be normal-ordered with respect to the quasiparticle magnon operators.
${}^{[2n]}\mathcal{R}_{0\mathrm{M}}$ merely contribute to elastic (Rayleigh) scatterings and
are henceforth omitted.

   With a tacit understanding of site indices being used as $i,k\in\mathrm{A}$ and
$j,l\in\mathrm{B}$, various spin interactions are written in terms of HP bosons as
{\color{white}blank blank blank blank blank}
\begin{widetext}
\begin{align}
   &
   \bm{S}_{i}\cdot\bm{S}_{j}
  =-S^{2}
   +S
   \left(
    a_{i}^{\dagger}a_{i}
   +b_{j}^{\dagger}b_{j}
   +a_{i}b_{j}
   +a_{i}^{\dagger}b_{j}^{\dagger}
   \right)
   -\left[
    a_{i}^{\dagger}a_{i}b_{j}^{\dagger}b_{j}
    +\frac{1}{4}
    \left(
     a_{i}^{\dagger}a_{i}a_{i}b_{j}
    +a_{i}^{\dagger}b_{j}^{\dagger}b_{j}^{\dagger}b_{j}
    +\mathrm{H.c.}
    \right)
   \right]
   +O(S^{-1}),
   \allowdisplaybreaks
   \\
   &
   \bm{S}_{i}\cdot\bm{S}_{k}
  =S^{2}
   +S
   \left(
   -a_{i}^{\dagger}a_{i}
   -a_{k}^{\dagger}a_{k}
   +a_{i}a_{k}^{\dagger}
   +a_{i}^{\dagger}a_{k}
   \right)
   +\left[
    a_{i}^{\dagger}a_{i}a_{k}^{\dagger}a_{k}
    -\frac{1}{4}
    \left(
     a_{i}^{\dagger}a_{i}a_{i}a_{k}^{\dagger}
    +a_{i}^{\dagger}a_{k}^{\dagger}a_{k}a_{k}
    +\mathrm{H.c.}
    \right)
   \right]
   +O(S^{-1}),
   \allowdisplaybreaks
   \\
   &
   \bm{S}_{j}\cdot\bm{S}_{l}
  =S^{2}
   +S
   \left(
   -b_{j}^{\dagger}b_{j}
   -b_{l}^{\dagger}b_{l}
   +b_{j}b_{l}^{\dagger}
   +b_{j}^{\dagger}b_{l}
   \right)
   +\left[
    b_{j}^{\dagger}b_{j}b_{l}^{\dagger}b_{l}
    -\frac{1}{4}
    \left(
     b_{j}^{\dagger}b_{j}b_{j}b_{l}^{\dagger}
    +b_{j}^{\dagger}b_{l}^{\dagger}b_{l}b_{l}
    +\mathrm{H.c.}
    \right)
   \right]
   +O(S^{-1}),
   \allowdisplaybreaks
   \\
   &
   i\bm{S}_{i}\cdot(\bm{S}_{j}\times\bm{S}_{k})
  =S^{2}
   \left(
    a_{i}^{\dagger}b_{j}^{\dagger}
   +a_{i}^{\dagger}a_{k}
   +a_{k}b_{j}
   -\mathrm{H.c.}
   \right)
   +S
   \left[
    a_{i}b_{j}a_{k}^{\dagger}a_{k}
   +a_{i}a_{k}^{\dagger}b_{j}^{\dagger}b_{j}
   +a_{i}^{\dagger}a_{i}a_{k}^{\dagger}b_{j}^{\dagger}
   \vphantom{\frac{1}{4}}
   \right.
   \allowdisplaybreaks
   \nonumber \\
   &\qquad
   \left.
   +\frac{1}{4}
    \left(
     a_{i}^{\dagger}a_{i}a_{i}b_{j}
    +a_{i}b_{j}^{\dagger}b_{j}b_{j}
    +a_{i}a_{k}^{\dagger}a_{k}^{\dagger}a_{k}
    +a_{i}^{\dagger}a_{i}a_{i}a_{k}^{\dagger}
    +a_{k}^{\dagger}a_{k}^{\dagger}a_{k}b_{j}^{\dagger}
    +a_{k}^{\dagger}b_{j}^{\dagger}b_{j}^{\dagger}b_{j}
    \right)
   -\mathrm{H.c.}
   \right]
   +O(S^{0}),
   \allowdisplaybreaks
   \\
   &
   i\bm{S}_{j}\cdot(\bm{S}_{k}\times\bm{S}_{l})
  =S^{2}
   \left(
    a_{k}^{\dagger}b_{j}^{\dagger}
   +b_{j}^{\dagger}b_{l}
   +a_{k}b_{l}
   -\mathrm{H.c.}
   \right)
   +S
   \left[
    a_{k}b_{j}b_{l}^{\dagger}b_{l}
   +a_{k}^{\dagger}a_{k}b_{j}b_{l}^{\dagger}
   +b_{j}^{\dagger}b_{j}a_{k}^{\dagger}b_{l}
   \vphantom{\frac{1}{4}}
   \right.
   \allowdisplaybreaks
   \nonumber \\
   &\qquad
   \left.
   +\frac{1}{4}
    \left(
     a_{k}^{\dagger}a_{k}a_{k}b_{j}
    +a_{k}b_{j}^{\dagger}b_{j}b_{j}
    +b_{j}b_{l}^{\dagger}b_{l}^{\dagger}b_{l}
    +b_{j}^{\dagger}b_{j}b_{j}b_{l}^{\dagger}
    +a_{k}^{\dagger}a_{k}^{\dagger}a_{k}b_{l}^{\dagger}
    +a_{k}^{\dagger}b_{l}^{\dagger}b_{l}^{\dagger}b_{l}
    \right)
   -\mathrm{H.c.}
   \right]
   +O(S^{0}),
   \allowdisplaybreaks
   \\
   &
   (\bm{S}_{i}\cdot\bm{S}_{j})(\bm{S}_{k}\cdot\bm{S}_{l})
  =S^{4}
   -S^{3}
   \left(
    a_{i}^{\dagger}a_{i}
   +b_{j}^{\dagger}b_{j}
   +a_{i}b_{j}
   +a_{i}^{\dagger}b_{j}^{\dagger}
   +a_{k}^{\dagger}a_{k}
   +b_{l}^{\dagger}b_{l}
   +a_{k}b_{l}
   +a_{k}^{\dagger}b_{l}^{\dagger}
   \right)
   \allowdisplaybreaks
   \nonumber \\
   &\qquad
   +S^{2}
   \left[
    a_{i}^{\dagger}a_{i}a_{k}^{\dagger}a_{k}
   +a_{i}^{\dagger}a_{i}b_{l}^{\dagger}b_{l}
   +a_{i}^{\dagger}a_{i}a_{k}b_{l}
   +a_{i}^{\dagger}a_{i}a_{k}^{\dagger}b_{l}^{\dagger}
   +b_{j}^{\dagger}b_{j}a_{k}^{\dagger}a_{k}
   +b_{j}^{\dagger}b_{j}b_{l}^{\dagger}b_{l}
   +b_{j}^{\dagger}b_{j}a_{k}b_{l}
   +b_{j}^{\dagger}b_{j}a_{k}^{\dagger}b_{l}^{\dagger}
   \vphantom{\frac{1}{4}}
   \right.
   \allowdisplaybreaks
   \nonumber \\
   &\qquad
   +a_{i}b_{j}a_{k}^{\dagger}a_{k}
   +a_{i}b_{j}b_{l}^{\dagger}b_{l}
   +a_{i}b_{j}a_{k}b_{l}
   +a_{i}b_{j}a_{k}^{\dagger}b_{l}^{\dagger}
   +a_{i}^{\dagger}b_{j}^{\dagger}a_{k}^{\dagger}a_{k}
   +a_{i}^{\dagger}b_{j}^{\dagger}b_{l}^{\dagger}b_{l}
   +a_{i}^{\dagger}b_{j}^{\dagger}a_{k}b_{l}
   +a_{i}^{\dagger}b_{j}^{\dagger}a_{k}^{\dagger}b_{l}^{\dagger}
   \allowdisplaybreaks
   \nonumber \\
   &\qquad
   \left.
   +a_{i}^{\dagger}a_{i}b_{j}^{\dagger}b_{j}
   +\frac{1}{4}
    \left(
     a_{i}^{\dagger}a_{i}a_{i}b_{j}
    +a_{i}^{\dagger}b_{j}^{\dagger}b_{j}^{\dagger}b_{j}
    +\mathrm{H.c.}
    \right)
   +a_{k}^{\dagger}a_{k}b_{l}^{\dagger}b_{l}
   +\frac{1}{4}
    \left(
     a_{k}^{\dagger}a_{k}a_{k}b_{l}
    +a_{k}^{\dagger}b_{l}^{\dagger}b_{l}^{\dagger}b_{l}
    +\mathrm{H.c.}
    \right)
   \right]
   +O(S^{1}),
   \allowdisplaybreaks
   \\
   &
   (\bm{S}_{i}\cdot\bm{S}_{k})(\bm{S}_{j}\cdot\bm{S}_{l})
  =S^{4}
   +S^{3}
   \left(
   -a_{i}^{\dagger}a_{i}
   -a_{k}^{\dagger}a_{k}
   +a_{i}a_{k}^{\dagger}
   +a_{i}^{\dagger}a_{k}
   -b_{j}^{\dagger}b_{j}
   -b_{l}^{\dagger}b_{l}
   +b_{j}^{\dagger}b_{l}
   +b_{j}b_{l}^{\dagger}
   \right)
   \allowdisplaybreaks
   \nonumber \\
   &\qquad
   +S^{2}
   \left[
    a_{i}^{\dagger}a_{i}b_{j}^{\dagger}b_{j}
   +a_{i}^{\dagger}a_{i}b_{l}^{\dagger}b_{l}
   -a_{i}^{\dagger}a_{i}b_{j}^{\dagger}b_{l}
   -a_{i}^{\dagger}a_{i}b_{j}b_{l}^{\dagger}
   +a_{k}^{\dagger}a_{k}b_{j}^{\dagger}b_{j}
   +a_{k}^{\dagger}a_{k}b_{l}^{\dagger}b_{l}
   -a_{k}^{\dagger}a_{k}b_{j}^{\dagger}b_{l}
   -a_{k}^{\dagger}a_{k}b_{j}b_{l}^{\dagger}
   \vphantom{\frac{1}{4}}
   \right.
   \allowdisplaybreaks
   \nonumber \\
   &\qquad
   -a_{i}a_{k}^{\dagger}b_{j}^{\dagger}b_{j}
   -a_{i}a_{k}^{\dagger}b_{l}^{\dagger}b_{l}
   +a_{i}a_{k}^{\dagger}b_{j}^{\dagger}b_{l}
   +a_{i}a_{k}^{\dagger}b_{j}b_{l}^{\dagger}
   -a_{i}^{\dagger}a_{k}b_{j}^{\dagger}b_{j}
   -a_{i}^{\dagger}a_{k}b_{l}^{\dagger}b_{l}
   +a_{i}^{\dagger}a_{k}b_{j}^{\dagger}b_{l}
   +a_{i}^{\dagger}a_{k}b_{j}b_{l}^{\dagger}
   \allowdisplaybreaks
   \nonumber \\
   &\qquad
   \left.
   +a_{i}^{\dagger}a_{i}a_{k}^{\dagger}a_{k}
   -\frac{1}{4}
    \left(
     a_{i}^{\dagger}a_{i}a_{i}a_{k}^{\dagger}
    +a_{i}^{\dagger}a_{k}^{\dagger}a_{k}a_{k}
    +\mathrm{H.c.}
    \right)
   +b_{j}^{\dagger}b_{j}b_{l}^{\dagger}b_{l}
   -\frac{1}{4}
    \left(
     b_{j}^{\dagger}b_{j}^{\dagger}b_{j}b_{l}
    +b_{j}b_{l}^{\dagger}b_{l}^{\dagger}b_{l}
    +\mathrm{H.c.}
    \right)
   \right]
   +O(S^{1}).
\end{align}
\end{widetext}

\noindent
Via the Bogoliubov transformation \eqref{E:BT}, the two-magnon (2M)- and four-magnon (4M)-mediated
Raman scattering operators read
\vspace{-1mm}
\begin{align}
   &
   \sum_{n=1}^{p/2}{}^{[2n]}\mathcal{R}_{2\mathrm{M}}
  =\sum_{l_{-},l_{-}'}
   {}^{[p]}W_{l_{-}l_{-}'}^{(1)}
   \alpha_{l_{-}}^{-\dagger}\alpha_{l_{-}'}^{-}
  +\sum_{l_{+},l_{-}'}
   {}^{[p]}W_{l_{+}l_{-}'}^{(2)}
   \alpha_{l_{+}}^{+\dagger}\alpha_{l_{-}'}^{-\dagger}
   \allowdisplaybreaks
   \nonumber \\[-1pt]
   &\qquad\qquad\quad
  +\sum_{l_{-},l_{+}'}
   {}^{[p]}W_{l_{-}l_{+}'}^{(3)}
   \alpha_{l_{-}}^{-}\alpha_{l_{+}'}^{+}
  +\sum_{l_{+},l_{+}'}
   {}^{[p]}W_{l_{+}l_{+}'}^{(4)}
   \alpha_{l_{+}}^{+\dagger}\alpha_{l_{+}'}^{+}
   \label{E:R2M=W(1)toW(4)}
\end{align}
with
$
   {}^{[p]}W_{l_{+}l_{-}'}^{(2)}
  ={}^{[p]}W_{l_{-}'l_{+}}^{(3)\,*}
$
and
\vspace{-1mm}
\begin{align}
   &
   \sum_{n=1}^{p/2}{}^{[2n]}\mathcal{R}_{4\mathrm{M}}
  =\sum_{l_{-},l_{-}',l_{-}'',l_{-}'''}
    {}^{[p]}X_{l_{-} l_{-}' l_{-}'' l_{-}'''}^{(1)}
    \alpha_{l_{-}}^{-\dagger}
    \alpha_{l_{-}'}^{-\dagger}
    \alpha_{l_{-}''}^{-}
    \alpha_{l_{-}'''}^{-}
   \allowdisplaybreaks
   \nonumber \\[-1pt]
   &\qquad\qquad\quad
  +\sum_{l_{-},l_{+}',l_{-}'',l_{-}'''}
    {}^{[p]}X_{l_{-} l_{+}' l_{-}'' l_{-}'''}^{(2)}
    \alpha_{l_{-}}^{-\dagger}
    \alpha_{l_{+}'}^{+}
    \alpha_{l_{-}''}^{-}
    \alpha_{l_{-}'''}^{-}
   \allowdisplaybreaks
   \nonumber \\[-1pt]
   &\qquad\qquad\quad
  +\sum_{l_{+},l_{-}',l_{-}'',l_{-}'''}
    {}^{[p]}X_{l_{+} l_{-}' l_{-}'' l_{-}'''}^{(3)}
    \alpha_{l_{+}}^{+\dagger}
    \alpha_{l_{-}'}^{-\dagger}
    \alpha_{l_{-}''}^{-\dagger}
    \alpha_{l_{-}'''}^{-}
   \allowdisplaybreaks
   \nonumber \\[-1pt]
   &\qquad\qquad\quad
  +\sum_{l_{+},l_{+}',l_{-}'',l_{-}'''}
    {}^{[p]}X_{l_{+} l_{+}'l_{-}'' l_{-}''' }^{(4)}
    \alpha_{l_{+}}^{+\dagger}
    \alpha_{l_{+}'}^{+}
    \alpha_{l_{-}''}^{-\dagger}
    \alpha_{l_{-}'''}^{-}
   \allowdisplaybreaks
   \nonumber \\[-1pt]
   &\qquad\qquad\quad
  +\sum_{l_{+},l_{+}',l_{+}'',l_{-}'''}
    {}^{[p]}X_{l_{+} l_{+}' l_{+}'' l_{-}'''}^{(5)}
    \alpha_{l_{+}}^{+\dagger}
    \alpha_{l_{+}'}^{+}
    \alpha_{l_{+}''}^{+}
    \alpha_{l_{-}'''}^{-}
   \allowdisplaybreaks
   \nonumber \\[-1pt]
   &\qquad\qquad\quad
  +\sum_{l_{+},l_{+}',l_{-}'',l_{+}'''}
    {}^{[p]}X_{l_{+} l_{+}' l_{-}'' l_{+}'''}^{(6)}
    \alpha_{l_{+}}^{+\dagger}
    \alpha_{l_{+}'}^{+\dagger}
    \alpha_{l_{-}''}^{-\dagger}
    \alpha_{l_{+}'''}^{+}
   \allowdisplaybreaks
   \nonumber \\[-1pt]
   &\qquad\qquad\quad
  +\sum_{l_{+},l_{+}',l_{-}'',l_{-}'''}
    {}^{[p]}X_{l_{+} l_{+}' l_{-}'' l_{-}'''}^{(7)}
    \alpha_{l_{+}}^{+\dagger}
    \alpha_{l_{+}'}^{+\dagger}
    \alpha_{l_{-}''}^{-\dagger}
    \alpha_{l_{-}'''}^{-\dagger}
   \allowdisplaybreaks
   \nonumber \\[-1pt]
   &\qquad\qquad\quad
  +\sum_{l_{+},l_{+}',l_{-}'',l_{-}'''}
    {}^{[p]}X_{l_{+} l_{+}' l_{-}'' l_{-}'''}^{(8)}
    \alpha_{l_{+}}^{+}
    \alpha_{l_{+}'}^{+}
    \alpha_{l_{-}''}^{-}
    \alpha_{l_{-}'''}^{-}
   \allowdisplaybreaks
   \nonumber \\[-1pt]
   &\qquad\qquad\quad
  +\sum_{l_{+},l_{+}',l_{+}'',l_{+}'''}
    {}^{[p]}X_{l_{+} l_{+}' l_{+}'' l_{+}'''}^{(9)}
    \alpha_{l_{+}}^{+\dagger}
    \alpha_{l_{+}'}^{+\dagger}
    \alpha_{l_{+}''}^{+}
    \alpha_{l_{+}'''}^{+}
   \label{E:R4M=X(1)toX(9)}
\end{align}
with
$
   {}^{[p]}X_{l_{-} l_{+}' l_{-}'' l_{-}'''}^{(2)}
  ={}^{[p]}X_{l_{+}' l_{-}''' l_{-}'' l_{-}}^{(3)\,*}
$,
$
   {}^{[p]}X_{l_{+} l_{+}' l_{+}'' l_{-}'''}^{(5)}
  ={}^{[p]}X_{l_{+}'' l_{+}' l_{-}''' l_{+}}^{(6)\,*}
$,
and
$
   {}^{[p]}X_{l_{+} l_{+}' l_{-}'' l_{-}'''}^{(7)}
  ={}^{[p]}X_{l_{+} l_{+}' l_{-}'' l_{-}'''}^{(8)\,*}
$.
The coefficients 
$
   {}^{[p]}W_{l_{\sigma}l_{\sigma'}'}^{(m)}
$ and 
$
   {}^{[p]}X_{l_{\sigma}l_{\sigma'}'l_{\sigma''}''l_{\sigma'''}'''}^{(m)}
$
are numerically obtained.

\section*{S3. Green's Function Formalism}
\vspace*{-1mm}
Putting
$
   \alpha(t)
  \equiv
   e^{ i\mathcal{H}t/\hbar}
   \alpha
   e^{-i\mathcal{H}t/\hbar}
$
for any operator $\alpha$,
we introduce the $l$($1\leq l \leq 4$)-magnon ($l$M) Green functions (GFs)
for the corrected ground state $|0\rangle$,
\vspace*{-1mm}
\begin{subequations}
\begin{align}
   &
   G^{k_{\tau}}_{l_{\sigma}}(t)
  \equiv
   -i
   \langle 0|
   \mathcal{T}
    \alpha_{k_{\tau}}^{\tau}(t)
    \alpha_{l_{\sigma}}^{\sigma\dagger}
   |0\rangle,
   \label{E:1to4MGF(t)a}
   \allowdisplaybreaks
   \\
   &
   G^{k_{\sigma}k_{\sigma'}'}_{l_{\sigma}l_{\sigma'}'}(t)
  \equiv
   -i
   \langle 0|
    \mathcal{T}
    \alpha_{k_{\sigma}}^{\sigma}(t)\alpha_{k_{\sigma'}'}^{\sigma'}(t)
    \alpha_{l_{\sigma}}^{\sigma\dagger}\alpha_{l_{\sigma'}'}^{\sigma'\dagger}
   |0\rangle,
   \label{E:1to4MGF(t)b}
   \allowdisplaybreaks
   \\
   &
   G^{k_{\sigma}k_{\sigma}'k_{\bar{\sigma}}''}
    _{l_{\sigma}l_{\sigma}'l_{\bar{\sigma}}''}(t)
  \equiv
   -i
   \langle 0|
    \mathcal{T}
    \alpha_{k_{\sigma} }^{\sigma}(t)
    \alpha_{k_{\sigma}'}^{\sigma}(t)
    \alpha_{k_{\bar{\sigma}}''}^{\bar{\sigma}}(t)
    \alpha_{l_{\sigma} }^{\sigma\dagger}
    \alpha_{l_{\sigma}'}^{\sigma\dagger}
    \alpha_{l_{\bar{\sigma}}''}^{\bar{\sigma}\dagger}
   |0\rangle,
   \label{E:1to4MGF(t)c}
   \allowdisplaybreaks
   \\
   &
   G^{k_{\sigma}k_{\sigma}'k_{\bar{\sigma}}''k_{\bar{\sigma}}'''}
    _{l_{\sigma}l_{\sigma}'l_{\bar{\sigma}}''l_{\bar{\sigma}}'''}(t)
   \allowdisplaybreaks
   \nonumber \\[-1mm]
   &
  \equiv
   -i
   \langle 0|
    \mathcal{T}
    \alpha_{k_{\sigma} }^{\sigma}(t)
    \alpha_{k_{\sigma}'}^{\sigma}(t)
    \alpha_{k_{\bar{\sigma}}'' }^{\bar{\sigma}}(t)
    \alpha_{k_{\bar{\sigma}}'''}^{\bar{\sigma}}(t)
    \alpha_{l_{\sigma} }^{\sigma\dagger}
    \alpha_{l_{\sigma}'}^{\sigma\dagger}
    \alpha_{l_{\bar{\sigma}}'' }^{\bar{\sigma}\dagger}
    \alpha_{l_{\bar{\sigma}}'''}^{\bar{\sigma}\dagger}
   |0\rangle,
   \label{E:1to4MGF(t)d}
\end{align}
\end{subequations}
where $\sigma(\equiv -\bar{\sigma})$, $\sigma'$, and $\tau$ take $\pm$ and $\mathcal{T}$
denotes the time-ordering operator with respect to the unperturbed magnon operators emergent
in \eqref{E:HBLdiag},
and define their Fourier transforms as
\vspace*{-2mm}
\begin{subequations}
\begin{align}
   &
   G^{k_{\tau}}_{l_{\sigma}}(\omega)
  \equiv
   \int_{-\infty}^{\infty}dt\, e^{i\omega t}
   G^{k_{\tau}}_{l_{\sigma}}(t),
   \label{E:1to4MGF(w)a}
   \allowdisplaybreaks
   \\
   &
   G^{k_{\sigma}k_{\sigma'}'}_{l_{\sigma}l_{\sigma'}'}(\omega)
  \equiv
   \int_{-\infty}^{\infty}dt\, e^{i\omega t}
   G^{k_{\sigma}k_{\sigma'}'}_{l_{\sigma}l_{\sigma'}'}(t),
   \label{E:1to4MGF(w)b}
   \allowdisplaybreaks 
   \\
   &
   G^{k_{\sigma}k_{\sigma}'k_{\bar{\sigma}}''}
    _{l_{\sigma}l_{\sigma}'l_{\bar{\sigma}}''}(\omega)
  \equiv
   \int_{-\infty}^{\infty}dt\, e^{i\omega t}
   G^{k_{\sigma}k_{\sigma}'k_{\bar{\sigma}}''}
    _{l_{\sigma}l_{\sigma}'l_{\bar{\sigma}}''}(t),
   \label{E:1to4MGF(w)c}
   \allowdisplaybreaks
   \\
   &
   G^{k_{\sigma}k_{\sigma}'k_{\bar{\sigma}}''k_{\bar{\sigma}}'''}
    _{l_{\sigma}l_{\sigma}'l_{\bar{\sigma}}''l_{\bar{\sigma}}'''}(\omega)
  \equiv
   \int_{-\infty}^{\infty}dt\, e^{i\omega t}
   G^{k_{\sigma}k_{\sigma}'k_{\bar{\sigma}}''k_{\bar{\sigma}}'''}
    _{l_{\sigma}l_{\sigma}'l_{\bar{\sigma}}''l_{\bar{\sigma}}'''}(t).
   \label{E:1to4MGF(w)d}
\end{align}
\end{subequations}
$\mathcal{H}_{\mathrm{BL}}$ is nothing but the $O(S^0)$ Hartree-Fock Hamiltonian and
the residual $O(S^0)$ interaction $:\mathcal{H}^{(0)}:$ has no effect on the 1M GFs,
i.e.,
$
   \langle 0|
    \mathcal{T}
    \alpha_{k_{\tau}}^{\tau}(t)
    \alpha_{l_{\sigma}}^{\sigma \dagger}
   |0\rangle
  ={}_{\mathrm{BL}}\langle 0|
    \mathcal{T}
    \alpha_{k_{\tau}}^{\tau}(t)_{\mathrm{BL}}
    \alpha_{l_{\sigma}}^{\sigma \dagger}
   |0\rangle_{\mathrm{BL}}
  \equiv
   iG^{k_{\tau}}_{l_{\sigma}}(t)_{\mathrm{BL}}
$
with
$
   \alpha(t)_{\mathrm{BL}}
  \equiv
   e^{ i\mathcal{H}_{\mathrm{BL}}t/\hbar}
   \alpha
   e^{-i\mathcal{H}_{\mathrm{BL}}t/\hbar}
$.
Then the Dyson equation for the 1M GFs reduces to
\vspace*{-1mm}
\begin{align}
   &
   G^{k_{\tau}}_{l_{\sigma}}(\omega)
  =G^{k_{\tau}}_{l_{\sigma}}(\omega)_{\mathrm{BL}}
  =\frac{\hbar\delta_{k_{\tau},l_{\sigma}}}
        {\hbar\omega-\varepsilon_{l_{\sigma}}^{\sigma}+i\eta},
\end{align}
where $\eta$ is understood to be infinitesimal.
Next we consider the ladder-approximation Bethe-Salpeter equations for the 2M GFs
\vspace{-1mm}
\begin{subequations}
\begin{align}
   G^{k_{+}k_{-}'}_{l_{+}l_{-}'}(\omega)
  =&G^{k_{+}k_{-}'}_{l_{+}l_{-}'}(\omega)_{\mathrm{BL}}
  -\frac{J}{\hbar}\sum_{p_{+},p_{-}'}\sum_{q_{+},q_{-}'}
   G^{k_{+}k_{-}'}_{p_{+}p_{-}'}(\omega)_{\mathrm{BL}}
   \allowdisplaybreaks
   \nonumber \\[-2pt]
  &\times
    \sum_{i\in\mathrm{A}}\sum_{j\in\mathrm{B}}l_{i,j}
    V_{ij;p_{+}q_{+}p_{-}'q_{-}'}^{(4)}
   G^{q_{+}q_{-}'}_{l_{+}l_{-}'}(\omega),
   \label{E:2lBS+-}
   \allowdisplaybreaks
   \\
   G^{k_{+}k_{+}'}_{l_{+}l_{+}'}(\omega)
  =&G^{k_{+}k_{+}'}_{l_{+}l_{+'}}(\omega)_{\mathrm{BL}}
  -\frac{J}{\hbar}\sum_{p_{+},p_{+}'}\sum_{q_{+},q_{+}'}
   G^{k_{+}k_{+}'}_{p_{+}p_{+}'}(\omega)_{\mathrm{BL}}
   \allowdisplaybreaks
   \nonumber \\[-2pt]
  &\times
    \sum_{i\in\mathrm{A}}\sum_{j\in\mathrm{B}}l_{i,j}
    V_{ij;p_{+}p_{+}'q_{+}q_{+}'}^{(9)}
   G^{q_{+}q_{+}'}_{l_{+}l_{+}'}(\omega),
   \label{E:2lBS++}
   \allowdisplaybreaks
   \\
   G^{k_{-}k_{-}'}_{l_{-}l_{-}'}(\omega)
  =&
   G^{k_{-}k_{-}'}_{l_{-}l_{-}'}(\omega)_{\mathrm{BL}}
  -\frac{J}{\hbar}\sum_{p_{-},p_{-}'}\sum_{q_{-},q_{-}'}
   G^{k_{-}k_{-}'}_{p_{-}p_{-}'}(\omega)_{\mathrm{BL}}
   \allowdisplaybreaks
   \nonumber \\[-2pt]
  &\times
    \sum_{i\in\mathrm{A}}\sum_{j\in\mathrm{B}}l_{i,j}
    V_{ij;p_{-}p_{-}'q_{-}q_{-}'}^{(1)}
   G^{q_{-}q_{-}'}_{l_{-}l_{-}'}(\omega),
   \label{E:2lBS--}
\end{align}
\end{subequations}
denoting the unperturbed 2M GFs by
\vspace{-1mm}
\begin{align}
   &
   G^{k_{\sigma}k_{\sigma'}'}_{l_{\sigma}l_{\sigma'}'}(\omega)_{\mathrm{BL}}
   \allowdisplaybreaks
   \nonumber \\
   &\quad
  \equiv
   \int_{-\infty}^{\infty}\frac{dt}{i}e^{i\omega t}
   {}_{\mathrm{BL}}\langle 0|
    \mathcal{T}
    \alpha_{k_{\sigma} }^{\sigma}(t)_{\mathrm{BL}}
    \alpha_{k_{\sigma'}'}^{\sigma'}(t)_{\mathrm{BL}}
    \alpha_{l_{\sigma }}^{\sigma\dagger}
    \alpha_{l_{\sigma'}'}^{\sigma'\dagger}
   |0\rangle_{\mathrm{BL}}
   \allowdisplaybreaks
   \nonumber \\
   &\quad
  =i\int_{-\infty}^{\infty}dt\, e^{i\omega t}
   \left[
    G^{k_{\sigma} }_{l_{\sigma} }(t)_{\mathrm{BL}}
    G^{k_{\sigma'}'}_{l_{\sigma'}'}(t)_{\mathrm{BL}}
   +G^{k_{\sigma} }_{l_{\sigma'}'}(t)_{\mathrm{BL}}
    G^{k_{\sigma'}'}_{l_{\sigma} }(t)_{\mathrm{BL}}
   \right]
   \allowdisplaybreaks
   \nonumber \\
   &\quad
  =\frac{\hbar
         (\delta_{k_{\sigma},l_{\sigma}}\delta_{k_{\sigma'}',l_{\sigma'}'}
         +\delta_{k_{\sigma},l_{\sigma'}'}\delta_{k_{\sigma'}',l_{\sigma}})}
        {\hbar\omega
         -\varepsilon_{l_{\sigma}}^{\sigma}-\varepsilon_{l_{\sigma'}'}^{\sigma'}+i\eta}.
\end{align}
We solve the eigenequations obtained from \eqref{E:2lBS+-}--\eqref{E:2lBS--},
\vspace{-1mm}
\begin{align}
   \mathcal{V}^{\sigma\sigma'}
   \bm{g}^{\sigma\sigma'}_{\lambda}
  =\hbar\varOmega_{\lambda}^{\sigma\sigma'}
   \bm{g}^{\sigma\sigma'}_{\lambda},
\end{align}
where
$\bm{g}^{\sigma\sigma'}_{\lambda}$ are the column vectors of dimension $L_\sigma L_{\sigma'}$
whose $[(k_{\sigma'}'\!-1)L_{\sigma}\!+k_{\sigma}]$-elements are given by
\vspace{-1mm}
\begin{align}
   \left[
    \bm{g}^{\sigma\sigma'}_{\lambda}
   \right]_{(k_{\sigma'}'-1)L_{\sigma}+k_{\sigma}}
  \equiv
   g^{k_{\sigma}k_{\sigma'}'}_{\lambda}
  \equiv
   \langle 0|
    \alpha_{k_{\sigma}}^{\sigma}
    \alpha_{k_{\sigma'}'}^{\sigma'}
   |\lambda\rangle,
\end{align}
whereas $\mathcal{V}^{\sigma\sigma'}$ are
the matrices of dimension $L_{\sigma}L_{\sigma'}\times L_{\sigma}L_{\sigma'}$
whose
$[(k_{\sigma'}'\!-1)L_{\sigma}\!+k_{\sigma},
  (l_{\sigma'}'\!-1)L_{\sigma}\!+l_{\sigma}]$-elements are given by
\vspace{-1mm}
\begin{subequations}
\begin{align}
   &
   \left[\mathcal{V}^{+-}\right]_{(k_{-}'-1)L_{+}+k_{+},(l_{-}'-1)L_{+}+l_{+}}
   \allowdisplaybreaks
   \nonumber \\[-1pt]
   &\quad
  =\delta_{k_{+},l_{+}}\delta_{k_{-}',l_{-}'}
   \left(
    \varepsilon_{l_{+}}^{+}+\varepsilon_{l_{-}'}^{-}
   \right)
  -J\sum_{i\in\mathrm{A}}\sum_{j\in\mathrm{B}}l_{i,j}
   V_{ij;k_{+}l_{+}k_{-}'l_{-}'}^{(4)},
   \label{E:eigenEQ2MGF_+-}
   \allowdisplaybreaks
   \\
   &
   \left[\mathcal{V}^{++}\right]_{(k_{+}'-1)L_{+}+k_{+},(l_{+}'-1)L_{+}+l_{+}}
   \allowdisplaybreaks
   \nonumber \\[-1pt]
   &\quad
  =\delta_{k_{+},l_{+}}\delta_{k_{+}',l_{+}'}
   \left(
    \varepsilon_{l_{+}}^{+}+\varepsilon_{l_{+}'}^{+}
   \right)
  -2J\sum_{i\in\mathrm{A}}\sum_{j\in\mathrm{B}}l_{i,j}
    V_{ij;k_{+}k_{+}'l_{+}l_{+}'}^{(9)},
   \label{E:eigenEQ2MGF_++}
   \allowdisplaybreaks
   \\
   &
   \left[\mathcal{V}^{--}\right]_{(k_{-}'-1)L_{-}+k_{-},(l_{-}'-1)L_{-}+l_{-}}
   \allowdisplaybreaks
   \nonumber \\[-1pt]
   &\quad
  =\delta_{k_{-},l_{-}}\delta_{k_{-}',l_{-}'}
   \left(
    \varepsilon_{l_{-}}^{-}+\varepsilon_{l_{-}'}^{-}
   \right)
  -2J\sum_{i\in\mathrm{A}}\sum_{j\in\mathrm{B}}l_{i,j}
    V_{ij;k_{-}k_{-}'l_{-}l_{-}'}^{(1)}.
   \label{E:eigenEQ2MGF_--}
\end{align}
\end{subequations}
The Lehmann representation \cite{FandWgreenfunction} of the 2M GFs reads
\vspace{-1mm}
\begin{align}
   G^{k_{\sigma}k_{\sigma'}'}_{l_{\sigma}l_{\sigma'}'}(\omega)
  =\sum_{\lambda=0}^{L_{\sigma}L_{\sigma'}-1}
   \frac{\hbar
          g^{k_{\sigma}k_{\sigma'}'}_{\lambda}
         \left(
          g^{l_{\sigma}l_{\sigma'}'}_{\lambda}
         \right)^{*}}
        {\hbar\omega-\hbar\varOmega_{\lambda}^{\sigma\sigma'}+i\eta}.
\end{align}
\vspace*{-12mm}
\begin{widetext}
   Likewise, we consider the three-leg-ladder analogs \cite{B034313,B052503} of
the Bethe-Salpeter equations
for the 3M GFs
\vspace*{-1mm}
\begin{subequations}
\begin{align}
   G^{k_{+}k_{+}'k_{-}''}_{l_{+}l_{+}'l_{-}''}(\omega)
  =&
   G^{k_{+}k_{+}'k_{-}''}_{l_{+}l_{+}'l_{-}''}(\omega)_{\mathrm{BL}}
  -\frac{J}{\hbar}
   \sum_{p_{+},p_{+}',p_{-}''}\,\sum_{q_{+},q_{+}',q_{-}''}
   G^{k_{+}k_{+}'k_{-}''}_{p_{+}p_{+}'p_{-}''}(\omega)_{\mathrm{BL}}
   \allowdisplaybreaks
   \nonumber \\[-1mm]
  &\times
   \sum_{i\in\mathrm{A}}\sum_{j\in\mathrm{B}}l_{i,j}
   \left(
     \delta_{p_{-}'',q_{-}''}
     V_{ij;p_{+}p_{+}'q_{+}q_{+}'}^{(9)}
    +\frac{1}{2}\delta_{p_{+},q_{+}}
     V_{ij;p_{+}'q_{+}'p_{-}''q_{-}''}^{(4)}
   +\frac{1}{2}\delta_{p_{+}',q_{+}'}
     V_{ij;p_{+}q_{+}p_{-}''q_{-}''}^{(4)}
   \right)
   G^{q_{+}q_{+}'q_{-}''}_{l_{+}l_{+}'l_{-}''}(\omega),
   \label{E:3lBS++-}
   \allowdisplaybreaks
   \\
   G^{k_{-}k_{-}'k_{+}''}_{l_{-}l_{-}'l_{+}''}(\omega)
  =&
   G^{k_{-}k_{-}'k_{+}''}_{l_{-}l_{-}'l_{+}''}(\omega)_{\mathrm{BL}}
  -\frac{J}{\hbar}
   \sum_{p_{-},p_{-}',p_{+}''}\,\sum_{q_{-},q_{-}',q_{+}''}
   G^{k_{-}k_{-}'k_{+}''}_{p_{-}p_{-}'p_{+}''}(\omega)_{\mathrm{BL}}
   \allowdisplaybreaks
   \nonumber \\[-1mm]
  &\times
   \sum_{i\in\mathrm{A}}\sum_{j\in\mathrm{B}}l_{i,j}
   \left(
    \delta_{p_{+}'',q_{+}''}
     V_{ij;p_{-}p_{-}'q_{-}q_{-}'}^{(1)}
   +\frac{1}{2}\delta_{p_{-},q_{-}}
     V_{ij;p_{+}''q_{+}''p_{-}'q_{-}'}^{(4)}
   +\frac{1}{2}\delta_{p_{-}',q_{-}'}
     V_{ij;p_{+}''q_{+}''p_{-}q_{-}}^{(4)}
   \right)
   G^{q_{-}q_{-}'q_{+}''}_{l_{-}l_{-}'l_{+}''}(\omega),
   \label{E:3lBS--+}
\end{align}
\end{subequations}
denoting the unperturbed 3M GFs by
\vspace*{-1mm}
\begin{align}
   G^{k_{\sigma}k_{\sigma}'k_{\bar{\sigma}}''}_{l_{\sigma}l_{\sigma}'l_{\bar{\sigma}}''}
    (\omega)_{\mathrm{BL}}
   &
  \equiv
   \int_{-\infty}^{\infty}\frac{dt}{i}e^{i\omega t}
   {}_{\mathrm{BL}}\langle 0|
    \mathcal{T}
    \alpha_{k_{\sigma} }^{\sigma}(t)_{\mathrm{BL}}
    \alpha_{k_{\sigma}'}^{\sigma}(t)_{\mathrm{BL}}
    \alpha_{k_{\bar{\sigma}}''}^{\bar{\sigma}}(t)_{\mathrm{BL}}
    \alpha_{l_{\sigma} }^{\sigma\dagger}
    \alpha_{l_{\sigma}'}^{\sigma\dagger}
    \alpha_{l_{\bar{\sigma}}''}^{\bar{\sigma}\dagger}
   |0\rangle_{\mathrm{BL}}
   \allowdisplaybreaks
   \nonumber \\
   &
  =-\int_{-\infty}^{\infty}dt\,e^{i\omega t}
   \left[
    G^{k_{\sigma} }_{l_{\sigma} }(t)_{\mathrm{BL}}
    G^{k_{\sigma}'}_{l_{\sigma}'}(t)_{\mathrm{BL}}
   +G^{k_{\sigma} }_{l_{\sigma}'}(t)_{\mathrm{BL}}
    G^{k_{\sigma}'}_{l_{\sigma} }(t)_{\mathrm{BL}}
   \right]
   G^{k_{\bar{\sigma}}''}_{l_{\bar{\sigma}}''}(t)_{\mathrm{BL}}
   \allowdisplaybreaks
   \nonumber \\
   &
  =\frac{\hbar
         ( \delta_{k_{\sigma},l_{\sigma}}\delta_{k_{\sigma}',l_{\sigma}'}
           +\delta_{k_{\sigma},l_{\sigma}'}\delta_{k_{\sigma}',l_{\sigma}})
          \delta_{k_{\bar{\sigma}}'',l_{\bar{\sigma}}''}}
        {\hbar\omega
         -\varepsilon_{l_{\sigma}}^{\sigma}
         -\varepsilon_{l_{\sigma}'}^{\sigma}
         -\varepsilon_{l_{\bar{\sigma}}''}^{\bar{\sigma}}
         +i\eta}.
\end{align}
We solve the eigenequations obtained from \eqref{E:3lBS++-} and \eqref{E:3lBS--+},
\vspace*{-1mm}
\begin{align}
   \mathcal{V}^{\sigma\sigma\bar{\sigma}}
   \bm{g}^{\sigma\sigma\bar{\sigma}}_{\lambda}
  =\hbar\varOmega_{\lambda}^{\sigma\sigma\bar{\sigma}}
   \bm{g}^{\sigma\sigma\bar{\sigma}}_{\lambda},
\end{align}
where $\bm{g}^{\sigma\sigma\bar{\sigma}}_{\lambda}$ are
the column vectors of dimension $L_{\sigma}^{2} L_{\bar{\sigma}}$ whose
$[(k_{\bar{\sigma}}''-1)L_{\sigma}^{2}+(k_{\sigma}'-1)L_{\sigma}+k_{\sigma}]$-elements are given by
\vspace*{-1mm}
\begin{align}
   &
   \left[
    \bm{g}^{\sigma\sigma\bar{\sigma}}_{\lambda}
   \right]_{(k_{\bar{\sigma}}''-1)L_{\sigma}^{2}+(k_{\sigma}'-1)L_{\sigma}+k_{\sigma}}
   \allowdisplaybreaks
  \equiv
   g^{k_{\sigma}k_{\sigma}'k_{\bar{\sigma}}''}_{\lambda}
  \equiv
   \langle 0|
    \alpha_{k_{\sigma} }^{\sigma}
    \alpha_{k_{\sigma}'}^{\sigma}
    \alpha_{k_{\bar{\sigma}}''}^{\bar{\sigma}}
   |\lambda\rangle,
\end{align}
while $\mathcal{V}^{\sigma\sigma\bar{\sigma}}$ are the matrices of dimension
$L_{\sigma}^{2}L_{\bar{\sigma}}\times L_{\sigma}^{2}L_{\bar{\sigma}}$
whose
$[(k_{\bar{\sigma}}''-1)L_{\sigma}^{2}+(k_{\sigma}'-1)L_{\sigma}+k_{\sigma},
  (l_{\bar{\sigma}}''-1)L_{\sigma}^{2}+(l_{\sigma}'-1)L_{\sigma}+l_{\sigma}]$-elements
are given by
\vspace*{-1mm}
\begin{subequations}
\begin{align}
  &
   \left[\mathcal{V}^{++-}\right]
   _{(k_{-}''-1)L_{+}^{2}+(k_{+}'-1)L_{+}+k_{+},
     (l_{-}''-1)L_{+}^{2}+(l_{+}'-1)L_{+}+l_{+}}
   \allowdisplaybreaks
   \nonumber \\
  &\quad=
   \delta_{k_{+},l_{+}}\delta_{k_{+}',l_{+}'}\delta_{k_{-}'',l_{-}''}
   \left(
    \varepsilon_{l_{+}}^{+}+\varepsilon_{l_{+}'}^{+}+\varepsilon_{l_{-}''}^{-}
   \right)
   -J\sum_{i\in\mathrm{A}}\sum_{j\in\mathrm{B}}l_{i,j}
   \left(
    2\delta_{k_{-}'',l_{-}''}
    V_{ij;k_{+}k_{+}'l_{+}l_{+}'}^{(9)}
   +\delta_{k_{+},l_{+}}
    V_{ij;k_{+}'l_{+}'k_{-}''l_{-}''}^{(4)}
   +\delta_{k_{+}',l_{+}'}
    V_{ij;k_{+}l_{+}k_{-}''l_{-}''}^{(4)}
   \right),
   \allowdisplaybreaks
   \\
  &
   \left[\mathcal{V}^{--+}\right]
   _{(k_{+}''-1)L_{-}^{2}+(k_{-}'-1)L_{-}+k_{-},
     (l_{+}''-1)L_{-}^{2}+(l_{-}'-1)L_{-}+l_{-}}
   \allowdisplaybreaks
   \nonumber \\
  &\quad=
   \delta_{k_{-},l_{-}}\delta_{k_{-}',l_{-}'}\delta_{k_{+}'',l_{+}''}
   \left(
    \varepsilon_{l_{-}}^{-}+\varepsilon_{l_{-}'}^{-}+\varepsilon_{l_{+}''}^{+}
   \right)
   -J\sum_{i\in\mathrm{A}}\sum_{j\in\mathrm{B}}l_{i,j}
   \left(
    2\delta_{k_{+}'',l_{+}''}
    V_{ij;k_{-}k_{-}'l_{-}l_{-}'}^{(1)}
   +\delta_{k_{-},l_{-}}
    V_{ij;k_{+}''l_{+}''k_{-}'l_{-}'}^{(4)}
   +\delta_{k_{-}',l_{-}'}
    V_{ij;k_{+}''l_{+}''k_{-}l_{-}}^{(4)}
   \right).
\end{align}
\end{subequations}
The Lehmann representation of the 3M GFs reads
\vspace*{-1mm}
\begin{align}
   G^{k_{\sigma}k_{\sigma}'k_{\bar{\sigma}}''}
    _{l_{\sigma}l_{\sigma}'l_{\bar{\sigma}}''}(\omega)
  =\sum_{\lambda=0}^{L_{\sigma}^{2}L_{\bar{\sigma}}-1}
   \frac{\hbar g_{\lambda}^{k_{\sigma}k_{\sigma}'k_{\bar{\sigma}}''}
         \left(g_{\lambda}^{l_{\sigma}l_{\sigma}'l_{\bar{\sigma}}''}\right)^{*}}
        {\hbar\omega-\hbar\varOmega_{\lambda}^{\sigma\sigma\bar{\sigma}}+i\eta}.
\end{align}

   Since any perturbative renormalization is hardly tractable for
more-than-3M GFs,
we decompose them into less-than-4M GFs \cite{C054326}.
The 4M GFs can be approximated by the 2M GFs on one hand,
\vspace*{-1mm}
\begin{align}
   &
   2iG^{k_{+}k_{+}'k_{-}''k_{-}'''}_{l_{+}l_{+}'l_{-}''l_{-}'''}(t)
  =2\langle 0|\mathcal{T}
    \alpha_{k_{+}}^{+}(t)\alpha_{k_{+}'}^{+}(t)
    \alpha_{k_{-}''}^{-}(t)\alpha_{k_{-}'''}^{-}(t)
    \alpha_{l_{+}}^{+\dagger}\alpha_{l_{+}'}^{+\dagger}
    \alpha_{l_{-}''}^{-\dagger}\alpha_{l_{-}'''}^{-\dagger}
   |0\rangle
  \allowdisplaybreaks
  \nonumber \\
   &\quad
  \simeq
   \langle 0|\mathcal{T}
    \alpha_{k_{+}}^{+}(t)\alpha_{k_{-}''}^{-}(t)
    \alpha_{l_{+}}^{+\dagger}\alpha_{l_{-}''}^{-\dagger}
   |0\rangle
   \langle 0|\mathcal{T}
    \alpha_{k_{+}'}^{+}(t)\alpha_{k_{-}'''}^{-}(t)
    \alpha_{l_{+}'}^{+\dagger}\alpha_{l_{-}'''}^{-\dagger}
   |0\rangle
  +\langle 0|\mathcal{T}
    \alpha_{k_{+}}^{+}(t)\alpha_{k_{-}''}^{-}(t)
    \alpha_{l_{+}}^{+\dagger}\alpha_{l_{-}'''}^{-\dagger}
   |0\rangle
   \langle 0|\mathcal{T}
    \alpha_{k_{+}'}^{+}(t)\alpha_{k_{-}'''}^{-}(t)
    \alpha_{l_{+}'}^{+\dagger}\alpha_{l_{-}''}^{-\dagger}
   |0\rangle
   \allowdisplaybreaks
   \nonumber \\[-1pt]
   &\quad
  +\langle 0|\mathcal{T}
    \alpha_{k_{+}}^{+}(t)\alpha_{k_{-}''}^{-}(t)
    \alpha_{l_{+}'}^{+\dagger}\alpha_{l_{-}''}^{-\dagger}
   |0\rangle
   \langle 0|\mathcal{T}
    \alpha_{k_{+}'}^{+}(t)\alpha_{k_{-}'''}^{-}(t)
    \alpha_{l_{+}}^{+\dagger}\alpha_{l_{-}'''}^{-\dagger}
   |0\rangle
  +\langle 0|\mathcal{T}
    \alpha_{k_{+}}^{+}(t)\alpha_{k_{-}''}^{-}(t)
    \alpha_{l_{+}'}^{+\dagger}\alpha_{l_{-}'''}^{-\dagger}
   |0\rangle
   \langle 0|\mathcal{T}
    \alpha_{k_{+}'}^{+}(t)\alpha_{k_{-}'''}^{-}(t)
    \alpha_{l_{+}}^{+\dagger}\alpha_{l_{-}''}^{-\dagger}
   |0\rangle
   \allowdisplaybreaks
   \nonumber \\[-1pt]
   &\quad
  +\langle 0|\mathcal{T}
    \alpha_{k_{+}}^{+}(t)\alpha_{k_{+}'}^{+}(t)
    \alpha_{l_{+}}^{+\dagger}\alpha_{l_{+}'}^{+\dagger}
   |0\rangle
   \langle 0|\mathcal{T}
    \alpha_{k_{-}''}^{-}(t)\alpha_{k_{-}'''}^{-}(t)
    \alpha_{l_{-}''}^{-\dagger}\alpha_{l_{-}'''}^{-\dagger}
   |0\rangle,
   \label{E:4MGFsimeq2+2}
\end{align}
and by the 3M and 1M GFs on the other hand,
\vspace*{-1mm}
\begin{align}
   &
   4i
   G^{k_{+}k_{+}'k_{-}''k_{-}'''}_{l_{+}l_{+}'l_{-}''l_{-}'''}(t)
  =4\langle 0|\mathcal{T}
    \alpha_{k_{+}}^{+}(t)
    \alpha_{k_{+}'}^{+}(t)
    \alpha_{k_{-}''}^{-}(t)
    \alpha_{k_{-}'''}^{-}(t)
    \alpha_{l_{+}}^{+\dagger}
    \alpha_{l_{+}'}^{+\dagger}
    \alpha_{l_{-}''}^{-\dagger}
    \alpha_{l_{-}'''}^{-\dagger}
   |0\rangle
   \allowdisplaybreaks
   \nonumber \\
   &\quad
  \simeq
   \langle 0|\mathcal{T}
    \alpha_{k_{+}}^{+}(t)
    \alpha_{l_{+}}^{+\dagger}
   |0\rangle
   \langle 0|\mathcal{T}
    \alpha_{k_{+}'}^{+}(t)
    \alpha_{k_{-}''}^{-}(t)
    \alpha_{k_{-}'''}^{-}(t)
    \alpha_{l_{+}'}^{+\dagger}
    \alpha_{l_{-}''}^{-\dagger}
    \alpha_{l_{-}'''}^{-\dagger}
   |0\rangle
  +\langle 0|\mathcal{T}
    \alpha_{k_{+}}^{+}(t)
    \alpha_{l_{+}'}^{+\dagger}
   |0\rangle
   \langle 0|\mathcal{T}
    \alpha_{k_{+}'}^{+}(t)
    \alpha_{k_{-}''}^{-}(t)
    \alpha_{k_{-}'''}^{-}(t)
    \alpha_{l_{+}}^{+\dagger}
    \alpha_{l_{-}''}^{-\dagger}
    \alpha_{l_{-}'''}^{-\dagger}
   |0\rangle
   \allowdisplaybreaks
   \nonumber \\[-2pt]
   &\quad
  +\langle 0|\mathcal{T}
    \alpha_{k_{+}'}^{+}(t)
    \alpha_{l_{+}}^{+\dagger}
   |0\rangle
   \langle 0|\mathcal{T}
    \alpha_{k_{+}}^{+}(t)
    \alpha_{k_{-}''}^{-}(t)
    \alpha_{k_{-}'''}^{-}(t)
    \alpha_{l_{+}'}^{+\dagger}
    \alpha_{l_{-}''}^{-\dagger}
    \alpha_{l_{-}'''}^{-\dagger}
   |0\rangle
  +\langle 0|\mathcal{T}
    \alpha_{k_{+}'}^{+}(t)
    \alpha_{l_{+}'}^{+\dagger}
   |0\rangle
   \langle 0|\mathcal{T}
    \alpha_{k_{+}}^{+}(t)
    \alpha_{k_{-}''}^{-}(t)
    \alpha_{k_{-}'''}^{-}(t)
    \alpha_{l_{+}}^{+\dagger}
    \alpha_{l_{-}''}^{-\dagger}
    \alpha_{l_{-}'''}^{-\dagger}
   |0\rangle
   \allowdisplaybreaks
   \nonumber \\[-2pt]
   &\quad
  +\langle 0|\mathcal{T}
    \alpha_{k_{-}''}^{-}(t)
    \alpha_{l_{-}''}^{-\dagger}
   |0\rangle
   \langle 0|\mathcal{T}
    \alpha_{k_{+}}^{+}(t)
    \alpha_{k_{+}'}^{+}(t)
    \alpha_{k_{-}'''}^{-}(t)
    \alpha_{l_{+}}^{+\dagger}
    \alpha_{l_{+}'}^{+\dagger}
    \alpha_{l_{-}'''}^{-\dagger}
   |0\rangle
  +\langle 0|\mathcal{T}
    \alpha_{k_{-}''}^{-}(t)
    \alpha_{l_{-}'''}^{-\dagger}
   |0\rangle
   \langle 0|\mathcal{T}
    \alpha_{k_{+}}^{+}(t)
    \alpha_{k_{+}'}^{+}(t)
    \alpha_{k_{-}'''}^{-}(t)
    \alpha_{l_{+}}^{+\dagger}
    \alpha_{l_{+}'}^{+\dagger}
    \alpha_{l_{-}''}^{-\dagger}
   |0\rangle
   \allowdisplaybreaks
   \nonumber \\[-2pt]
   &\quad
  +\langle 0|\mathcal{T}
    \alpha_{k_{-}'''}^{-}(t)
    \alpha_{l_{-}''}^{-\dagger}
   |0\rangle
   \langle 0|\mathcal{T}
    \alpha_{k_{+}}^{+}(t)
    \alpha_{k_{+}'}^{+}(t)
    \alpha_{k_{-}''}^{-}(t)
    \alpha_{l_{+}}^{+\dagger}
    \alpha_{l_{+}'}^{+\dagger}
    \alpha_{l_{-}'''}^{-\dagger}
   |0\rangle
  +\langle 0|\mathcal{T}
    \alpha_{k_{-}'''}^{-}(t)
    \alpha_{l_{-}'''}^{-\dagger}
   |0\rangle
   \langle 0|\mathcal{T}
    \alpha_{k_{+}}^{+}(t)
    \alpha_{k_{+}'}^{+}(t)
    \alpha_{k_{-}''}^{-}(t)
    \alpha_{l_{+}}^{+\dagger}
    \alpha_{l_{+}'}^{+\dagger}
    \alpha_{l_{-}''}^{-\dagger}
   |0\rangle.
   \tag{\ref{E:4MGFsimeq2+2}$'$}
   \label{E:4MGFsimeq3+1}
\end{align}
Since the coefficient of the Raman correlation function
${}^{[p]}X_{l_{+}l_{+}'l_{-}''l_{-}'''}$ is symmetric with respect to the replacements
$l_{+} \leftrightarrow l_{+}'$ and $l_{-}'' \leftrightarrow l_{-}'''$,
the 4M GFs can be expressed as
\vspace*{-1mm}
\begin{align}
   &
   G^{k_{+}k_{+}'k_{-}''k_{-}'''}_{l_{+}l_{+}'l_{-}''l_{-}'''}(t)
  \simeq
   2iG^{k_{+}k_{-}''}_{l_{+}l_{-}''}(t)
     G^{k_{+}'k_{-}'''}_{l_{+}'l_{-}'''}(t)
  +\frac{i}{2}
     G^{k_{+}k_{+}'}_{l_{+}l_{+}'}(t)
     G^{k_{-}''k_{-}'''}_{l_{-}''l_{-}'''}(t),
   \label{E:4MGF22}
   \allowdisplaybreaks
   \\
   &
   G^{k_{+}k_{+}'k_{-}''k_{-}'''}_{l_{+}l_{+}'l_{-}''l_{-}'''}(t)
  \simeq
   iG^{k_{+}k_{+}'k_{-}''}_{l_{+}l_{+}'l_{-}''}(t)
    G^{k_{-}'''}_{l_{-}'''}(t)
  +iG^{k_{-}''k_{-}'''k_{+}}_{l_{-}''l_{-}'''l_{+}}(t)
    G^{k_{+}'}_{l_{+}'}(t),
   \tag{\ref{E:4MGF22}$'$}
   \label{E:4MGF31}
\end{align}
and their Fourier transforms are given by
\vspace*{-1mm}
\begin{align}
   G^{k_{+}k_{+}'k_{-}''k_{-}'''}_{l_{+}l_{+}'l_{-}''l_{-}'''}(\omega)
  &\simeq
   i\int_{-\infty}^{\infty}dt\, e^{i\omega t}
   \left[
    2
     G^{k_{+}k_{-}''}_{l_{+}l_{-}''}(t)
     G^{k_{+}'k_{-}'''}_{l_{+}'l_{-}'''}(t)
   +\frac{1}{2}
     G^{k_{+}k_{+}'}_{l_{+}l_{+}'}(t)
     G^{k_{-}''k_{-}'''}_{l_{-}''l_{-}'''}(t)
   \right]
   \allowdisplaybreaks
   \nonumber \\
  &=i\int_{-\infty}^{\infty}\frac{ds}{2\pi}
   \left[
    2G^{k_{+}k_{-}''}_{l_{+}l_{-}''}(s)
    G^{k_{+}'k_{-}'''}_{l_{+}'l_{-}'''}(\omega-s)
   +\frac{1}{2}G^{k_{+}k_{+}'}_{l_{+}l_{+}'}(s)
    G^{k_{-}''k_{-}'''}_{l_{-}''l_{-}'''}(\omega-s)
   \right]
   \allowdisplaybreaks
   \nonumber \\
  &=2\sum_{\lambda,\lambda'=0}^{L_{+}L_{-}-1}
    \frac{\hbar
          g^{k_{+}k_{-}''}_{\lambda}
          \left(g^{l_{+}l_{-}''}_{\lambda}\right)^{*}
          g^{k_{+}'k_{-}'''}_{\lambda'}
          \left(g^{l_{+}'l_{-}'''}_{\lambda'}\right)^{*}}
         {\hbar\omega-\hbar\varOmega_{\lambda}^{+-}-\hbar\varOmega_{\lambda'}^{+-}+i\eta}
   +\frac{1}{2}\sum_{\lambda=0}^{L_{+}^{2}-1}\sum_{\lambda'=0}^{L_{-}^{2}-1}
    \frac{\hbar
          g^{k_{+}k_{+}'}_{\lambda}
          \left(g^{l_{+}l_{+}'}_{\lambda}\right)^{*}
          g^{k_{-}''k_{-}'''}_{\lambda'}
          \left(g^{l_{-}''l_{-}'''}_{\lambda'}\right)^{*}}
         {\hbar\omega-\hbar\varOmega_{\lambda}^{++}-\hbar\varOmega_{\lambda'}^{--}+i\eta},
   \label{E:4MGF22freq}
   \allowdisplaybreaks
   \\
   G^{k_{+}k_{+}'k_{-}''k_{-}'''}_{l_{+}l_{+}'l_{-}''l_{-}'''}(\omega)
  &\simeq
   i\int_{-\infty}^{\infty}dt\, e^{i\omega t}
   \left[
    G^{k_{+}k_{+}'k_{-}''}_{l_{+}l_{+}'l_{-}''}(t)
    G^{k_{-}'''}_{l_{-}'''}(t)
   +G^{k_{-}''k_{-}'''k_{+}}_{l_{-}''l_{-}'''l_{+}}(t)
    G^{k_{+}'}_{l_{+}'}(t)
   \right]
   \allowdisplaybreaks
   \nonumber \\
  &=i\int_{-\infty}^{\infty}\frac{ds}{2\pi}
   \left[
    G^{k_{+}k_{+}'k_{-}''}_{l_{+}l_{+}'l_{-}''}(s)
    G^{k_{-}'''}_{l_{-}'''}(\omega-s)
   +G^{k_{-}''k_{-}'''k_{+}}_{l_{-}''l_{-}'''l_{+}}(s)
    G^{k_{+}'}_{l_{+}'}(\omega-s)
   \right]
   \allowdisplaybreaks
   \nonumber \\
  &=\sum_{\lambda=0}^{L_{+}^{2}L_{-}-1}
   \frac{\hbar
         g^{k_{+}k_{+}'k_{-}''}_{\lambda}
         \left(g^{l_{+}l_{+}'l_{-}''}_{\lambda}\right)^{*}
         \delta_{k_{-}''',l_{-}'''}}
        {\hbar\omega-\hbar\varOmega_{\lambda}^{++-}
         -\varepsilon_{l_{-}'''}^{-}+i\eta}
  +\sum_{\lambda=0}^{L_{-}^{2}L_{+}-1}
   \frac{\hbar
         g^{k_{-}''k_{-}'''k_{+}}_{\lambda}
         \left(g^{l_{-}''l_{-}'''l_{+}}_{\lambda}\right)^{*}
         \delta_{k_{+}',l_{+}'}}
        {\hbar\omega-\hbar\varOmega_{\lambda}^{--+}
         -\varepsilon_{l_{+}'}^{+}+i\eta}.
   \tag{\ref{E:4MGF22freq}$'$}
   \label{E:4MGF31freq}
\end{align}
\end{widetext}
\vspace*{-12mm}

\vspace*{-5mm}
\section*{S4. Irreducible Decomposition of Raman Operators}
\vspace*{-1mm}
   The 2D Raman operator reads
\begin{align}
   {}^{[p]}\mathcal{R}
  =\sum_{\mu,\nu=x,y}
   e_{\mathrm{in}}^{\mu}
   {}^{[p]}\mathcal{R}^{\mu\nu}
   e_{\mathrm{sc}}^{\nu *},
   \tag{\ref{E:[p]R_munu}}
   \\[-8mm] \nonumber
\end{align}
where
$
   \bm{e}_{\mathrm{in}}
  \equiv
   (e_{\mathrm{in}}^{x},e_{\mathrm{in}}^{y},0)
$ and
$
   \bm{e}_{\mathrm{sc}}
  \equiv
   (e_{\mathrm{sc}}^{x},e_{\mathrm{sc}}^{y},0)
$
are the unit vectors
indicating the polarizations of incident and scattered photons, respectively,
while ${}^{[p]}\mathcal{R}^{\mu\nu}$ is the $(\mu,\nu)$-element of
${}^{[p]}\mathcal{R}$ in Cartesian coordinates.
We introduce four matrices
\begin{align}
   &
   \mathbf{\Xi}_{1}
  \equiv
   \left[
   \begin{array}{cc}
    0 & 1 \\
    1 & 0
   \end{array}
   \right],
   \ 
   \mathbf{\Xi}_{2}
  \equiv
   \left[
   \begin{array}{cc}
    0 & 1 \\
   -1 & 0
   \end{array}
   \right],
   \allowdisplaybreaks
   \nonumber \\[-1pt]
   &
   \mathbf{\Xi}_{3}
  \equiv
   \left[
   \begin{array}{cc}
    1 & 0 \\
    0 & -1
   \end{array}
   \right],
   \ 
   \mathbf{\Xi}_{0}
  \equiv
   \left[
   \begin{array}{cc}
    1 & 0 \\
    0 & 1
   \end{array}
   \right]
   \label{E:PM}
\end{align}
with their Hilbert-Schmidt inner products satisfying
\begin{align}
   \mathrm{Tr}
   \left[
    {}^{\mathrm{t}}\mathbf{\Xi}_{i}\mathbf{\Xi}_{j}
   \right]
  =\sum_{\mu,\nu=x,y}
   \Xi_{i}^{\mu\nu}\Xi_{j}^{\mu\nu}
  =2\delta_{i,j}
   \\[-8mm] \nonumber
\end{align}
to rewrite the Raman operator \eqref{E:[p]R_munu} into
\vspace*{-1mm}
\begin{align}
   &
   {}^{[p]}\mathcal{R}
  =\sum_{i=0}^{3}
   E_{\mathbf{\Xi}_{i}}\,
   {}^{[p]}\mathcal{R}_{\mathbf{\Xi}_{i}},\ 
   E_{\mathbf{\Xi}_{i}}
  \equiv
   \sum_{\mu,\nu=x,y}
   e_{\mathrm{in}}^{\mu}
   \Xi_{i}^{\mu\nu}
   e_{\mathrm{sc}}^{\nu\,*};
   \allowdisplaybreaks
   \label{E:PMBF&RV}
   \\
   &
   E_{\mathbf{\Xi}_{1}}
  =e_{\mathrm{in}}^{x}e_{\mathrm{sc}}^{y\,*}
  +e_{\mathrm{in}}^{y}e_{\mathrm{sc}}^{x\,*},\ 
   E_{\mathbf{\Xi}_{2}}
  =e_{\mathrm{in}}^{x}e_{\mathrm{sc}}^{y\,*}
  -e_{\mathrm{in}}^{y}e_{\mathrm{sc}}^{x\,*},
   \allowdisplaybreaks
   \nonumber \\[-1pt]
   &
   E_{\mathbf{\Xi}_{3}}
  =e_{\mathrm{in}}^{x}e_{\mathrm{sc}}^{x\,*}
  -e_{\mathrm{in}}^{y}e_{\mathrm{sc}}^{y\,*},\ 
   E_{\mathbf{\Xi}_{0}}
  =e_{\mathrm{in}}^{x}e_{\mathrm{sc}}^{x\,*}
  +e_{\mathrm{in}}^{y}e_{\mathrm{sc}}^{y\,*},
   \allowdisplaybreaks
   \tag{\ref{E:PMBF&RV}a}
   \\
   &
   {}^{[p]}\mathcal{R}_{\mathbf{\Xi}_{1}}
  =\frac{{}^{[p]}\mathcal{R}^{xy}+{}^{[p]}\mathcal{R}^{yx}}{2},\ 
   {}^{[p]}\mathcal{R}_{\mathbf{\Xi}_{2}}
  =\frac{{}^{[p]}\mathcal{R}^{xy}-{}^{[p]}\mathcal{R}^{yx}}{2},
   \allowdisplaybreaks
   \nonumber \\[-1pt]
   &
   {}^{[p]}\mathcal{R}_{\mathbf{\Xi}_{3}}
  =\frac{{}^{[p]}\mathcal{R}^{xx}-{}^{[p]}\mathcal{R}^{yy}}{2},\ 
   {}^{[p]}\mathcal{R}_{\mathbf{\Xi}_{0}}
  =\frac{{}^{[p]}\mathcal{R}^{xx}+{}^{[p]}\mathcal{R}^{yy}}{2}.
   \allowdisplaybreaks
   \tag{\ref{E:PMBF&RV}b}
\end{align}
\noindent
$\!\!\!\ $
We recall the irreducible decomposition of the Raman operator for an arbitrary point symmetry group
$\mathbf{P}$,
\begin{align}
   {}^{[p]}\mathcal{R}
  =\mathop{{\sum_{i}}'}\sum_{\mu=1}^{d_{\varXi_{i}}^{\mathbf{P}}}
   E_{\varXi_{i}:\mu}^{\mathbf{P}}\,
   {}^{[p]}\mathcal{R}_{\varXi_{i}:\mu}^{\mathbf{P}},
   \tag{\ref{E:[p]R_Xi}}
   \\[-8mm] \nonumber
\end{align}
to consider 2D lattices of ${\mathbf{C}}_{n\mathrm{v}}$ point symmetry in general.
The polarization-vector basis functions $E_{\varXi_{i}:\mu}^{{\mathbf{C}}_{n\mathrm{v}}}$
relevant to Raman scattering have their equivalent in $E_{\mathbf{\Xi}_{i}}$'s
[cf. \eqref{E:PSBF} and (\ref{E:PMBF&RV}a)].
We list their correspondence relations in Table \ref{T:IRREP}.

   Depolarization of the Loudon-Fleury second-order Raman response, such as \eqref{E:[2]I(w)LP}, is
the consequence of ${}^{[2]}\mathcal{R}$ containing one and only multidimensional irreducible
representation \cite{K214411}.
In Table \ref{T:IRREP}, neither $\mathbf{C}_{2\mathrm{v}}$ nor $\mathbf{C}_{4\mathrm{v}}$
meets this criterion, while all the rest do.
Let us investigate the $\mathbf{C}_{n\mathrm{v}}$ symmetry operations on $\mathbf{\Xi}_i$
intending to reveal the possible dimensionality $d_{\varXi_{i}}^{\mathbf{C}_{n\mathrm{v}}}$.
We denote the matrix representation for a point symmetry operation $P\in\mathbf{P}$ by
$\mathcal{P}$.
Setting $P$ to the rotation $C_{n}^{z}\in\mathbf{C}_{n\mathrm{v}}$,
\begin{align}
   \mathcal{P}
  =\left[
   \begin{array}{rr}
    \cos{\displaystyle\frac{2\pi}{n}} & \sin{\displaystyle\frac{2\pi}{n}} \\
    \vspace{-2.5mm} & \\
   -\sin{\displaystyle\frac{2\pi}{n}} & \cos{\displaystyle\frac{2\pi}{n}} \\
   \end{array}
   \right],
   \\[-8mm] \nonumber
\end{align}
\begin{table}[htb]
\vspace*{2mm}
\caption{Irreducible representations of the point symmetry groups
         $\mathbf{P}=\mathbf{C}_{n\mathrm{v}}$ 
         corresponding to the basis matrices $\mathbf{\Xi}_{i}$,
         which are doubly underlined when they are Raman-active symmetry species
         within the Loudon-Fleury second-order perturbation scheme,
         while they are singly underlined when it is not until we employ
         the Shastry-Shraiman fourth-order perturbation scheme that
         they become Raman-active symmetry species,
         where we specify a particular lattice to the point symmetry groups each,
         because it depends on the lattice shape which symmetry species is Raman active.
         In the case of $\mathbf{P}=\mathbf{C}_{2\mathrm{v}}$,
         not only $\mathbf{\Xi}_{0}$ but also $\mathbf{\Xi}_{3}$
         may belong to the $\mathrm{A}_{1}$ symmetry species and
         the coefficients of their linear combination depend on further details of
         the lattice.}
\centering
\begin{tabular*}{\linewidth}{@{\extracolsep{\fill}}l  cccc}
\hline\hline
 & & & & 
\vspace{-3mm}
\\
$\mathbf{P}$ &
$\mathbf{\Xi}_{0}$ &
$\mathbf{\Xi}_{1}$ &
$\mathbf{\Xi}_{2}$ &
$\mathbf{\Xi}_{3}$ \\
\hline
 & & & & 
\vspace{-3mm}
\\
$\mathbf{C}_{2\mathrm{v}}$ (ladder) &
$\underline{\underline{\mathrm{A}_{1}:1}}$ &
$\mathrm{A}_{2}:1$ &
$\mathrm{A}_{2}:1$ &
$\underline{\underline{\mathrm{A}_{1}:1}}$
\\[2mm]
$\mathbf{C}_{3\mathrm{v}}$ (kagome) &
$\underline{\mathrm{A}_{1}:1}$ &
$\underline{\underline{\mathrm{E}:1}}$ &
$\underline{\mathrm{A}_{2}:1}$ &
$\underline{\underline{\mathrm{E}:2}}$
\\[2mm]
$\mathbf{C}_{4\mathrm{v}}$ (square) &
$\underline{\mathrm{A}_{1}:1}$ &
$\mathrm{B}_{2}:1$ &
$\mathrm{A}_{2}:1$ &
$\underline{\underline{\mathrm{B}_{1}:1}}$
\\[2mm]
$\mathbf{C}_{6\mathrm{v}}$ (triangular) &
$\underline{\mathrm{A}_{1}:1}$ &
$\underline{\underline{\mathrm{E}_{2}:1}}$ &
$\mathrm{A}_{2}:1$ &
$\underline{\underline{\mathrm{E}_{2}:2}}$
\\[2mm]
$\mathbf{C}_{6\mathrm{v}}$ (honeycomb) &
$\underline{\mathrm{A}_{1}:1}$ &
$\underline{\underline{\mathrm{E}_{2}:1}}$ &
$\underline{\mathrm{A}_{2}:1}$ &
$\underline{\underline{\mathrm{E}_{2}:2}}$
\\[2mm]
$\mathbf{C}_{5\mathrm{v}}$ (Penrose) &
$\underline{\mathrm{A}_{1}:1}$ &
$\underline{\underline{\mathrm{E}_{2}:1}}$ &
$\underline{\mathrm{A}_{2}:1}$ &
$\underline{\underline{\mathrm{E}_{2}:2}}$
\\[2mm]
$\mathbf{C}_{8\mathrm{v}}$ (Ammann-Beenker) &
$\underline{\mathrm{A}_{1}:1}$ &
$\underline{\underline{\mathrm{E}_{2}:1}}$ &
$\underline{\mathrm{A}_{2}:1}$ &
$\underline{\underline{\mathrm{E}_{2}:2}}$
\\[2mm]
\hline\hline
\end{tabular*}
\vspace*{-8mm}
\label{T:IRREP}
\end{table}
we obtain
\begin{subequations}
\begin{align}
   \mathcal{P}^{-1}\mathbf{\Xi}_{0}\mathcal{P}
  &=\mathbf{\Xi}_{0},
   \label{E:Xi0C}
   \allowdisplaybreaks
   \\
   \mathcal{P}^{-1}\mathbf{\Xi}_{1}\mathcal{P}
  &=\mathbf{\Xi}_{1}\cos \frac{4\pi}{n}
   -\mathbf{\Xi}_{3}\sin \frac{4\pi}{n},
   \label{E:Xi1C}
   \allowdisplaybreaks
   \\
   \mathcal{P}^{-1}\mathbf{\Xi}_{2}\mathcal{P}
  &=\mathbf{\Xi}_{2},
   \label{E:Xi2C}
   \allowdisplaybreaks
   \\
   \mathcal{P}^{-1}\mathbf{\Xi}_{3}\mathcal{P}
  &=\mathbf{\Xi}_{3}\cos \frac{4\pi}{n}
   +\mathbf{\Xi}_{1}\sin \frac{4\pi}{n},
   \label{E:Xi3C}
\end{align}
\end{subequations}
while setting $P$ to the mirror operation
$\sigma_{\alpha_l}^{\mathrm{v}}\in\mathbf{C}_{n\mathrm{v}}$ with
$\alpha_l\equiv l\pi/n$ $(l=1,2,\cdots,n)$ being
the angle between the normal vector and $x$ axis,
\begin{align}
   \mathcal{P}
  =\left[
   \begin{array}{rr}
    \cos2\alpha_l &  \sin2\alpha_l
   \\
    \sin2\alpha_l & -\cos2\alpha_l
   \end{array}
   \right]
  =\left[
   \begin{array}{rr}
    \cos{\displaystyle\frac{2\pi l}{n}} &  \sin{\displaystyle\frac{2\pi l}{n}} \\
    \vspace{-2.5mm} & \\
    \sin{\displaystyle\frac{2\pi l}{n}} & -\cos{\displaystyle\frac{2\pi l}{n}} \\
   \end{array}
   \right],
\end{align}
\\

\vspace*{-9mm}
\noindent
we have
\vspace*{-1mm}
\begin{subequations}
\begin{align}
   \mathcal{P}^{-1}\mathbf{\Xi}_{0}\mathcal{P}
  &=\mathbf{\Xi}_{0},
   \label{E:Xi0S}
   \allowdisplaybreaks
   \\[-2pt]
   \mathcal{P}^{-1}\mathbf{\Xi}_{1}\mathcal{P}
  &=-\mathbf{\Xi}_{1}\cos\frac{4\pi l}{n}
    +\mathbf{\Xi}_{3}\sin\frac{4\pi l}{n},
   \label{E:Xi1S}
   \allowdisplaybreaks
   \\[-2pt]
   \mathcal{P}^{-1}\mathbf{\Xi}_{2}\mathcal{P}
  &=-\mathbf{\Xi}_{2},
   \label{E:Xi2S}
   \allowdisplaybreaks
   \\[-2pt]
   \mathcal{P}^{-1}\mathbf{\Xi}_{3}\mathcal{P}
  &=\mathbf{\Xi}_{3}\cos\frac{4\pi l}{n}
   +\mathbf{\Xi}_{1}\sin\frac{4\pi l}{n}.
   \label{E:Xi3S}
\end{align}
\end{subequations}
$\mathbf{\Xi}_{0}$ and $\mathbf{\Xi}_{2}$ each correspond to a 1D irreducible
representation for any point symmetry group $\mathbf{C}_{n\mathrm{v}}$, whereas
$\mathbf{\Xi}_{1}$ and $\mathbf{\Xi}_{3}$ span a 2D irreducible representation
unless $n=2$ or $n=4$.
No quasiperiodic lattice in two dimensions belongs to either of $\mathbf{C}_{2\mathrm{v}}$
and $\mathbf{C}_{4\mathrm{v}}$, and therefore, depolarization of ${}^{[2]}I(\omega)$,
which we observe in Figs. \ref{F:PenroseGF}(a) and \ref{F:PenroseGF}(a$'$),
is common to all 2D quasiperiodic lattices.

\section*{S5. Configuration-Interaction Formalism}
{\color{black}
   In order to give a precise description of multimagnon-mediated inelastic light scatterings,
we consider interactions between the up-to-4M basis states
\vspace*{-1mm}
\begin{subequations}
\begin{align}
   &
   |\mathrm{0M}\rangle
  \equiv
   |0\rangle_{\mathrm{BL}},
   \label{E:CIBasisSets0M}
   \allowdisplaybreaks
   \\[-1pt]
   &
   |\mathrm{2M}\rangle^{l_{+}}_{l_{-}}
  \equiv
   \alpha_{l_{+}}^{+ \dagger}\alpha_{l_{-}}^{- \dagger}|0\rangle_{\mathrm{BL}}
   \ (1 \leq l_{\sigma} \leq L_{\sigma}),
   \label{E:CIBasisSets2M}
   \allowdisplaybreaks
   \\[-2pt]
   &
   |\mathrm{4M}\rangle^{l_{+}l_{+}'}_{l_{-}l_{-}'}
  \equiv
   \frac{\alpha_{l_{+}}^{+\dagger}\alpha_{l_{+}'}^{+\dagger}
         \alpha_{l_{-}}^{-\dagger}\alpha_{l_{-}'}^{-\dagger}|0\rangle_{\mathrm{BL}}}
        {\sqrt{1+\delta_{l_{+},l_{+}'}}\sqrt{1+\delta_{l_{-},l_{-}'}}}
   \ (1 \leq l_{\sigma} \leq l_{\sigma}' \leq L_{\sigma}).
   \label{E:CIBasisSets4M}
\end{align}
\end{subequations}
}

\vspace*{-11mm}
\noindent
{\color{black}
Considering that the Hamiltonian as well as the Raman operator commutes with the total
magnetization, any other 2M and 4M states, changing the total magnetization, are ineffective in
the ground-state Raman response.}
The up-to-$O(S^0)$ 2M-4M-configuration-interaction (CI) Hamiltonian is formally written as
{\color{white}blank blank blank blank blank}
\vspace*{-9mm}
\begin{widetext}
\vspace*{-5mm}
\begin{align}
   \mathcal{H}
  =\left[
   \vphantom{
   \begin{array}{c ccc ccc}
\biggl[    & & & & & & \\
    \langle 0\mathrm{M}|\mathcal{H}|0\mathrm{M}\rangle &
    \langle 0\mathrm{M}|\mathcal{H}|2\mathrm{M}\rangle^{1}_{1} &
    \!\!\!\cdots\!\!\! &
    \langle 0\mathrm{M}|\mathcal{H}|2\mathrm{M}\rangle^{L_{+}}_{L_{-}} &
    \langle 0\mathrm{M}|\mathcal{H}|4\mathrm{M}\rangle^{11}_{11} &
    \!\!\!\cdots\!\!\! &
    \langle 0\mathrm{M}|\mathcal{H}|4\mathrm{M}\rangle^{L_{+}L_{+}}_{L_{-}L_{-}}
    \\
    {}^{1}_{1}\langle 2\mathrm{M}|\mathcal{H}|0\mathrm{M}\rangle &
    {}^{1}_{1}\langle 2\mathrm{M}|\mathcal{H}|2\mathrm{M}\rangle^{1}_{1} &
    \!\!\!\cdots\!\!\! &
    {}^{1}_{1}\langle 2\mathrm{M}|\mathcal{H}|2\mathrm{M}\rangle^{L_{+}}_{L_{-}} &
    {}^{1}_{1}\langle 2\mathrm{M}|\mathcal{H}|4\mathrm{M}\rangle^{11}_{11} &
    \!\!\!\cdots\!\!\! &
    {}^{1}_{1}\langle 2\mathrm{M}|\mathcal{H}|4\mathrm{M}\rangle^{L_{+}L_{+}}_{L_{-}L_{-}}
    \\
    \vdots &
    \vdots &
    \!\!\!\ddots\!\!\! &
    \vdots &
    \vdots &
    \!\!\!\ddots\!\!\! &
    \vdots
    \\
    {}^{L_{+}}_{L_{-}}\!\langle 2\mathrm{M}|\mathcal{H}|0\mathrm{M}\rangle \!\!\!\!&
    {}^{L_{+}}_{L_{-}}\!\langle 2\mathrm{M}|\mathcal{H}|2\mathrm{M}\rangle^{1}_{1} &
    \!\!\!\cdots\!\!\! &
    {}^{L_{+}}_{L_{-}}\!\langle 2\mathrm{M}|\mathcal{H}|2\mathrm{M}\rangle^{L_{+}}_{L_{-}} &
    {}^{L_{+}}_{L_{-}}\!\langle 2\mathrm{M}|\mathcal{H}|4\mathrm{M}\rangle^{11}_{11} &
    \!\!\!\cdots\!\!\! &
    {}^{L_{+}}_{L_{-}}\!\langle 2\mathrm{M}|
     \mathcal{H}
    |4\mathrm{M}\rangle^{L_{+}L_{+}}_{L_{-}L_{-}}
    \\ 
    {}^{11}_{11}\langle 4\mathrm{M}|\mathcal{H}|0\mathrm{M}\rangle \!\!\!\!&
    {}^{11}_{11}\langle 4\mathrm{M}|\mathcal{H}|2\mathrm{M}\rangle^{1}_{1} &
    \!\!\!\cdots\!\!\! &
    {}^{11}_{11}\langle 4\mathrm{M}|\mathcal{H}|2\mathrm{M}\rangle^{L_{+}}_{L_{-}} &
    {}^{11}_{11}\langle 4\mathrm{M}|\mathcal{H}|4\mathrm{M}\rangle^{11}_{11} &
    \!\!\!\cdots\!\!\! &
    {}^{11}_{11}\langle 4\mathrm{M}|
     \mathcal{H}
    |4\mathrm{M}\rangle^{L_{+}L_{+}}_{L_{-}L_{-}}
    \\
    \vdots &
    \vdots &
    \!\!\!\ddots\!\!\! &
    \vdots &
    \vdots &
    \!\!\!\ddots\!\!\! &
    \vdots
    \\
    {}^{L_{+}L_{+}}_{L_{-}L_{-}}\!\langle 4\mathrm{M}|\mathcal{H}|0\mathrm{M}\rangle \!\!&
    {}^{L_{+}L_{+}}_{L_{-}L_{-}}\!\langle 4\mathrm{M}|
     \mathcal{H}
    |2\mathrm{M}\rangle^{1}_{1} \!\!&
    \!\!\!\cdots\!\!\! &
    {}^{L_{+}L_{+}}_{L_{-}L_{-}}\!\langle 4\mathrm{M}|
     \mathcal{H}
    |2\mathrm{M}\rangle^{L_{+}}_{L_{-}} \!\!&
    {}^{L_{+}L_{+}}_{L_{-}L_{-}}\!\langle 4\mathrm{M}|
     \mathcal{H}
    |4\mathrm{M}\rangle^{11}_{11} \!\!&
    \!\!\!\cdots\!\!\! &
    {}^{L_{+}L_{+}}_{L_{-}L_{-}}\!\langle 4\mathrm{M}|
     \mathcal{H}
    |4\mathrm{M}\rangle^{L_{+}L_{+}}_{L_{-}L_{-}}
   \end{array}
   }
   \begin{array}{|c | ccc | ccc|}
    \hline
    \vphantom{\Bigl[}
    \langle 0\mathrm{M}|\mathcal{H}|0\mathrm{M}\rangle &
    \langle 0\mathrm{M}|\mathcal{H}|2\mathrm{M}\rangle^{1}_{1} &
    \!\!\!\cdots\!\!\! &
    \langle 0\mathrm{M}|\mathcal{H}|2\mathrm{M}\rangle^{L_{+}}_{L_{-}} &
    \langle 0\mathrm{M}|\mathcal{H}|4\mathrm{M}\rangle^{11}_{11} &
    \!\!\!\cdots\!\!\! &
    \langle 0\mathrm{M}|\mathcal{H}|4\mathrm{M}\rangle^{L_{+}L_{+}}_{L_{-}L_{-}}
    \\ \hline
    \vphantom{\Bigl[}
    {}^{1}_{1}\langle 2\mathrm{M}|\mathcal{H}|0\mathrm{M}\rangle &
    {}^{1}_{1}\langle 2\mathrm{M}|\mathcal{H}|2\mathrm{M}\rangle^{1}_{1} &
    \!\!\!\cdots\!\!\! &
    {}^{1}_{1}\langle 2\mathrm{M}|\mathcal{H}|2\mathrm{M}\rangle^{L_{+}}_{L_{-}} &
    {}^{1}_{1}\langle 2\mathrm{M}|\mathcal{H}|4\mathrm{M}\rangle^{11}_{11} &
    \!\!\!\cdots\!\!\! &
    {}^{1}_{1}\langle 2\mathrm{M}|\mathcal{H}|4\mathrm{M}\rangle^{L_{+}L_{+}}_{L_{-}L_{-}}
    \\
    \vdots &
    \vdots &
    \!\!\!\ddots\!\!\! &
    \vdots &
    \vdots &
    \!\!\!\ddots\!\!\! &
    \vdots
    \\
    \vphantom{\Bigl[}
    {}^{L_{+}}_{L_{-}}\!\langle 2\mathrm{M}|\mathcal{H}|0\mathrm{M}\rangle \!\!\!\!&
    {}^{L_{+}}_{L_{-}}\!\langle 2\mathrm{M}|\mathcal{H}|2\mathrm{M}\rangle^{1}_{1} &
    \!\!\!\cdots\!\!\! &
    {}^{L_{+}}_{L_{-}}\!\langle 2\mathrm{M}|\mathcal{H}|2\mathrm{M}\rangle^{L_{+}}_{L_{-}} &
    {}^{L_{+}}_{L_{-}}\!\langle 2\mathrm{M}|\mathcal{H}|4\mathrm{M}\rangle^{11}_{11} &
    \!\!\!\cdots\!\!\! &
    {}^{L_{+}}_{L_{-}}\!\langle 2\mathrm{M}|
     \mathcal{H}
    |4\mathrm{M}\rangle^{L_{+}L_{+}}_{L_{-}L_{-}}
    \\ \hline
    \vphantom{\Bigl[}
    {}^{11}_{11}\langle 4\mathrm{M}|\mathcal{H}|0\mathrm{M}\rangle \!\!\!\!&
    {}^{11}_{11}\langle 4\mathrm{M}|\mathcal{H}|2\mathrm{M}\rangle^{1}_{1} &
    \!\!\!\cdots\!\!\! &
    {}^{11}_{11}\langle 4\mathrm{M}|\mathcal{H}|2\mathrm{M}\rangle^{L_{+}}_{L_{-}} &
    {}^{11}_{11}\langle 4\mathrm{M}|\mathcal{H}|4\mathrm{M}\rangle^{11}_{11} &
    \!\!\!\cdots\!\!\! &
    {}^{11}_{11}\langle 4\mathrm{M}|
     \mathcal{H}
    |4\mathrm{M}\rangle^{L_{+}L_{+}}_{L_{-}L_{-}}
    \\
    \vdots &
    \vdots &
    \!\!\!\ddots\!\!\! &
    \vdots &
    \vdots &
    \!\!\!\ddots\!\!\! &
    \vdots
    \\
    \vphantom{\Bigl[}
    {}^{L_{+}L_{+}}_{L_{-}L_{-}}\!\langle 4\mathrm{M}|\mathcal{H}|0\mathrm{M}\rangle \!\!&
    {}^{L_{+}L_{+}}_{L_{-}L_{-}}\!\langle 4\mathrm{M}|
     \mathcal{H}
    |2\mathrm{M}\rangle^{1}_{1} \!\!&
    \!\!\cdots\!\! &
    {}^{L_{+}L_{+}}_{L_{-}L_{-}}\!\langle 4\mathrm{M}|
     \mathcal{H}
    |2\mathrm{M}\rangle^{L_{+}}_{L_{-}} \!\!&
    {}^{L_{+}L_{+}}_{L_{-}L_{-}}\!\langle 4\mathrm{M}|
     \mathcal{H}
    |4\mathrm{M}\rangle^{11}_{11} \!\!&
    \!\!\cdots\!\! &
    {}^{L_{+}L_{+}}_{L_{-}L_{-}}\!\langle 4\mathrm{M}|
     \mathcal{H}
    |4\mathrm{M}\rangle^{L_{+}L_{+}}_{L_{-}L_{-}}\!\!
    \\
    \hline
   \end{array}
   \right]
   \label{E:CIHgeneral}
\end{align}
with $\mathcal{H}=\mathcal{H}_{\mathrm{BL}}+:\mathcal{H}^{(0)}:$.
With the ladder-approximation Bethe-Salpeter equation formalism
[cf. \eqref{E:2lBS+-}--\eqref{E:2lBS--} for 2M GFs and
     \eqref{E:3lBS++-} and \eqref{E:3lBS--+} for 3M GFs] in mind,
we retain only the magnon-number-conserving interactions
$V_{ij;l_{-}l_{-}'l_{-}''l_{-}'''}^{(1)}$,
$V_{ij;l_{+}l_{+}'l_{-}''l_{-}'''}^{(4)}$, and
$V_{ij;l_{+}l_{+}'l_{+}''l_{+}'''}^{(9)}$
in our CI scheme.
Then \eqref{E:CIHgeneral} reduces to  the block-diagonal Hamiltonian
\vspace*{-2mm}
\begin{align}
   &
   \mathcal{H}
  -\sum_{m=0}^{2}E^{(m)}
  =\left[
   \vphantom{
   \begin{array}{c ccc ccc}
\biggl[    & & & & & & \\
   \cline{1-1}
    \multicolumn{1}{|c|}{0} &
     &
     &
     &
     &
     & 
   \\ \cline{1-4}
     & 
    \multicolumn{1}{|c}{{}^{1}_{1}\langle 2\mathrm{M}|\mathcal{H}|2\mathrm{M}\rangle^{1}_{1}} &
    \!\!\!\cdots\!\!\! &
    \multicolumn{1}{c|}{{}^{1}_{1}\langle 2\mathrm{M}|\mathcal{H}|2\mathrm{M}\rangle^{L_{+}}_{L_{-}}} &
     &
     &
   \\ 
     &
    \multicolumn{1}{|c}{\vdots} &
    \!\!\!\ddots\!\!\! &
    \multicolumn{1}{c|}{\vdots} &
     &
     &
   \\
     &
    \multicolumn{1}{|c}{{}^{L_{+}}_{L_{-}}\!\langle 2\mathrm{M}|\mathcal{H}|2\mathrm{M}\rangle^{1}_{1}} &
    \!\!\!\cdots\!\!\! &
    \multicolumn{1}{c|}{{}^{L_{+}}_{L_{-}}\!\langle 2\mathrm{M}|\mathcal{H}|2\mathrm{M}\rangle^{L_{+}}_{L_{-}}} &
     &
     &
   \\ \cline{2-4}\cline{5-7}
     &
     &
     &
     &
    \multicolumn{1}{|c}{{}^{11}_{11}\langle 4\mathrm{M}|\mathcal{H}|4\mathrm{M}\rangle^{11}_{11}} &
    \!\!\!\cdots\!\!\! &
    \multicolumn{1}{c|}{{}^{11}_{11}\langle 4\mathrm{M}|\mathcal{H}|4\mathrm{M}\rangle^{L_{+}L_{+}}_{L_{-}L_{-}}}
   \\ 
    & & & &
    \multicolumn{1}{|c}{\vdots} &
    \!\!\!\ddots\!\!\! &
    \multicolumn{1}{c|}{\vdots}
   \\
     & & & &
    \multicolumn{1}{|c}{{}^{L_{+}L_{+}}_{L_{-}L_{-}}\!\langle 4\mathrm{M}|\mathcal{H}|4\mathrm{M}\rangle^{11}_{11}} &
    \!\!\!\cdots\!\!\! &
    \multicolumn{1}{c|}{{}^{L_{+}L_{+}}_{L_{-}L_{-}}\!\langle 4\mathrm{M}|\mathcal{H}|4\mathrm{M}\rangle^{L_{+}L_{+}}_{L_{-}L_{-}}}
   \\ \cline{5-7}
   \end{array}
   }
   \begin{array}{c ccc ccc}
   \cline{1-1}
    \multicolumn{1}{|c|}{0} &
     &
     &
     &
     &
     & 
   \\ \cline{1-4}
   \vphantom{\Bigl[}
     & 
    \multicolumn{1}{|c}{{}^{1}_{1}\langle 2\mathrm{M}|\mathcal{H}|2\mathrm{M}\rangle^{1}_{1}} &
    \!\!\!\cdots\!\!\! &
    \multicolumn{1}{c|}{{}^{1}_{1}\langle 2\mathrm{M}|\mathcal{H}|2\mathrm{M}\rangle^{L_{+}}_{L_{-}}} &
     &
     &
   \\ 
     &
    \multicolumn{1}{|c}{\vdots} &
    \!\!\!\ddots\!\!\! &
    \multicolumn{1}{c|}{\vdots} &
     &
     &
   \\
   \vphantom{\Bigl[}
     &
    \multicolumn{1}{|c}{{}^{L_{+}}_{L_{-}}\!\langle 2\mathrm{M}|\mathcal{H}|2\mathrm{M}\rangle^{1}_{1}} &
    \!\!\!\cdots\!\!\! &
    \multicolumn{1}{c|}{{}^{L_{+}}_{L_{-}}\!\langle 2\mathrm{M}|\mathcal{H}|2\mathrm{M}\rangle^{L_{+}}_{L_{-}}} &
     &
     &
   \\ \cline{2-4}\cline{5-7}
   \vphantom{\Bigl[}
     &
     &
     &
     &
    \multicolumn{1}{|c}{{}^{11}_{11}\langle 4\mathrm{M}|\mathcal{H}|4\mathrm{M}\rangle^{11}_{11}} &
    \!\!\!\cdots\!\!\! &
    \multicolumn{1}{c|}{{}^{11}_{11}\langle 4\mathrm{M}|\mathcal{H}|4\mathrm{M}\rangle^{L_{+}L_{+}}_{L_{-}L_{-}}}
   \\ 
    & & & &
    \multicolumn{1}{|c}{\vdots} &
    \!\!\!\ddots\!\!\! &
    \multicolumn{1}{c|}{\vdots}
   \\
   \vphantom{\Bigl[}
     & & & &
    \multicolumn{1}{|c}{{}^{L_{+}L_{+}}_{L_{-}L_{-}}\!\langle 4\mathrm{M}|\mathcal{H}|4\mathrm{M}\rangle^{11}_{11}} &
    \!\!\!\cdots\!\!\! &
    \multicolumn{1}{c|}{{}^{L_{+}L_{+}}_{L_{-}L_{-}}\!\langle 4\mathrm{M}|\mathcal{H}|4\mathrm{M}\rangle^{L_{+}L_{+}}_{L_{-}L_{-}}}
   \\ \cline{5-7}
   \end{array}
   \right];
   \allowdisplaybreaks
   \label{E:CIHV(1)V(4)V(9)}
   \\
   &
   {}^{k_{+}}_{k_{-}}\langle\mathrm{2M}|\mathcal{H}|\mathrm{2M}\rangle^{l_{+}}_{l_{-}}
  =\delta_{k_{+},l_{+}}\delta_{k_{-},l_{-}}
   \left(\varepsilon_{l_{+}}^{+}+\varepsilon_{l_{-}}^{-}\right)
  -J
   \sum_{i\in\mathrm{A}}\sum_{j\in\mathrm{B}}
   l_{i,j}
   V_{ij;k_{+}l_{+}k_{-}l_{-}}^{(4)},
   \allowdisplaybreaks
   \tag{\ref{E:CIHV(1)V(4)V(9)}b}
   \label{E:CIHV(1)V(4)V(9)2M}
   \\
   &
   {}^{k_{+}k_{+}'}_{k_{-}k_{-}'}\langle\mathrm{4M}|
   \mathcal{H}
   |\mathrm{4M}\rangle^{l_{+}l_{+}'}_{l_{-}l_{-}'}
  = \delta_{k_{+} ,l_{+} }
    \delta_{k_{+}',l_{+}'}
    \delta_{k_{-} ,l_{-} }
    \delta_{k_{-}',l_{-}'}
   \left(
     \varepsilon_{l_{+} }^{+}
    +\varepsilon_{l_{+}'}^{+}
    +\varepsilon_{l_{-} }^{-}
    +\varepsilon_{l_{-}'}^{-}
    \right)
   \allowdisplaybreaks
   \nonumber \\
   &
  -J
   \sum_{i\in\mathrm{A}}\sum_{j\in\mathrm{B}}
   l_{i,j}
   \left[
    \vphantom{\frac{\delta_{k_{+},l_{+}}\delta_{k_{+}',l_{+}'}}
             {\sqrt{1+\delta_{k_{-},k_{-}'}}\sqrt{1+\delta_{l_{-},l_{-}'}}}}
    \frac{\delta_{k_{+},l_{+}}\delta_{k_{+}',l_{+}'}}
         {\sqrt{1+\delta_{k_{-},k_{-}'}}\sqrt{1+\delta_{l_{-},l_{-}'}}}
   \right.
   \left(
     V_{ij;k_{-}k_{-}'l_{-}l_{-}'}^{(1)}
    +V_{ij;k_{-}k_{-}'l_{-}'l_{-}}^{(1)}
    +V_{ij;k_{-}'k_{-}l_{-}l_{-}'}^{(1)}
    +V_{ij;k_{-}'k_{-}l_{-}'l_{-}}^{(1)}
   \right)
   \allowdisplaybreaks
   \nonumber \\
   &\quad
   +\frac{\delta_{k_{-},l_{-}}\delta_{k_{-}',l_{-}'}}
         {\sqrt{1+\delta_{k_{+},k_{+}'}}\sqrt{1+\delta_{l_{+},l_{+}'}}}
   \left(
     V_{ij;k_{+}k_{+}l_{+}'l_{+}'}^{(9)}
    +V_{ij;k_{+}k_{+}'l_{+}'l_{+}}^{(9)}
    +V_{ij;k_{+}'k_{+}l_{+}l_{+}'}^{(9)}
    +V_{ij;k_{+}'k_{+}l_{+}'l_{+}}^{(9)}
   \right)
   \allowdisplaybreaks
   \nonumber \\
   &\quad
   +\frac{1}{\sqrt{1+\delta_{k_{+},k_{+}'}}
             \sqrt{1+\delta_{l_{+},l_{+}'}}
             \sqrt{1+\delta_{k_{-},k_{-}'}}
             \sqrt{1+\delta_{l_{-},l_{-}'}}}
   \allowdisplaybreaks
   \nonumber \\
   &\quad\times
   \left(
     V_{ij;k_{+}l_{+}k_{-}l_{-}}^{(4)}
     \delta_{k_{+}',l_{+}'}\delta_{k_{-}',l_{-}'}
    +V_{ij;k_{+}l_{+}k_{-}l_{-}'}^{(4)}
     \delta_{k_{+}',l_{+}'}\delta_{k_{-}',l_{-}}
    +V_{ij;k_{+}l_{+}k_{-}'l_{-}}^{(4)}
     \delta_{k_{+}',l_{+}'}\delta_{k_{-},l_{-}'}
    +V_{ij;k_{+}l_{+}k_{-}'l_{-}'}^{(4)}
     \delta_{k_{+}',l_{+}'}\delta_{k_{-},l_{-}}
   \right.
   \allowdisplaybreaks
   \nonumber \\
   &\quad\quad
    +V_{ij;k_{+}l_{+}'k_{-}l_{-}}^{(4)}
     \delta_{k_{+}',l_{+}}\delta_{k_{-}',l_{-}'}
    +V_{ij;k_{+}l_{+}'k_{-}l_{-}'}^{(4)}
     \delta_{k_{+}',l_{+}}\delta_{k_{-}',l_{-}}
    +V_{ij;k_{+}l_{+}'k_{-}'l_{-}}^{(4)}
     \delta_{k_{+}',l_{+}}\delta_{k_{-},l_{-}'}
    +V_{ij;k_{+}l_{+}'k_{-}'l_{-}'}^{(4)}
     \delta_{k_{+}',l_{+}}\delta_{k_{-},l_{-}}
   \allowdisplaybreaks
   \nonumber \\
   &\quad\quad
    +V_{ij;k_{+}'l_{+}k_{-}l_{-}}^{(4)}
     \delta_{k_{+},l_{+}'}\delta_{k_{-}',l_{-}'}
    +V_{ij;k_{+}'l_{+}k_{-}l_{-}'}^{(4)}
     \delta_{k_{+},l_{+}'}\delta_{k_{-}',l_{-}}
    +V_{ij;k_{+}'l_{+}k_{-}'l_{-}}^{(4)}
     \delta_{k_{+},l_{+}'}\delta_{k_{-},l_{-}'}
    +V_{ij;k_{+}'l_{+}k_{-}'l_{-}'}^{(4)}
     \delta_{k_{+},l_{+}'}\delta_{k_{-},l_{-}}
   \allowdisplaybreaks
   \nonumber \\
   &\quad\quad
  \left.
   \left.
    +V_{ij;k_{+}'l_{+}'k_{-}l_{-}}^{(4)}
     \delta_{k_{+},l_{+}}\delta_{k_{-}',l_{-}'}
    +V_{ij;k_{+}'l_{+}'k_{-}l_{-}'}^{(4)}
     \delta_{k_{+},l_{+}}\delta_{k_{-}',l_{-}}
    +V_{ij;k_{+}'l_{+}'k_{-}'l_{-}}^{(4)}
     \delta_{k_{+},l_{+}}\delta_{k_{-},l_{-}'}
    +V_{ij;k_{+}'l_{+}'k_{-}'l_{-}'}^{(4)}
     \delta_{k_{+},l_{+}}\delta_{k_{-},l_{-}}
   \right)
   \vphantom{\frac{\delta_{k_{+},l_{+}}\delta_{k_{+}',l_{+}'}}
            {\sqrt{1+\delta_{k_{-},k_{-}'}}\sqrt{1+\delta_{l_{-},l_{-}'}}}}
  \right].
   \allowdisplaybreaks
   \tag{\ref{E:CIHV(1)V(4)V(9)}c}
   \label{E:CIHV(1)V(4)V(9)4M}
\end{align}
Note that the 2M and 4M sectors are of dimension
$\prod_{\sigma=\pm}L_\sigma \times \prod_{\sigma=\pm}L_\sigma$ and
$\prod_{\sigma=\pm}[L_\sigma(L_\sigma+1)/2] \times \prod_{\sigma=\pm}[L_\sigma(L_\sigma+1)/2]$,
respectively.
We diagonalize the 2M-4M-CI magnon-number-conserving block-diagonal Hamiltonian
\eqref{E:CIHV(1)V(4)V(9)} to obtain the variationally corrected eigenstates $|\nu\rangle$ and
eigenvalues $\varepsilon_\nu$
[$\nu=0,1,\cdots,L_+L_- + L_+(L_+ + 1)L_-(L_- + 1)/4\equiv N_{\mathrm{CI}}-1$].
\vspace{6pt}
\end{widetext}

\vspace*{-24pt}
   If we discard the 4M basis states \eqref{E:CIBasisSets4M} in \eqref{E:CIHV(1)V(4)V(9)},
the resultant 2M-CI findings for ${}^{[p]}I_{2\mathrm{M}}(\omega)$ are exactly the same as
the ladder-approximation Bethe-Salpeter calculations.
There is a complete correspondence of the 2M sector \eqref{E:CIHV(1)V(4)V(9)2M} of
any CI magnon-number-conserving block-diagonal Hamiltonian with the 2M Bethe-Salpeter interaction
matrix $\mathcal{V}^{+-}$ \eqref{E:eigenEQ2MGF_+-}.
\eqref{E:2lBS++} and \eqref{E:2lBS--} are irrelevant to any calculation of
${}^{[p]}I_{2\mathrm{M}}(\omega)$ but necessary for calculating ${}^{[p]}I_{4\mathrm{M}}(\omega)$,
or more precisely, for decomposing the 4M GFs as \eqref{E:4MGF22} and \eqref{E:4MGF31}.
When we go beyond the Loudon-Fleury second-order perturbation theory and take a considerable
interest in multimagnon-mediated Raman intensities, the 2M-4M-CI scheme is much superior to
any tractable self-consistent GF formalism.

\section*{S6. Configuration-Interaction versus Green's Function Calculations of Raman Spectra}
\begin{figure*}[tb]
\centering
\includegraphics[width=\linewidth]{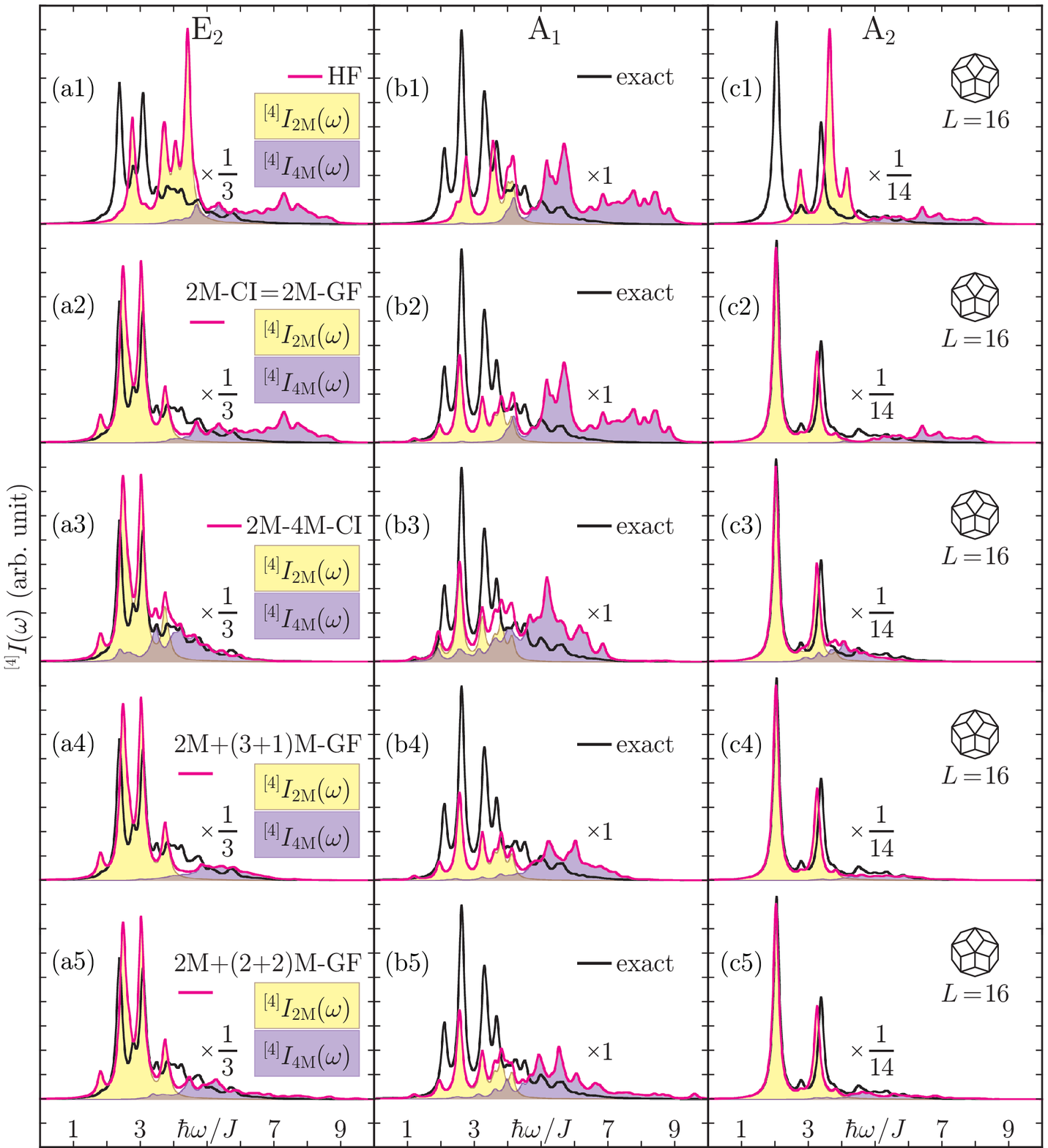}
\caption{CI and GF calculations of the Shastry-Shraiman fourth-order Raman intensities
         ${}^{[4]}I(\omega) \equiv \sum_{l=1}^2 {}^{[4]}I_{2l\mathrm{M}}(\omega)$ for
         the $L=16$ 2D Penrose lattice of ${\mathbf{C}}_{5\mathrm{v}}$ point symmetry
         in comparison with the exact solutions,
         where the perturbation parameter $t/(U-\hbar\omega_{\mathrm{in}})$ is set to $9/10$
         and every spectral line is Lorentzian-broadened by a width of $0.1J$.
         From the top to the bottom, the calculational schemes employed are
         the Hartree-Fock approximation [(a1) to (c1)],
         which retains only the magnon vacuum in \eqref{E:CIBasisSets0M} and substitutes
         the eigenstates and eigenvalues of $\mathcal{H}_{\mathrm{BL}}$ \eqref{E:HBLdiag}
         for the basis states $|\nu\rangle$ and their energies $\varepsilon_\nu$,
         with $N_{\mathrm{CI}}$ reducing to $L$, in \eqref{E:[p]I2lM(w)CI},
         2M-CI [(a2) to (c2)], which is equivalent to solving the 2M Bethe-Salpeter equation
         \eqref{E:2lBS+-},
         2M-4M-CI [(a3) to (c3)],
         $2\mathrm{M}+4\mathrm{M}$ [approximated by \eqref{E:4MGF31}]-GF [(a4) to (c4)], and
         $2\mathrm{M}+4\mathrm{M}$ [approximated by \eqref{E:4MGF22}]-GF [(a5) to (c5)].
         The pure symmetry components are extracted from three polarization combinations,
         \eqref{E:[4]I(w)LP}
         with $\phi_-=0,\pi/2$ and
         \eqref{E:[p]I(w)CP}
         with $\sigma_{\mathrm{in}}\sigma_{\mathrm{sc}}=-1$.
         All the SW calculations of $I_{2l\mathrm{M}}(\omega)$ each
         are distinguishably colored.}
\label{F:[4]I(w)PenroseL=16}
\end{figure*}

\begin{figure*}[tb]
\centering
\includegraphics[width=\linewidth]{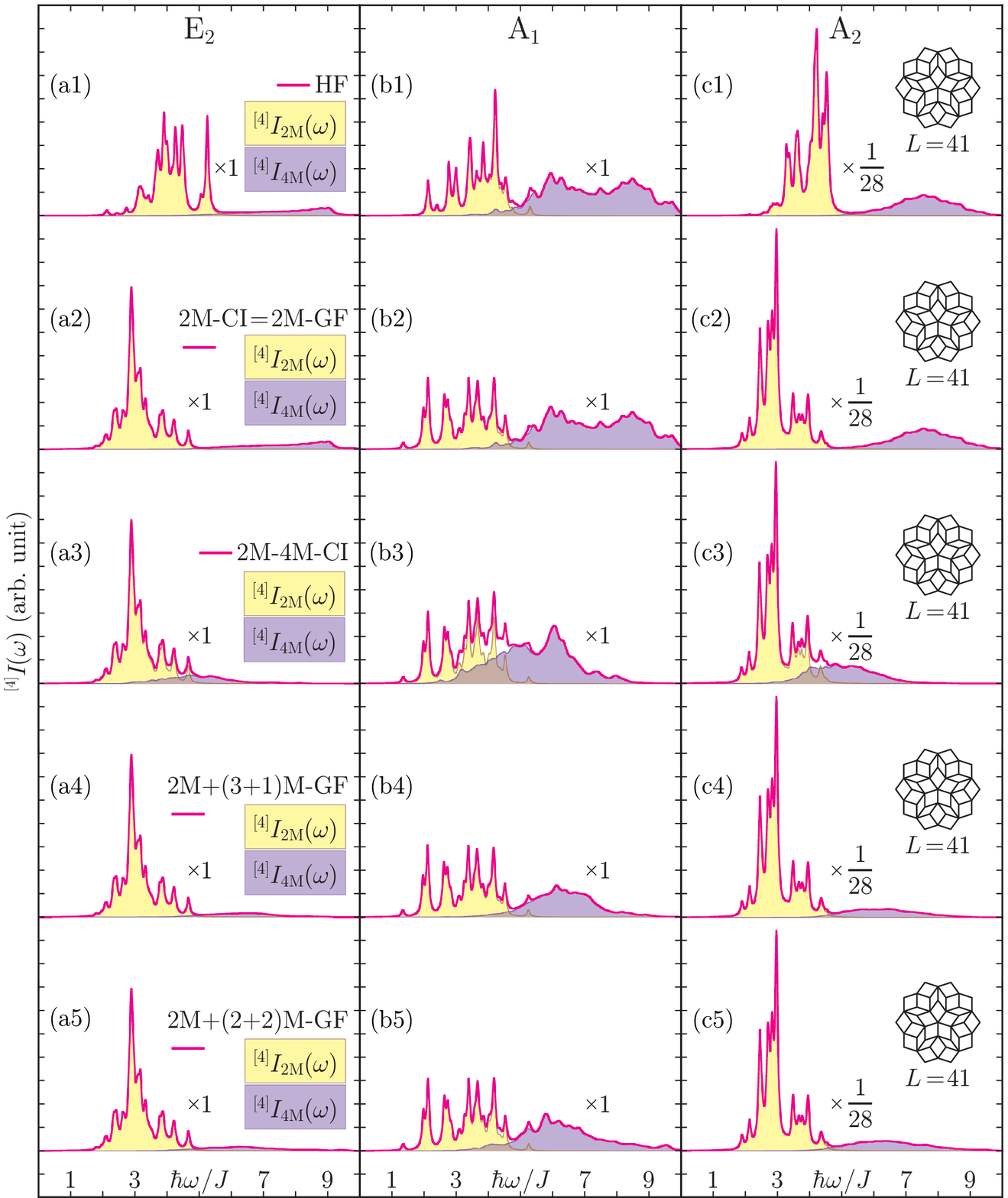}
\caption{The same as Fig. \ref{F:[4]I(w)PenroseL=16} for
         the $L=41$ 2D Penrose lattice of ${\mathbf{C}}_{5\mathrm{v}}$ point symmetry
         without any exact solution available.}
\label{F:[4]I(w)PenroseL=41}
\end{figure*}

\begin{figure*}[tb]
\centering
\includegraphics[width=\linewidth]{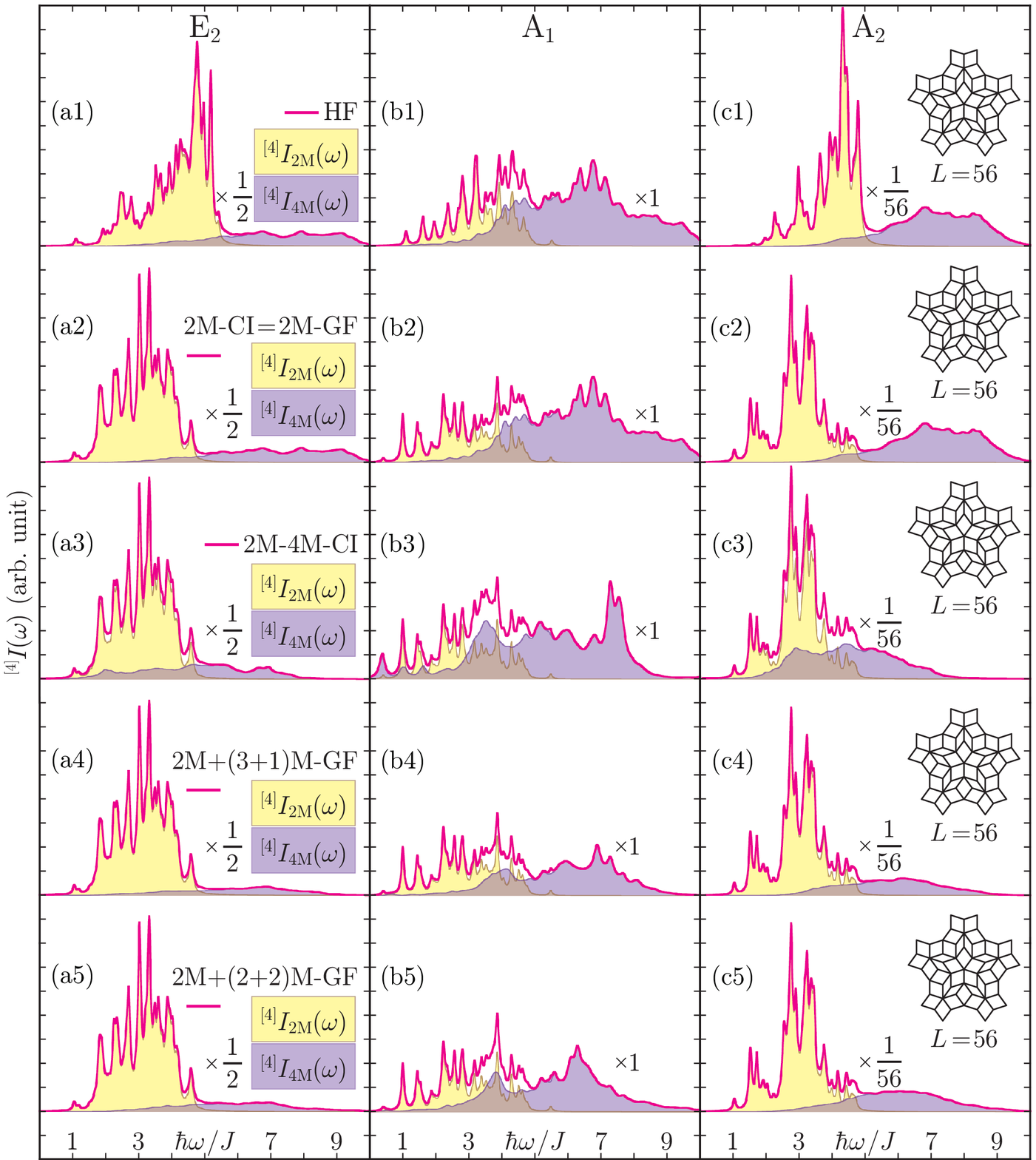}
\caption{The same as Fig. \ref{F:[4]I(w)PenroseL=16} for
         the $L=56$ 2D Penrose lattice of ${\mathbf{C}}_{5\mathrm{v}}$ point symmetry
         without any exact solution available.}
\label{F:[4]I(w)PenroseL=56}
\end{figure*}

\begin{figure*}[tb]
\centering
\includegraphics[width=\linewidth]{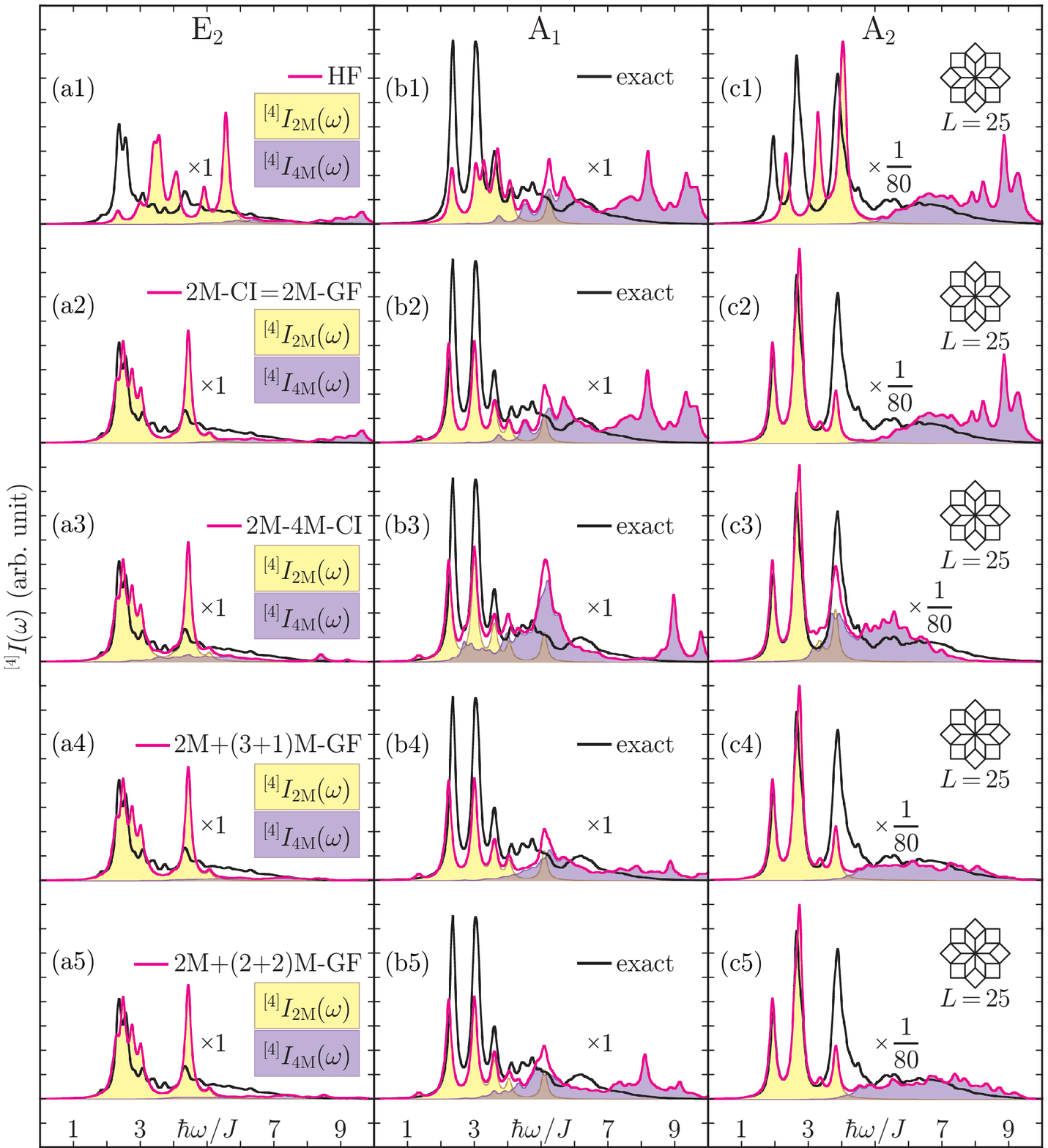}
\caption{The same as Fig. \ref{F:[4]I(w)PenroseL=16} for
         the $L=25$ 2D Ammann-Beenker lattice of ${\mathbf{C}}_{8\mathrm{v}}$ point symmetry.}
\label{F:[4]I(w)Ammann-BeenkerL=25}
\end{figure*}

\begin{figure*}[tb]
\centering
\includegraphics[width=\linewidth]{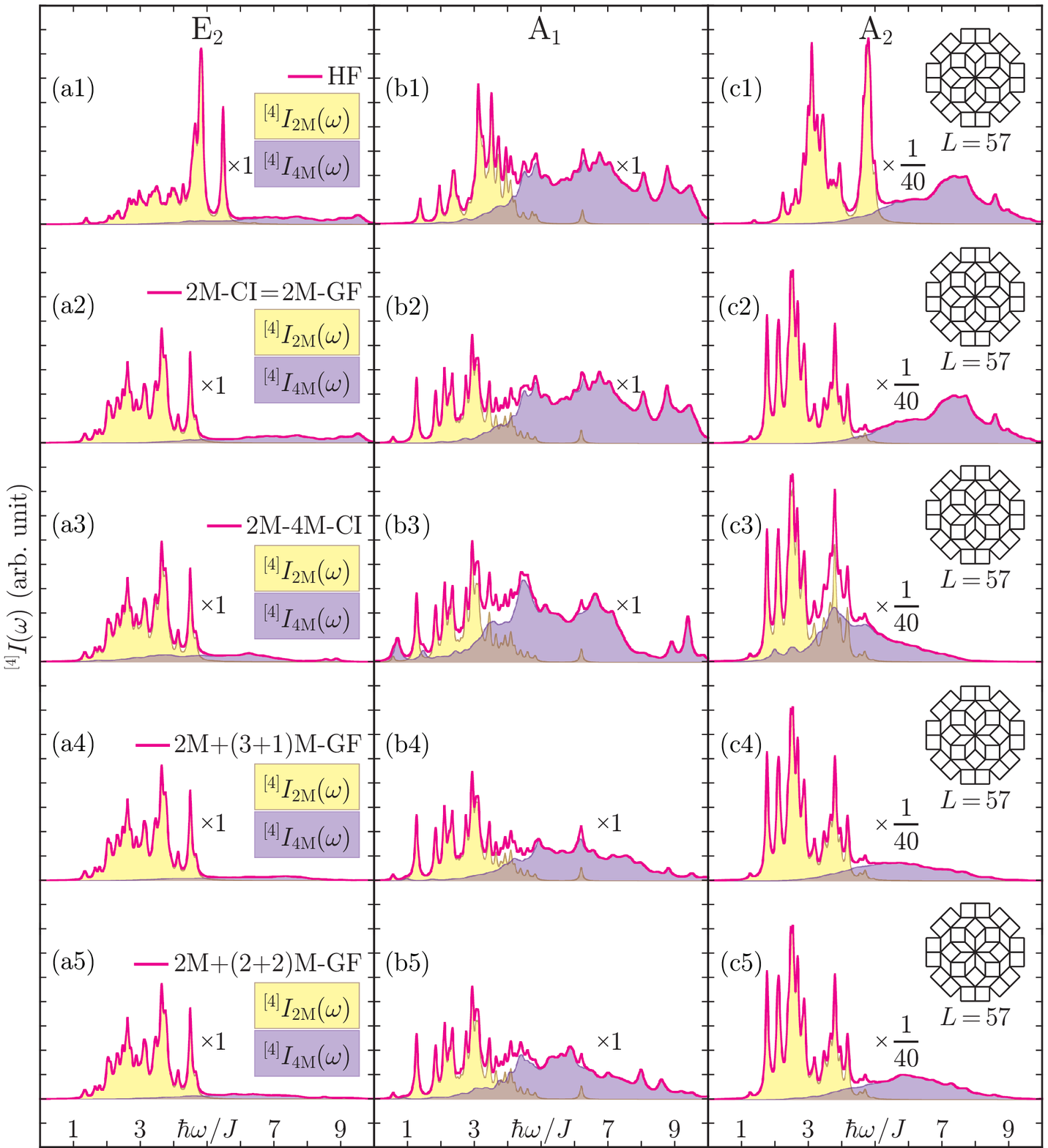}
\caption{The same as Fig. \ref{F:[4]I(w)Ammann-BeenkerL=25} for
         the $L=57$ 2D Ammann-Beenker lattice of ${\mathbf{C}}_{8\mathrm{v}}$ point symmetry
         without any exact solution available.}
\label{F:[4]I(w)Ammann-BeenkerL=57}
\end{figure*}

   We show the cluster-size and calculational-scheme dependences of Raman spectra in full detail
for both the Penrose (Figs. \ref{F:[4]I(w)PenroseL=16} to \ref{F:[4]I(w)PenroseL=56}) and
Ammann-Beenker (Figs. \ref{F:[4]I(w)Ammann-BeenkerL=25} and \ref{F:[4]I(w)Ammann-BeenkerL=57})
lattices,
intending to demonstrate the superiority of the 2M-4M-CI scheme over the others especially in
evaluating 4M-mediated scattering intensities.

   For the $L=16$ Penrose (Fig. \ref{F:[4]I(w)PenroseL=16}) and $L=25$ Ammann-Beenker
(Fig. \ref{F:[4]I(w)Ammann-BeenkerL=25}) clusters, all the perturbative and variational
calculations are compared with the exact solutions.
The Hartree-Fock approximation cannot reproduce any of the major peak positions.
The 2M-CI formulation, which is equivalent to the 2M Bethe-Salpeter equation, is good at
reproducing the low-energy peaks essentially of 2M character but poor in describing
higher-energy spectral weight.
The 2M-4M-CI formulation overcomes this drawback owing to its precise evaluation of
4M-mediated scattering intensities.
This is not the {\color{white} blank}
\newpage
\vspace*{-24pt}
\noindent
case with any of the GF findings.
Indeed both $(3+1)\mathrm{M}$ and $(2+2)\mathrm{M}$ approximate evaluations of the 4M Raman
correlation function ${}^{[p]}\mathcal{G}_{4\mathrm{M}}(t)$ lead to a reasonable reduction of
the high-energy excess spectral weight, but neither of them gives such a satisfactory description
of the intermediate-energy scattering bands of 2M-4M-mixed character as to be obtained through
the 2M-4M-CI scheme.
Such observations are common to both the Penrose and Ammann-Beenker lattices and hold good for
all the symmetry species but $\mathrm{A}_1$.
Any excess 4M scattering intensity for the $\mathrm{A}_1$ symmetry species present in
the SW calculations should rather be ascribed to the up-to-$O(S^0)$ expansion of $\mathcal{R}$
than otherwise.
Note that the exact diagonalization calculation of the Raman intensity \eqref{E:[p]I(w)}
is performed with the spin operator expressions \eqref{E:[2]R} and \eqref{E:[4]R},
whereas any SW calculation, whether in the GF description \eqref{E:[p]I2lM(w)GF}
or through the CI scheme \eqref{E:[p]I2lM(w)CI},
is performed with the up-to-$O(S^0)$ approximate vertices \eqref{E:RuptoO(S^0)} written
in terms of the magnon operators \eqref{E:R2M=W(1)toW(4)} and \eqref{E:R4M=X(1)toX(9)},
with the aim of revealing the scenario of inelastic light scattering.

   With increasing system size, the spectral shape and/or density change in a complicated manner,
but the balance in intensity sharing between 2M and 4M scatterings remains qualitatively the same.
The energy ranges in which 2M scattering intensities distribute for the three symmetry species
each remain almost unchanged from those of the smallest cluster calculated and this is essentially
the case with 4M scattering intensities as well.
Indeed, neither of the range of the eigenvalue distribution nor the specific heat curve is
sensitive to the system size for the $L\gtrsim 31$ Penrose and $L\gtrsim 33$ Ammann-Beenker lattices,
as was already shown in Fig. \ref{F:EV&C}.
Eigenvalues corresponding to the one or two highest coordination numbers have no serious effect
on bulk properties.
\clearpage
\clearpage
\clearpage
\clearpage
\clearpage

\clearpage

\addtolength{\footskip}{217.5mm}